\documentclass[rmp,reprint]{revtex4-2}
\usepackage{hyperref}
\usepackage{graphicx}
\usepackage{bm}
\usepackage{amsmath}
\usepackage{amssymb}
\usepackage{tikz}
\def\r{{\bm r}}
\def\x{{\bm x}}

\def\z{{\bm z}}
\def\k{{\bm k}}
\def\B{{\bm B}}
\def\E{{\bm E}}
\def\j{{\bm j}}
\def\v{{\bm v}}
\def\btheta{{\bm\theta}}

\def\hp{\hat\phi}
\def\tf{\tilde f}
\def\tL{\tilde L}
\def\BOmega{{\bm \Omega}}
\def\energy{{\cal E}}
\def\sech{{\rm sech}}
\def\ket#1{|#1\rangle}
\def\bra#1{\langle#1}
\def\a{{\bf a}}
\def\M{{\bf M}}
\newdimen\partwork
\ifx\affiliation\undefined 
\def\affiliation#1{\date{\normalsize #1\\ \today}}
\usepackage[margin=1in]{geometry}
\usepackage[sort&compress,numbers]{natbib}
\def\pamper{}\def\pampert{}
\def\partwidth#1{\includegraphics[width=#1\hsize]}
\else 
\def\twocoltrue{}
\def\pamper{\\&}\def\pampert{\\&\times{}}
\def\partwidth#1{\partwork=1.6\hsize\includegraphics[width=#1\partwork]}
\fi
\renewcommand{\selectlanguage}[1]{} 

\begin{document}

\title{Kinetic Solitary Electrostatic Structures in Collisionless
  Plasma: Phase-Space Holes}
\author{I H Hutchinson}

\affiliation{Plasma Science and Fusion Center,\\ Massachusetts Institute
  of Technology,\\ Cambridge, MA 02139, USA}

\ifx\altaffiliation\undefined\maketitle\fi 
\begin{abstract}
  The physics of isolated plasma potential structures sustained by a
  deficit of phase-space density on trapped orbits, commonly known as
  electron or ion holes, is reviewed. The principles of their
  equilibria are explained and illustrated, and contrasted with
  solitons. A literature review mostly prior to 2016 highlights
  the key historical developments of the field. Progress since,
  especially in hole acceleration, stability, and multi-dimensional
  effects, is summarized in more detail.
\end{abstract}
\ifx\altaffiliation\undefined\else\maketitle\fi  

\tableofcontents

\section{Introduction}
The aim of this review is to introduce and survey the physics of
long-lived nonlinear electrostatic plasma structures in which the
electric potentials are supported self-consistently by non-thermal
particle velocity distributions. It is now realized that these
phenomena occur very widely, often as the end state of kinetic
instabilities. They are coherent rather than turbulent, and persist
for long durations relative to typical instability timescales.

Modern satellite measurements observe solitary potential structures
almost everywhere in space plasmas; and scattering from them may
sometimes cause particle energization and precipitation.  Because of
their short spatial extent, their often rapid motion past an
observing satellite produces an electric field pulse of wide bandwidth
in the frequency domain. Pulsed nonlinear low density laboratory
experiments can also generate them.

There is a wide theoretical variety of such phenomena, which would
soon overwhelm the compass of any single review. Therefore to limit
the scope, I have adopted boundaries to the domain of interest that
are indicated in the paper's title.

``Kinetic'' limits the scope to phenomena that need a comprehensive
plasma description beyond the widely-used fluid or Maxwellian
approximations. Complicated velocity distributions of the particles,
having depleted regions of phase-space (so called holes) are intrinsic
to the present topic. Fluid solitons such as ion acoustic solitons
that are describable by purely fluid nonlinear equations have an
enormous literature and various book-length reviews already (see e.g.\
\citealp{Lamb1980}, \citealp{Dodd1982} and more recently
\citealp{Drazin1989}, \citealp{Infeld2000}). An elementary discussion
aiming to clarify these omitted phenomena from a kinetic perspective
is included for comparison with holes, but the soliton literature is
not reviewed.

``Solitary'' means, nevertheless, that the focus is on aperiodic
structures, not extended periodic waves. When periodic plasma waves have
substantial amplitudes, important nonlinear effects occur, such as
particle trapping and wave-wave coupling. There is an extensive
literature concerning periodic nonlinear waves that is not addressed
here. It is now increasingly appreciated, however, that many important
coherent non-linear plasma phenomena can only readily be understood in
the space domain $\z$, not the Fourier domain $\k$. More simply put,
our present concern is potential structures that vary only over a
limited spatial extent and connect to external regions that are
essentially uniform. These localized objects can be considered to be
isolated positive or negative potential humps, which will be referred
to as electron or ion holes and are our main focus, but the expression
arguably applies also to isolated double layers, referring to transitions
between two different externally uniform potentials. They are a
subsidiary interest and will be mentioned but not comprehensively
reviewed.

``Electrostatic'' phenomena predominate in both theory and
observation. However, a magnetic field $\B$ is often important for the
existence and stability of the holes, and while most analyses regard
the magnetic field as a uniform background, the presence of moving
holes can also induce measurable local perturbations to $\B$, which
will briefly be addressed.

Many plasmas are to an excellent approximation ``Collisionless'',
which is essential for the non-thermal structures to persist. The
equation governing the particles is then Vlasov's equation, which
expresses the invariance of the velocity distribution $f(\v)$ along
particle orbits.

Calling the phenomena ``Structures'' alludes to them being of a
persistent shape and size that varies only slowly relative to their
particle transit times. (However the literature also widely uses
instead the term solitary ``wave'', hence the acronym ESW for
electrostatic solitary waves, though the alternative phrase ``Time
Domain Structure'' TDS is arguably more precise.) In the rest frame of
the solitary structure, there are then generally considered to be
three ordered timescales: the transit or bounce time of the particles,
which is much shorter than the structure's evolution or acceleration
time, which is much shorter than the particle collision time (on which
the structure decays collisionally). Time variations of the potential
that are as fast as the particle transit or bounce frequency cannot be
ignored when analyzing the \emph{stability} of holes, but the
\emph{equilibria} are generally taken to be almost steady on the
transit time, which means kinetic plus potential energy is an
approximate constant of the motion, and in a one-dimensional situation
the equilibrium distribution is a function only of energy in the rest
frame of the structure.

\begin{figure}[ht]
  \includegraphics[width=0.48\hsize]{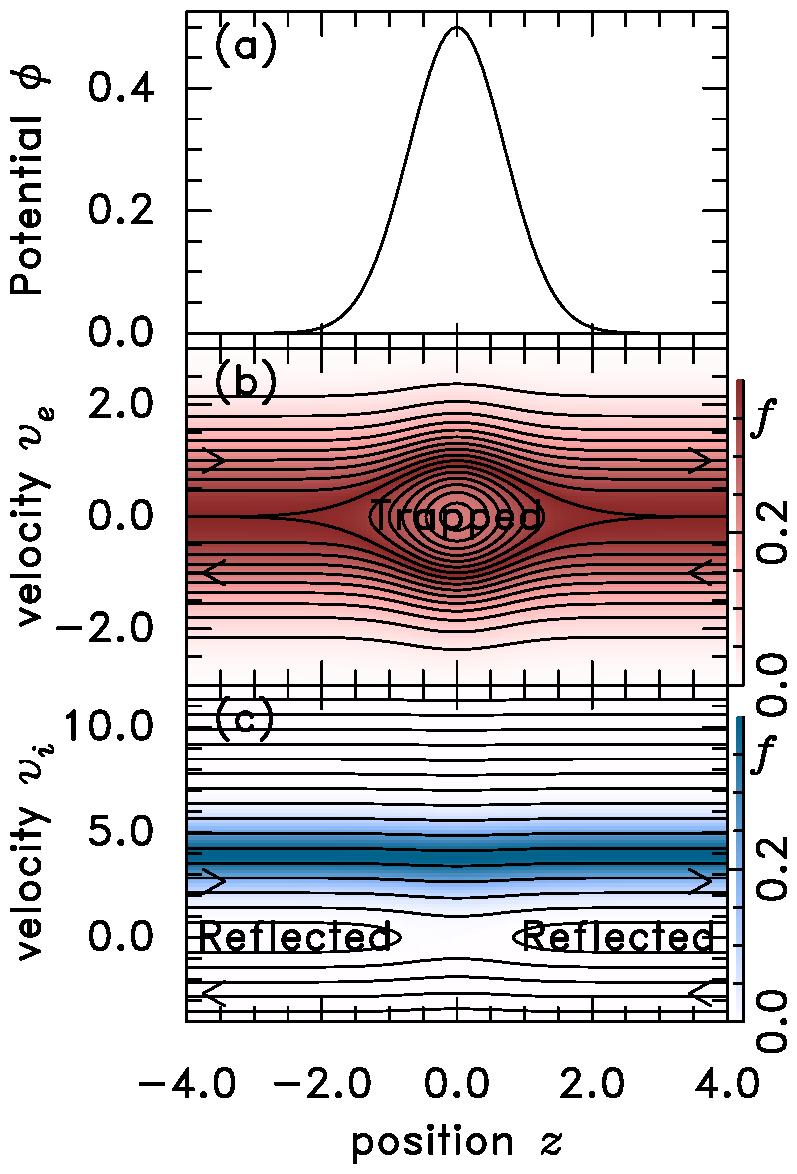}\hskip-1.5em(i)\hskip1em
  \includegraphics[width=0.48\hsize]{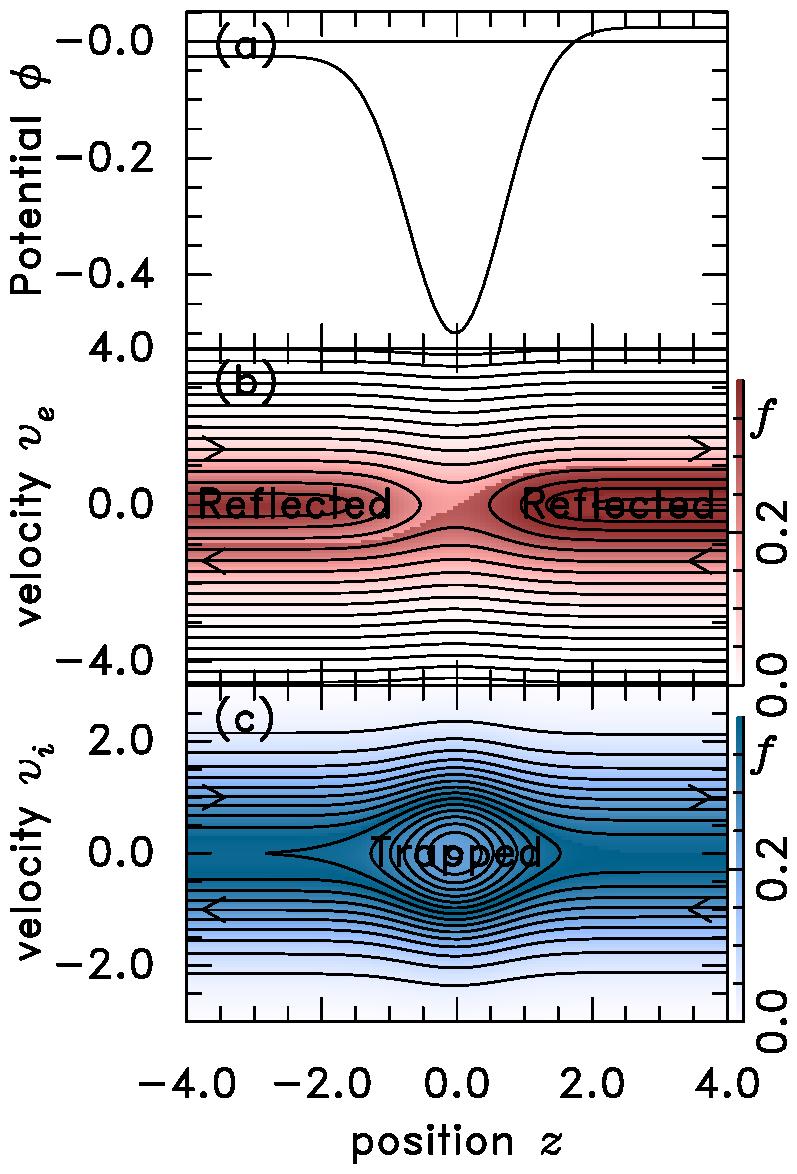}\hskip-1.5em(ii)
  \caption{Illustrative one dimensional electron hole (i), and ion
    hole (ii), showing (a) the assumed potential, (b) the electron
    phase space, and (c) the ion phase space. The electron and ion
    velocities are normalized respectively to their thermal velocities
    $\sqrt{T/m_e}$ and $\sqrt{T/m_i}$. On the contours of constant
    energy the distribution function $f(v)$ has a constant value
    indicated by the color. Particle orbits coincide with the
    contours. In (i) the external ion Maxwellian has a mean velocity shift of
    4. In (ii) there is no ion shift but a small electron shift, plus
    an asymmetry in $\phi$ between $z\to \pm \infty$.  \label{symholes}}
\end{figure}
By way of orientation, Figure \ref{symholes} illustrates the particle
orbits in one-dimensional phase-space (the $z,v$ plane) that arise
from a prescribed steady potential $\phi(z)$. In Fig.\
\ref{symholes}(i) producing a positive potential requires $f_e(v)$ in
the trapped region to have a depression relative to its distant value
at $v=0$. The acceleration induced by the gradients of the potential
cause the trapped orbits to circulate clockwise around a stagnation
point, forming a phase-space vortex. That depressed phase-space
density vortex is the electron ``hole''. Meanwhile, in this
illustration, the mean ion velocity is larger ($4v_{ti}$) in the rest
frame of the hole than its thermal speed
$v_{ti}=\sqrt{T/m_i}$. Therefore negligible numbers of ions are on
reflected orbits, but the density of ions at $z=0$ is slightly
enhanced near $z=0$ by the orbit expansion. When the ion shift is
large enough that the populated ion orbits are effectively straight
horizontal lines, a hole equilibrium would be well approximated by
regarding the ions as uniform neutralizing background. In Fig.\
\ref{symholes}(ii), by contrast, the ions are taken as unshifted in
the hole frame, sustaining the negative potential by their trapped
phase-space-density depression.  In this more complicated illustrative
case, the external electron distribution is slightly shifted, by
$-0.3 v_{te}$, giving rise to important asymmetric electron reflection
from the negative potential peak, and accompanying discontinuities in
$f_e(v)$. Such reflection generally induces a potential difference
across the hole, with important imbalanced forces causing
acceleration, which will be discussed later in detail.

\begin{figure}[htp]
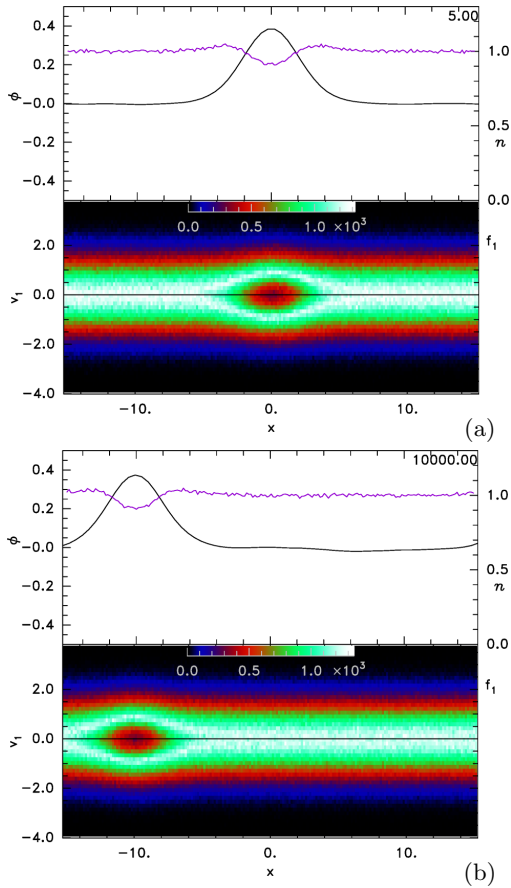

  {\partwidth{0.49}{phaseplot10}\hskip-2 em (a)}
  {\partwidth{0.49}{phaseplot10k}\hskip-2 em (b)}
  \caption{Two states of a 1-D particle in cell simulation showing
    negligible amplitude decay of a pre-formed electron hole:
    initially (a), after time 10,000$\omega_{pe}^{-1}$ (b). Upper
    frame: potential and density; lower frame: contours of phase-space
    density.\label{holepersistence}}
\end{figure}

To illustrate the long-lasting persistence of holes,
Fig. \ref{holepersistence} plots normalized $\phi$, $n_e$, and color
contours of $f_e(x,v)$ from a fully self-consistent particle in cell
(PIC) simulation of a pre-formed pure electron hole with 600,000
computational particles per cell. A tiny amplitude decay rate of only
approximately $5\times10^{-5}\omega_{pe}$ occurs. It is attributable
to the numerical collisionality arising from particle discreteness
noise.  Electron holes can persist stably with negligible evolution
other than slow spatial drift for immense durations in sufficiently
quiescent plasmas.

A guide to the contents of the article is as follows.  Section
\ref{1d} sets forth the key equations governing these localized
Vlasov-Poisson structures, and explains the scaling of space and time
to give the appropriate dimensionless variables in which it is most
convenient to work. Elementary examples of self-consistent
one-dimensional electron holes are discussed, illustrating the
physical principles that allow steady solitary structures to
exist. These principles were a major emphasis of a tutorial review by
the author \citet{Hutchinson2017}, originally presented in 2016. So
introductory development is curtailed in the present review and the
reader should consult that tutorial paper for a more comprehensive
introduction to the principles of the field, and its prospects at that
time. For the same reason, characteristic formation mechanisms of
electron holes, discussed in \citet{Hutchinson2017}, are omitted here,
but some videos of simulations of hole formation may be found at
\url{https://www.youtube.com/@ihhutchinson/playlists} and links in the
Supplementary Material.

Section \ref{solitonsec} shows how the pseudo-potential approach is
used to calculate solitary structures when the charge density as a
function of potential is known. Fluid solitons are compared with
electron and ion holes, and the important distinctions are explained. 

The limited length of the earlier tutorial obviated a thorough
critical review of prior literature. Therefore Section
\ref{Historical} seeks to expand substantially upon the historical
perspective from the field's beginnings in the 1950s. It aims to cite
the most influential papers, note why they were influential, and also
sometimes to provide a critical assessment, from today's perspective,
of their strengths and weaknesses. This approach risks not only
inadvertent omissions, but also apparent disrespect toward my
predecessors in the field, because it is not profitable to attempt to
cite every paper that bears on electron and ion hole theory. So let me
apologize up front for any unwelcome omissions or remarks concerning
the pioneering work of many creative authors that opened up the topic;
and simply plead that I am deeply conscious of their contributions,
and have attempted to state as fairly and respectfully as I can my
understanding of what the contributions are.

The subsequent sections cover largely conceptual developments in
electron and ion hole physics since 2016, and are organized topically
rather than historically. The idea of hole \emph{kinematics}, principally
momentum conservation, Section \ref{Kinematics}, has seen great
development since then. It addresses the behavior of a hole as a
robust compound entity that is able to accelerate in response to
imbalance of forces, while retaining approximately constant potential
shape in its rest frame. The forces treated within the kinematic
discussion can also induce asymmetry in the hole potential, most
notably a potential drop across the hole, as discussed in Section
\ref{asymmetry}.  The kinematic perspective underlies a rigorous
development of hole stability (Section \ref{Stability}) to motions
along the direction of the predominant electric field and particle
motion: generally parallel to the magnetic field. In addition to
uniform one-dimensional acceleration, holes can often be unstable to
parallel shifts that vary sinusoidally in the plane perpendicular to
the magnetic field. This phenomenon kinks the hole, and is called the
``transverse instability''. Recent stability analysis has revealed its
true underlying mechanisms. Holes are observed usually to be localized
in the transverse directions as well as the parallel direction. The
strength of the background magnetic field is critical in determining
stability and instability. Often, holes initially well approximated as
one-dimensional break up by the transverse instability into
multidimensional structures that are usually oblate, but sometimes even
prolate. Their resulting localized equilibria, once the break up is
complete, require multidimensional equilibrium calculations which are
the topic of Section \ref{Multidimensional}.

It is extremely difficult to observe electron holes in laboratory
experiments. The main problem is that electron holes have
characteristic extent of a few Debye lengths, and in moderate density
lab plasmas, the Debye length is uncomfortably short (typically a few tens
of micrometers). Diagnostics to measure the plasma potential with
sufficient spatial resolution, which so far are limited to electric
Langmuir type probes, have only rarely been possible.  Another problem
is the great difficulty of producing truly collisionless plasmas
either because of limitations of finite connection lengths, or high
Coulomb or neutral collision frequency even in toroidal
configurations. Nevertheless a handful of laboratory observations will
briefly be described in the historical review.

By contrast, space plasma measurements occur in deeply collisionless
plasmas with Debye lengths from tens to thousands of meters. And
satellites routinely encounter electron and ion holes. In the mid 1990s
diagnostic sampling speeds began to be sufficient to resolve in time (and
thereby space) the electric fields of holes, along a single track
through the hole. With the availability of data now from coordinated
multi-satellite missions, near-simultaneous multiple-point measurements
within a single hole have become possible. This enables direct
diagnosis of the three dimensional extent of holes, and provides
unambiguous and precise measurements of their velocity. At the same time (but
with much slower time resolution) many satellites obtain comprehensive
measurements of the background ion and electron velocity distribution
functions. These new measurements provide rich opportunities to
confront the theories of electron and ion holes and to use them to
help interpret the physics of the solar wind, the Earth's
magnetosphere, and many other heliospherical plasma regions.

It would be inappropriate in this paper to review extensively the
diagnostic instrumental techniques themselves, whether on Earth or in
space, but substantial effort has been made in the relevant sections
to relate the theoretical developments to observations that bear on
them. We are beginning to see very encouraging signs of agreement
between theory and experiment in critical tests of our understanding
of plasma holes.

Another important contribution, since the earliest years, to
understanding electron holes and indeed a host of other non-linear
phenomena, has come from kinetic plasma simulation. One dimensional
particle in cell and continuum Vlasov simulations have long given the
clearest picture of how electron holes arise as the natural result of
nonlinear trapping in two-stream instabilities, followed by the
merging of smaller holes into fewer or even just one deeper or longer
hole. Multidimensional simulations around the year 2000 were very
influential in showing the importance and observing the threshold of
the transverse instability, although it was not until much more
recently that rigorous analytical calculations have shown good
quantitative agreement with simulation in the frequencies,
wavelengths, and growth rates, and thereby identified the driving
mechanism of these instabilities. Numerical simulation
\emph{techniques} are not reviewed here, but the results of
simulations and their comparison with the phenomena will be discussed
within the previously outlined sections.

\section{Principles and formulation}
\label{1d}

\subsection{Governing Equations}
The Vlasov equation governs the particle velocity distribution $f(\v)$
in collisionless plasmas. In three dimensions it is written
\begin{equation}
  \label{Vlasov}
  {\partial f\over \partial t} + \v\cdot{\partial f\over \partial \x}
  +{q\over m}(\E+\v\wedge\B)\cdot{\partial f\over \partial \v}=0,
\end{equation}
and can be considered Boltzmann's equation including the Lorentz
acceleration arising from electromagnetic fields acting on particle
charge $q$ and mass $m$, but with the collision term on the right hand
side set to zero. A recent book about the Vlasov equation
\citet*{Bertrand2019} provides a comprehensive introduction and
reviews linear and nonlinear treatments. It also includes some
discussion of phase-space holes. Since $\v$ is the time rate of change
of position $\x$ and the acceleration ${q\over m}(\E+\v\wedge\B)$ is
the time rate of change of velocity $\v$, the left hand side is the
total convective derivative of $f$ along a particle's six-dimensional
phase-space $(\x,\v)$ orbit; and Vlasov's equation states that $f$
remains constant along phase-space orbits.

We shall adopt the usual convention in plasma physics to take
cartesian coordinates in which the background magnetic field $\B$ is
in the $\z$ direction. Particles then move freely along $z$
but are constrained to circular Larmor orbits in the transverse directions
$x$ and $y$. In strong enough magnetic fields, or when the plasma is
uniform in the transverse directions, the variation can be taken as
one-dimensional along $z$. Therefore in a uniform background magnetic
field we will consider $z$ to be the spatial coordinate of the
one-dimensional Vlasov equation
\begin{equation}
  \label{1dVlasov}
  {\partial f\over \partial t}+v {\partial f\over \partial z}
  -{q\over m} {d \phi\over dz} {\partial f\over \partial v}=0,
\end{equation}
where the electric force is expressed in terms of the electrostatic
potential $\phi$ and there is no component of the magnetic force along
$z$. The one-dimensional distribution $f(v)=f_\parallel(v_\parallel)\equiv\int f(\v) d^2\v_\perp$ is
then implied.

The potential is governed by Poisson's equation
\begin{equation}
  \label{Poisson}
  \nabla\cdot\E=-\nabla^2\phi={\rho\over\epsilon_0}={1\over\epsilon_0}\sum_{species} qn,
\end{equation}
where $\rho$ is the total charge density determined by the sum over
charge species of their charge densities $qn$. The particle densities
$n$ are the integrals over all velocities of the respective species
velocity distributions:
\begin{equation}
  \label{densitydef}
  n = \int f(\v)d^3\v. 
\end{equation}
The one-dimensional approximation reduces equations (\ref{Poisson})
and (\ref{densitydef}) to scalar differential and integral
expressions.  Equations (\ref{Vlasov}), (\ref{Poisson}) and
(\ref{densitydef}) or their one-dimensional forms constitute the time
dependent kinetic mathematical description of the collisionless
plasma, and must be solved self-consistently. 

No plasma in nature \emph{exactly} obeys the Vlasov equation. At
sufficiently fine scales in velocity or position, collisional or other
non-ideal effects become significant. Their influence is normally
represented by replacing the right hand side of equation \ref{Vlasov}
with a non-zero ``collision term''. Nevertheless, a pure Vlasov
treatment is an excellent approximation for almost all the purposes of
this review. A phase-space density modulation usually progresses with
time, governed by the Vlasov equation, to finer and finer scales by
``phase-mixing'', until dissipative effects (represented by the
collision term which we will not discuss) take over. This process
undergirds Landau damping, for example, which is taught in most plasma
textbooks. The nonlinear aspects ignoring collisions have been
analyzed in great mathematical detail, most famously by Villani. A
moderately approachable tutorial for his ideas is
\citet{Villani2014}. Landau damping questions in the present context
will be revisited briefly in section \ref{LandauD}.

\subsection{Non-dimensionalization}

A characteristic energy, effectively temperature measured in energy
units (so as to avoid repetitive factors of Boltzmann's constant), is
taken as $T_0$ (the same regardless of species).  If we measure
potential in units of $T_0$ then its dimensionless form $\hat \phi$ is
such that $\phi =\hat \phi T_0/e$ where $-e$ is the charge on an
electron. The velocity is normalized to the thermal speed; so
$v=\hat v \sqrt{T_0/m} = \hat v v_t$. The characteristic length over
which potential varies in a plasma is the Debye length, $\lambda_D$,
such that $\lambda_D^2=\epsilon_0T_0/n_{e\infty}e^2$, using the distant
background electron density $n_{e\infty}$. This is the natural length
normalization unit: $\x=\hat \x \lambda_D$. Substituting these into
equation (\ref{Vlasov}) it becomes for each particle species
\begin{equation}
  \label{Vscaled}
  {\partial f\over \partial \hat t}
  +\hat\v\cdot{\partial f\over \partial \hat\x}
  +{q\over e}\left[-\hat\nabla\hat\phi+\hat\v\wedge\hat\BOmega\right]
  \cdot {\partial f\over \partial \hat\v}=0,
\end{equation}
where $\hat t = t\sqrt{T_0/m}/\lambda_D $ is the normalized time and
$\hat\BOmega=(e\B/m)\lambda_D/\sqrt{T_0/m}$ is the normalized cyclotron
frequency (vector). This time scaling is written
$t=\hat t \lambda_D/v_t =\hat t/\omega_p$, where $\omega_p$ is the
plasma frequency:
$\omega_p^2=(T_0/m)/\lambda_D^2= n_{e\infty} e^2/\epsilon_0 m$. It is
important to note that the time scaling is \emph{different} for
different species because by these definitions
$\omega_{pi}^2=\omega_{pe}^2m_e/m_i$, even though the length scaling
is the same.

The dimensionless particle density $\hat n$ is normalized to the
(uniform) background density $n_\infty$ giving $n = \hat n n_\infty$.
The normalized distribution function $\hat f$ is then such that $\hat n=
\int f d^3\v/n_\infty=\int\hat f d^3\hat\v=\int \hat f
d^3\v/(T_0/m)^{3/2}$, so that $\hat f(\hat\v)=[(T_0/m)^{3/2}/n_\infty]f(\v)$
for each species. The one-dimensional distribution has instead
$\hat f(v) = [(T_0/m)^{1/2}/n_\infty]f(v)$. 
Poisson's equation (\ref{Poisson}) then becomes
\begin{equation}
  \label{Pscaled}
  -\hat \nabla^2\hat\phi=\sum_{species} \left({n_\infty q\over
    n_{e\infty}e}\hat n\right).
\end{equation}
Evidently for electrons $n_\infty q/n_{e\infty}e=-1$.  For a single
ion species, distant quasineutrality implies that
$n_\infty q/n_{e\infty}e=1$ (regardless of ion charge magnitude
$Z=q/e$). We can therefore regard these ratios as the normalized
charge $\hat q$ equal to $\pm1$. We shall not worry in our treatment about the
possibility that a dependence of $q$ on $Z$ (absent in $\hat q$)
remains in the normalized Vlasov equation (\ref{Vscaled}), but it
ought to be accommodated when necessary.

To avoid encumbering the notation, the hats on dimensionless
quantities are henceforth dropped and it is understood that we are
working in scaled units, unless otherwise indicated.

\subsection{Constancy on orbits}

Usually Vlasov's differential equation need not be solved
explicitly. Instead, its property of ensuring $f$ is constant along
orbits enables us, if $\phi$ is known for all prior time and space, to
obtain $f(\v)$ at any position and time by solving for the phase-space
orbit and following it backwards in time until some boundary condition
or initial condition is encountered that prescribes the value of $f$
on it. And this approach is almost universal in electron and ion hole
theory. However, the loop of equations consisting of $f$ being
given by Vlasov using $\phi(\x,t)$, $\phi$ given by Poisson using
$n(\x)$, $n$ being given by $f$, will not be satisfied for arbitrarily
chosen $\phi$ and $f$ conditions. A process is required to determine a
self-consistent set of parameters for which the equations are all satisfied.

\subsection{Steady One-Dimensional Equilibria}
\label{1dequil}

For the rest of this section we limit ourselves to one-dimensional
discussion of structures that are steady in their rest frame. An
immediate consequence of steadiness is that particle energy
$\energy(z,v)={1\over2}v^2+q\phi(z)$ (dimensionless units) is constant
along orbits, which then means $f(z,v)$ is a function only of
$\energy$ and (possibly) sign of $v$ (written $\sigma_v$), not of $z$
and $v$ separately. As a shorthand, we will sometimes write the
distribution function with an energy argument as $f(\energy)$ meaning
$f(z,\sigma_v\sqrt{2[\energy-q\phi(z)]})$. It follows that the density
$n(z)=\int f dv$ is a function of potential $n(z)=n(\phi(z))$, and
Poisson's equation takes on the special form
$-{d^2\phi\over dz^2}=\rho(\phi)$, where $\rho=\sum qn(\phi)$. The
standard technique to solve such a differential equation where the
inhomogeneous term is a function only of the dependent (not the
independent) variable, is to multiply by a factor $d\phi\over dz$ and
integrate, giving $-{1\over 2}(d\phi/dz)^2=\int \rho d\phi +C$. The
right hand side of this equation is often called the pseudo-potential,
written $V$. The integral of the inverse of square root of this
equation gives the implicit solution
\begin{equation}
  \label{Implicit}
    z(\phi)=\int {d\phi\over\sqrt{-2V} } + const.
\end{equation}
This is the obvious way to proceed if the $n(\phi)$ are known, which
they are if $f(\energy)$ is prescribed. It is also applicable to
fluid plasma equations where only densities $n(\phi)$ are considered,
as will be discussed in the next section.

To determine self-consistent steady kinetic equilibria, however, there
are two alternative ways proceed. We can either assume that
$f(\energy)$ is prescribed and find the self-consistent $\phi(z)$,
which is the inverse of the $z(\phi)$ just discussed, equation (\ref{Implicit}); or we can assume that the potential variation
$\phi(z)$ is prescribed and find the $f(\energy)$ (satisfying the
Vlasov and Poisson equations) that is required for self-consistency.

These two alternatives were introduced by \citet*{Bernstein1957}, who
called them the ``differential equation'' and ``integral equation''
approaches, referring to the type of equation that needs to be solved.
In the subsequent literature the first is often called the
pseudo-potential, classical-potential, or Sagdeev approach, and the
second the BGK approach. The insight \citeauthor{Bernstein1957}
brought was that although the distribution $f$ of passing particles
(with orbits extending to distant boundaries) is determined by
boundary conditions and constancy on orbits, the distribution on
trapped orbits (that cannot be tracked back to a boundary) is
determined by initial conditions of hole formation, and can for
equilibrium purposes be considered to be a free choice. Moreover, the
required self-consistent choice can be found for an essentially
arbitrarily prescribed potential $\phi(z)$ sequentially (over a
progressive series of potential maxima and minima). It requires that
Poisson's equation be satisfied by a combination of the known passing
distribution $f(\energy)$ for $\energy\ge 0$ plus initially unknown
distributions of trapped particles ($\energy < 0$) that can be found
by sequentially solving integral equations. The result is called a BGK
wave or mode.

Specifically, for a positive solitary potential structure $\phi(z)$,
such as is illustrated in Fig.\ \ref{symholes}(i), of maximum height
$\phi=\psi$, prescribed in a distant uniform background having
$\phi=0$, the density of the passing (untrapped) electrons is
$n_p(\phi)=\int_{\energy>0} f(\energy) dv$, where $f$ is known from
boundary conditions, while the trapped electron density is
$n_t(\phi)=\int_{-\phi<\energy<0} f(\energy) dv$ where $f$ is
unknown. The ion density $n(\phi)$ is entirely prescribed by boundary
conditions since no ion orbits are trapped. And the divergence term
$d^2\phi/dz^2$ is prescribed (and a function of $\phi$). Consequently
Poisson's equation $-{d^2\phi\over dz^2}=q_e(n_p+n_t-n_i)$ determines
  the required value of the trapped density $n_t(\phi)$. It will be
  correctly obtained if and only if the trapped $f$ satisfies an
  integral equation:
\begin{equation}
  \label{IntEq}
  n_t(\phi)=\int_{-\phi<\energy<0}f
  dv=2\int_{-\phi}^0f(\energy){d\energy\over \sqrt{2(\energy+\phi)}},  
\end{equation}
accounting for both signs of $v$ and trapped symmetry.
This is a form of Abel's integral equation, whose well known solution
for $-\psi < \energy < 0$ is
\begin{equation}
  \label{Abel}
  f(\energy)=f(0)+{1\over  \sqrt{2}\,\pi}
  \int_0^{-\energy} {dn_t\over d\phi} {d\phi \over \sqrt{-\energy-\phi}}.
\end{equation}
Obviously, the same argument applies for negative potential
structures, that is ion holes, exchanging the identities of electrons
and ions and the resulting charge signs. For an extended BGK spatial
profile of potential maxima and minima, the treatment alternates
between determining reflected electrons and reflected ions.

The two different starting points for obtaining self-consistent hole
equilibria --- the differential aka pseudo-potential, vs.\ the
integral aka BGK approaches --- have different practical and
mathematical weaknesses and strengths. But both require an extra
constraint to avoid singularities in trapped
$df/dv|_{\energy=0}$ at the trapping boundary $\energy=0$. Such
mathematical singularities are physically implausible, except as an
approximation of uncertain applicability. The differential approach
can easily specify an $f(v)$ without such singularities, but then
finds that, when $f$'s relative shape at $-\psi<\energy <0$ is fixed
(which is commonly assumed), the amplitude $\psi$ is set by the slope
$df/dv|_{\energy=0}$ chosen, for fixed passing and repelled-species
distributions. Conversely, the integral approach will encounter
singularities in the solved-for trapped $df/dv|_{\energy=0}$ unless
the potential near $\phi=0$ ($\phi(z)$ as $|z|\to \infty$) has the
form of an exponential decay with characteristic length equal to the
background plasma (generalized) Debye shielding length $\lambda$ such that
\begin{equation}\label{shielding}
  {1\over \lambda^2}=\left.dn_e\over d\phi\right|_{0}
+\left.dn_i\over d\phi\right|_{0}.
\end{equation}
These constraints are effectively the same, as is further explained,
for example, in references \citep{Hutchinson2016} section II and
\citep{Hutchinson2023} section III. However, this fact is subtle and
is not widely recognized; and some authors fallaciously maintain that
only the pseudo-potential approach is appropriate (see e.g.\ \citet{Hutchinson2023a,Schamel2023b}). Equation \ref{shielding}, though
necessary, is not always sufficient to avoid singularity.

Unfortunately the
situation is confused still more by an erroneous assertion about ion
holes \citep{Schamel1980}, that they cannot exist unless the species
temperature ratio $T_e/T_i=\theta\gtrsim 3.5$. This assertion has been
widely cited and repeated
(\cite{Bujarbarua1981,Schamel1982,Hudson1983,Pecseli1984,Johnsen1987,Buchanan1993,Griessmeier2002,Eliasson2006a,Schamel2018,Wang2021}),
despite published theoretical counter-examples such as
\citep{Chen2004}.  The correct form arising from the requirement that
$\lambda^2$ be positive is that $\theta<3.5$ when the value of
$dn_i/d\phi$ takes its most negative value, which for a shifted
passing Maxwellian is -0.285($\simeq$-1/3.5) at a shift velocity of
2.13. But in any case that shift places the hole velocity far out in
the wing of the background distribution, where the phase-space density
is only one tenth of its Maxwellian peak, and the amplitude of any ion
hole would be severely limited. It does not apply at all to more
typical holes in the bulk of the ion distribution. In short, no
absolute temperature ratio requirement applies to ion (or electron)
holes.

\subsection{Illustrative BGK equilibria}

\begin{figure}\center
  \includegraphics[height=.46\hsize]{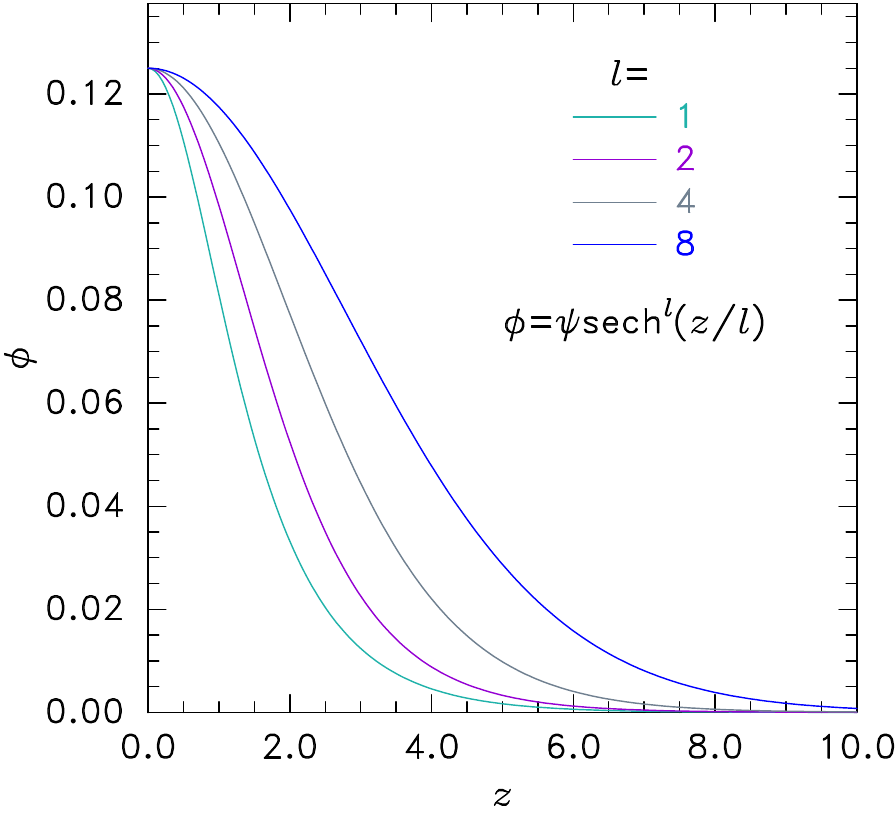}\hskip-1.5em(a)
  \includegraphics[height=.46\hsize]{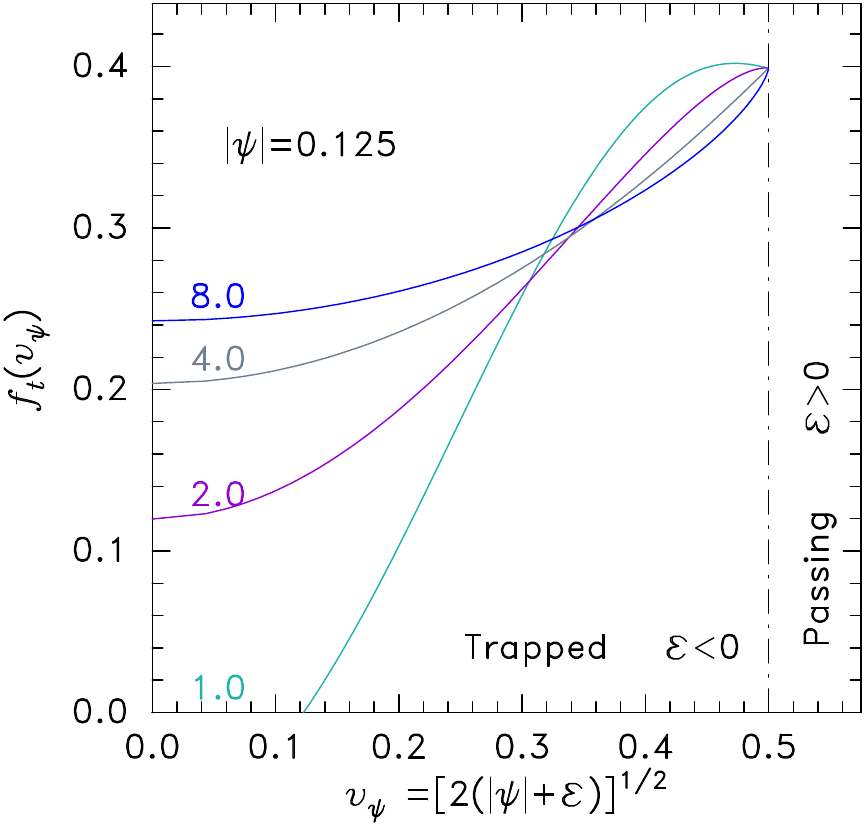}\hskip-1.em(b)
  \caption{Illustrative examples for (a) potential profiles
    $\phi=\psi\sech^l(z/l)$ of (b) the trapped distributions as a
    function of dimensionless velocity $v_\psi$ at the potential
    peak. Lines labeled with the value of $l$.  \label{sechln}}
\end{figure}

The integral equation BGK approach has been extensively discussed in
the literature; for expanded introductions see for example
\citep{Bernstein1957,Krasovsky2003,Hutchinson2017}; we here give a few
general examples of this approach to electron holes (ignoring ion
response) to illustrate its key features and restrictions. One simple
and convenient mathematical potential shape is to take
\begin{equation}
  \label{sechl}
  \phi(z) =\psi \sech^l(z/l\lambda).
\end{equation}
The contribution of $d^2\phi/dz^2$ to the required trapped particle
distribution \ref{Abel} can then be found
analytically\citep{Hutchinson2023}. Examples of distribution functions
are shown in figure \ref{sechln} for an unshifted Maxwellian
background distribution and negligible repelled particle response,
i.e.\ $\lambda=1$. For $l=2$ the slope $df_t/dv$ is zero at $\energy=0$, while
the more typical $l=4$ gives a profile that is a negative temperature
Maxwellian.  It turns out that values of $l$ greater than 4 produce a
weak slope singularity despite having the correct Debye shielding
asymptotic potential decay. 
For all potentials of this type, the required trapped
distribution falls monotonically at low $v_\psi$. 

As is clear from the figure, the lower the value of $l$, the narrower
the potential profile, and the deeper the hole in the distribution is
required to be.  An important limitation of allowable hole amplitude
$\psi$ for any prescribed shape like this is that the distribution cannot
physically be negative, yet nothing in the mathematics prevents
finding a solution with such a trapped $f(\energy)$; one must simply
avoid prescribing a $\psi$ that causes it. The illustrated case $l=1$
for $|\psi=0.125|$ violates this requirement, and is disallowed. Quite
generally the width in $z$ of a hole must exceed a minimum that is
proportional to $\psi^{1/4}$ \citep{Hutchinson2017}.
 
To explore a fuller range of possible hole shapes, a potential form
that has more adjustable parameters is needed. One way to implement
that, while still observing the asymptotic constraints at
$z\to\infty$, $\phi\to0$, is to use an inner form which is a Taylor
expansion of $\sqrt{\phi(z)}$ about the peak $z=0$ thus:
$\sqrt{\phi}=\sqrt{\psi}(1+c_2z^2+c_4z^4+c_6z^6)$; and then matching
it at some join position $z_j$ and potential $\phi_j$ to the
asymptotic form required. The reason for expanding $\sqrt{\phi}$
rather than $\phi$ itself, is that, as detailed in the Supplementary
Material, it is possible not merely to avoid singularities in $f'$ but
actually to prescribe its trapped value at and near $\energy=0$ by
requiring the variation for $z>z_j$ to be such that
\begin{equation}\label{epsphi}
 {\sqrt\phi_j\over 1+a\sqrt\phi_j}\exp(\pm(z- z_j)/2\lambda) = {\sqrt\phi\over 1+a\sqrt\phi},
\end{equation}
where $a$ is a constant depending on the prescribed $f'(0)$ value.  The other
adjustable specified parameters (in addition to $\psi$) used here are
$c_2$ and $\phi_j$. The equations that result from requiring $\phi$,
$d\phi/dz$ and $d^2\phi/dz^2$ to be continuous at the join can be
solved to find $c_4$, $c_6$, and $z_j$, giving a fully specified
$\phi(z)$ function. The Abel integral equation then is solved
(numerically) to find $f$.
\begin{figure}
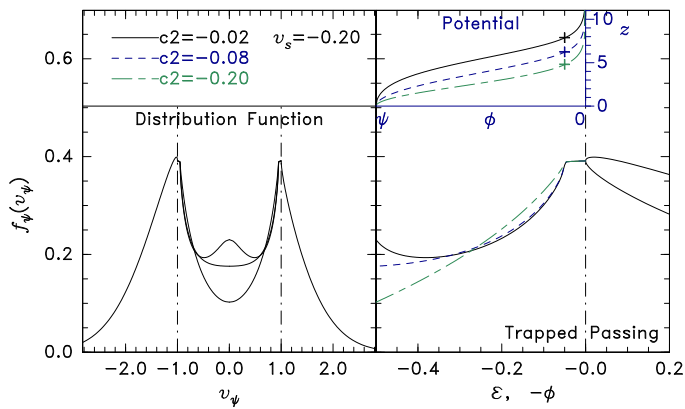
\center
  \partwidth{.65}{c2scan}
  \caption{Examples of the adjustable asymptotically matched potential profile and the resulting trapped distribution shapes. \label{c2scan}}
\end{figure}
Figure \ref{c2scan} shows example results varying the potential
curvature at the peak, keeping the join potential $\phi_j$ fixed (as
indicated by the cross), and setting $f'(0)=0$ (to lowest relevant
order). The entire distributions are plotted: at left versus velocity
$v_\psi$ measured at the peak of the potential (the deepest part of
the potential energy well), and at right versus energy.  The passing
distribution is chosen as a Maxwellian shifted by $v_s=-0.2$, to
illustrate asymmetry. Low curvature ($c_2=-0.02$) causes a
non-monotonic $f(v_\psi)$ having a rise at low $|v_\psi|$. As $|c_2|$
is increased, it disappears and the hole bottom becomes deeper. Note
the flat trapped region near $\energy=0$, and the different
$f(\energy)$ curves for the two velocity signs of passing particles
$\energy>0$.  The potential is plotted on its side on the same energy
scale to indicate the correspondence between $\phi$ (at some position
$z$) and the marginal trapped orbit energy that contributes to the
density integral. This case has $|\psi|=0.5$ but without encountering
negative trapped $f(v)$ limitation, because the hole $z$-width is large
enough.

\begin{figure}\center
  \includegraphics[width=0.8\hsize]{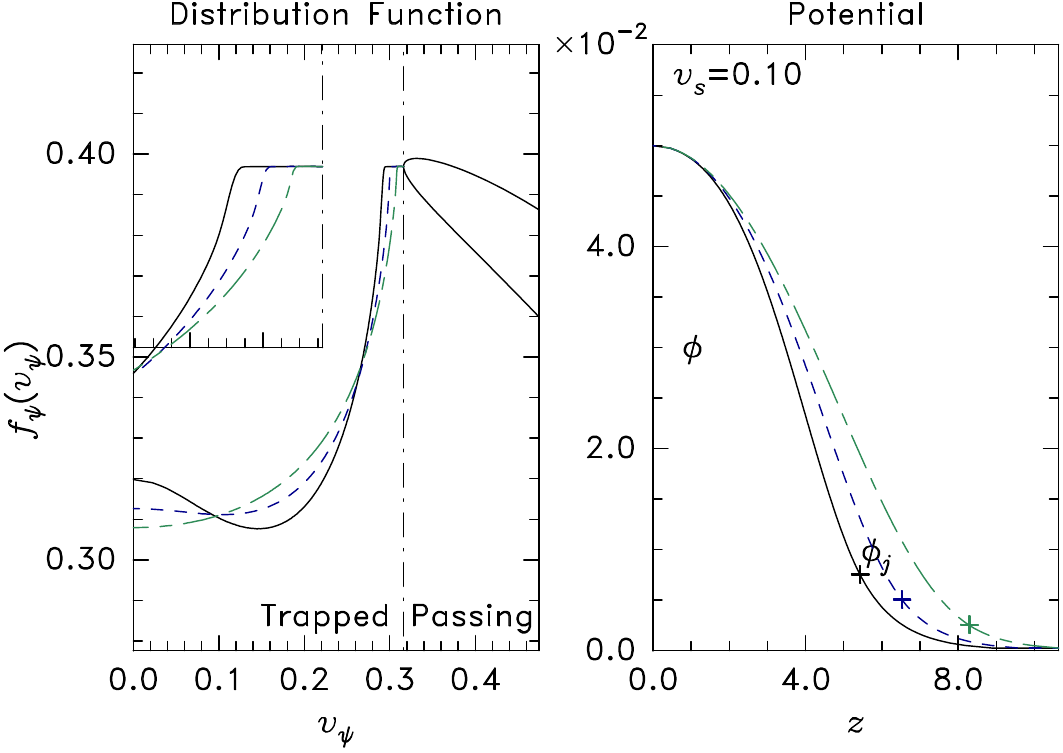}
  \caption{Variation of the join potential $\phi_j$ and the resulting
    shape for fixed central potential curvature $c_2=-0.05$.\label{pjscan}}
\end{figure}
Figure \ref{pjscan} explores several values of $\phi_j$ at fixed
$c_2=-0.05$, giving a detailed insert of the flat and join regions of
$f(v_\psi)$, and showing the potential shape trends, for $\psi=0.05$.
Requiring a small derivative $f'$ at the trapped/passing boundary
$\energy\to 0$, as these examples do, is physically appropriate for
more complicated situations (to be discussed later) where stochastic
trapping and detrapping of orbits occurs. Newly trapped orbits will
generally sustain phase space density approximately equal to the
adjacent passing value, down to a trapping depth determined by the
orbit energy diffusion. That depth might be represented approximately
by energy $-\phi_j$. Actually the passing distribution, when
asymmetric, also possesses divergent $v$-derivative at $\energy=0$ and
non-zero $\phi$. A modification of the passing velocity distribution
has been proposed \citep{Korn1996} to remove this singularity.
However, orbit diffusion will probably be less important for passing
orbits because they are being constantly renewed by influx from
distant $|z|$ regions, so it is justifiable to pay more attention to
avoiding the trapped distribution slope singularity, or even
constraining its finite value. Of course, all these mathematical
representations are just analytic conveniences and approximations.
Actual physical holes may have fine-scale distribution variation
arising from phase mixing but will have coarse-grained variation in
$v$ and $z$ that is smooth.

\section{One dimensional soliton and hole equilibria}
\label{solitonsec}

The main purpose of this section is to identify the characteristics of
one-dimensional fluid solitons that distinguish them from kinetic
structures such as electron and ion holes. This helps us to understand
kinetically how electrostatic structures work, and gives practical
guidance on identifying structures observed in nature.
In keeping with our overall approach, we treat a soliton in its own
rest frame right from the start.

In the process we will exemplify the pseudo-potential approach to
calculating solitary structure equilibria, but with highly simplified
particle distributions and consequent density behavior.

\subsection{Prototypical Solitons}

In a positive solitary potential structure like figure
\ref{symholes}(i), the potential curvature $\phi''=d^2\phi/dz^2$ at
its peak is negative. But in the wings of the structure,
approaching $\phi=0$, $\phi''$ must be positive, so as to bring the
potential derivative to zero, matching the uniform external
background. More generally for a single positive or negative peak
potential $\psi$, the signs of $\phi''$ must be opposite $\psi$ at
$\phi=\psi$, and the same as $\psi$ at $\phi=0$.

Obviously Poisson's equation $-d^2\phi/dz^2=\rho$ (scaled units)
requires equivalent statements about the charge density $\rho$: that
it has the same sign as $\psi$ at its peak and the opposite sign in
the wings.  Also the charge density must be zero in the background,
i.e.\ at $\phi=0$, in order for the potential there to be
uniform. Therefore the charge density must be a nonlinear
function of potential: starting at zero when $\phi=0$, initially
having sign opposite to $\psi$, that is $\psi d\rho/d\phi|_0<0$, then
reversing and becoming the same sign as $\psi$ at the potential
peak. At least one of the components of the plasma must therefore have
substantial curvature in its density dependence $n(\phi)$.

A \emph{single-velocity-stream} ($S$) component moving in a static
potential has velocity satisfying the conservation of energy
$\energy_S=v^2/2+q\phi=constant$. Provided $q\phi/\energy_S<1$, no
reflection occurs and the fluid continuity equation then immediately
yields
\begin{equation}\label{Stream}
  n_S(\phi)= {n_{S\infty}\over\sqrt{1-q\phi/\energy_S} },
\end{equation}
where $n_\infty$ denotes the density at $\phi=0$, which is $z=\infty$.
Notice that the slope $dn_S/d\phi$ has the sign of $q\phi$, so it
could not by itself be responsible for the correct $\phi''$ sign in the
potential wings near $\phi=0$. Also, $|dn_S/d\phi|$ becomes large as $\phi$
approaches $\energy_S/q$. Bear in mind that the stream energy
$\energy_S$ can be considered to set the speed of the structure in the
frame in which the stream is stationary.

By contrast, a \emph{broad distribution of velocities} ($B$),
approximately centered on zero, has density that depends on potential
with opposite slope sign. Qualitatively, when $q\phi$ is positive (repelling
particles) low energy particles are reflected before reaching
potential $\phi$; their absence reduces $n_B(\phi)$ below $n_{B\infty}$. When
$q\phi$ is negative (attracting particles), then provided that the
distribution on trapped (negative $\energy$) orbits remains fully
populated, the density increases with $\phi$ as the distribution
broadens through acceleration. The standard model of such a
distribution is the Maxwellian, which when unshifted from $v=0$ gives
the Boltzmann density distribution:
\begin{equation}\label{Boltzmann}
  n_B(\phi)=n_{B\infty}\exp(-q\phi/T).
\end{equation}
This dependence can also be regarded as the solution of the steady
isothermal fluid momentum equation.  Shifted Maxwellians or
alternative shapes change this expression, but the general principle
remains that provided the trapped orbits remain populated such that
$f(\energy)\gtrsim f(0)$, and
the distribution peak is shifted less than approximately its width,
thus remaining substantial at $\energy=0$, the slope $dn/d\phi$ has
the sign of $-q\phi$: opposite that for an unreflected single stream.

Fluid solitons work by balancing these two types of density behavior
against one another. Reduced to their simplest prototype, fluid
solitons have a stream component $S$, and a broad component $B$, plus
possibly an extra neutralizing uniform charge component $C$, which
might represent ions for a soliton moving at approximately electron
thermal speeds (so fast that ions cannot respond), or a background of
(approximately immobile) charged dust for ``dust ion acoustic''
solitons (see e.g.\ \citet{Mamun2005}).  These two or three components
give rise to a charge density that depends only on $\phi$:
\begin{equation}\label{solitoncharge}
  \rho(\phi) = {q_Sn_{S\infty}\over\sqrt{1-q_S\phi/\energy_S} }
  + q_Bn_{B\infty}\exp(-q_B\phi/T_B) +q_Cn_C,
\end{equation}
where $q_Sn_{S\infty}+q_Bn_{B\infty}+q_Cn_C=0$ to satisfy distant
neutrality. Its derivative is
\begin{equation}
{d\rho\over d\phi}= {1\over 2}{n_{S\infty}q_S^2\over
  \energy_S(1-q_S\phi/\energy_S)^{3/2}}-{n_{B\infty}q_B^2\exp(-q_B\phi/T_B)\over
  T_B},   
\end{equation}
which at $\phi=0$ is
$ {d\rho\over d\phi}={1\over 2} {n_{S\infty}q_S^2/\energy_S} -
{n_{B\infty}q_B^2/T_B}$; so
the required distant condition $d\rho/d\phi|_0<0$ is satisfied only if
\begin{equation}
  \energy_S>|n_{S\infty}q_S/n_{B\infty}q_B|T_B/2.
\end{equation}

\subsection{Pseudo-potential solitary structure analysis}

\begin{figure}[ht]
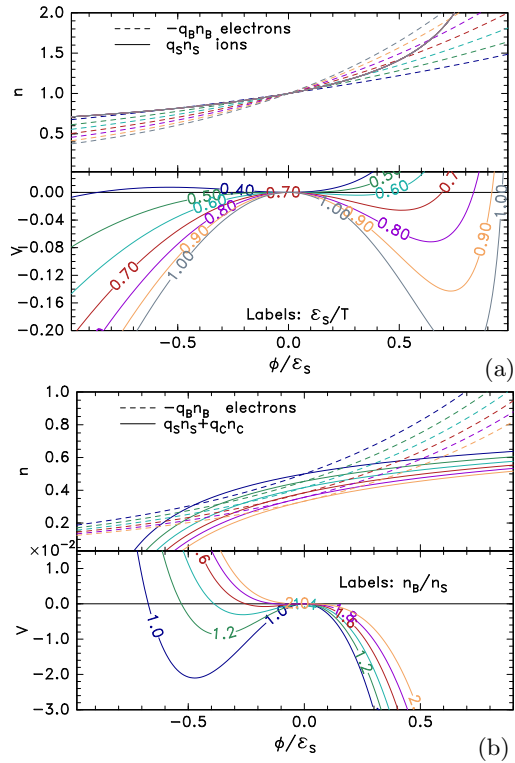

  \partwidth{0.48}{invfsphi}\hskip-1.em(a)
  \partwidth{0.48}{envfsphi}\hskip-1.5em(b)  
  \caption{Examples of particle densities $n(\phi)$ and the resulting
    pseudo-potential $V(\phi)$ for (a) ion acoustic solitons for a
    range of soliton speeds relative to the (ion) stream $\energy_S=v^2/2$, (b)
    electron acoustic solitons for a range of ratios of broad to stream
    densities $n_B/n_S$  (for the case $\energy_S=T_B$).\label{nvfsphi}}
\end{figure}

Pseudo-potential analysis can be applied to any one-dimensional
equilibrium where the charge density is a known function of potential,
whether solitons or holes. Then, because in Poisson's equation the
charge is a function of the dependent variable, $\phi$, but not of the
independent variable, $z$, a first integral can be found by
multiplying by $d\phi/dz$ and integrating to get
\begin{equation}\label{pseudopot}
  \begin{split}
  {1\over 2} \left(d\phi\over dz\right)^2&=-V(\phi)=-\int_0^\phi \rho
  d\phi\\
  &= 2\energy_S n_{S\infty}\left[\sqrt{1-{q_S\phi\over\energy_S}}-1\right]\pamper
  +T_Bn_{B\infty}\left[\exp\left(-{q_B\phi\over T_B}\right)-1\right] -q_Cn_C\phi.    
  \end{split}
\end{equation}
Because of analogies that will not be explained here, the quantity
$V(\phi)$ is called the pseudo-potential, classical potential, or
Sagdeev potential\footnote{The expression Sagdeev potential recognizes
  his extensive early contributions, using this technique, to fluid
  ion acoustic soliton theory that we are guided by in this subsection.},
regardless of the exact form of $\rho(\phi)$.
It is illustrated together with the densities that give rise to it, in
figure \ref{nvfsphi}. The pseudo-potential must be negative throughout
the potential range $0-\psi$ and satisfy the conditions $V(0)=0$,
$dV/d\phi|_{\phi=0}=0$ and $V(\psi)=0$. The conditions at $\phi=0$ are
satisfied by the chosen integration constant in eq.\ (\ref{pseudopot})
and distant neutrality $\rho(0)=0$.  The third condition, $d\phi/dz=0$
(i.e.\ $V=0$) at the potential peak $\phi=\psi$, for our simple fluid
soliton gives an equation relating the stream energy $\energy_S$
(equivalent to speed of the soliton with respect to the stream) and
$\psi$
\begin{equation}
  1-{q_S\psi\over\energy_S}=\left\{1
  -{T_Bn_{B\infty}\over 2\energy_S n_{S\infty}}
  \left[\exp\left(-{q_B\psi\over T_B}\right)-1\right]
  +{n_Cq_C\psi\over n_{S\infty}2\energy_S} \right\}^2.
\end{equation}
Whose solution is
\begin{equation}\label{Eofpsi}
  \energy_S={ {1\over4}T_B{n_{B\infty}\over n_{S\infty}}
\left\{
    \exp\left(-{q_B\psi\over T_B}\right)-1
      -{n_Cq_C\psi\over n_{B\infty}T_B}\right\}^2
    \over
    \exp\left(-{q_B\psi\over T_B}\right)-1
      +{q_B\psi\over T_B}
    },
\end{equation}
using $n_Cq_C+n_{S\infty}q_S=-n_{B\infty}q_B$.

\subsection{Ion Acoustic Solitons}
\begin{figure}[ht]
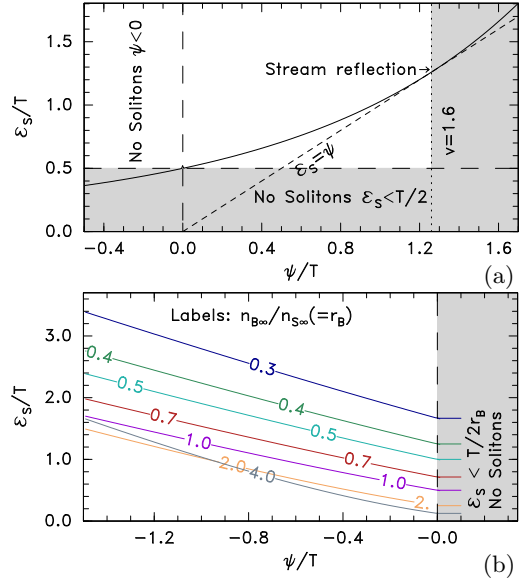
\center
  \partwidth{0.49}{ivsphi}\hskip-1.5em(a)
  \partwidth{0.49}{evsphi}\hskip-1.5em(b)
  \caption{Relation between stream energy and amplitude of (a) simple ion
    acoustic solitons, (b) simple electron acoustic solitons for
    various values of $n_{B\infty}/n_{S\infty}$.\label{evsphi}}
\end{figure}
Ion acoustic solitons are so called because they operate under the
same assumptions as linear sinusoidal ion acoustic waves. They take
the stream to be ions and the broad distribution to be electrons. They
do not require $n_C$ and have (most simply) $q_S=-q_B=1$ and
$n_{S\infty}=n_{B\infty}=1$. Figure \ref{nvfsphi}(a) shows the form of
their densities and pseudo-potentials. Equation \ref{Eofpsi} is then
exactly the classic result of \citet{Sagdeev1966} eq (39). It is
plotted in Figure \ref{evsphi}(a). These cold ion, Maxwellian
electron, equations possess no solution with $\psi<0$ because for
negative $\phi$ and $\energy_S>T_B/2$, the equation $d\rho/d\phi=0$
has no solution, meaning $d\rho/d\phi$ cannot change sign. So these
solitons are always positive. Also, at $\energy_S=1.26T_B$,
$v_h\simeq 1.59\sqrt{T_B/m_S}$, the potential $\psi$ reaches the same
value as the energy, thus beginning to reflect the stream and breaking
the assumptions governing the stream density.  Solitons with higher
speed and non-zero stream width cannot exist. More complicated
electron distributions, non-Maxwellian or even sums of multiple
Maxwellians of different temperatures or velocity shifts alter the
function $n_B(\phi)$. It can then sometimes be possible for negative
potential (``rarefactive'') ion acoustic solitons to exist \citep{Cairns1995}.

\subsection{Electron acoustic solitons}

Electron acoustic solitons, relate by analogy to a much more unusual
type of sinusoidal wave referred to as the electron acoustic mode (see
\citet*{Gary1985}), which arises from the electron response in a
distribution having hot and cold electron components. Reduced to the
simplest prototype, one can assume a cold stream $S$ and broad
distribution $B$ that are both electrons ($q_S=q_B=-1$), but of
different densities. Ions are a constant (in the simplest case) charge
neutralizing background $C$. Then the slope is
$d\rho/d\phi|_0=(e^2/T_B)[n_{S\infty}/2\energy_S - n_{B\infty}/T_B]$,
which to be negative requires
$\energy_S > (n_{S\infty}/n_{B\infty})T_B/2$. Figure \ref{nvfsphi}(b)
shows the resulting density and pseudo-potential shapes, for
different ratios of stream to broad distribution densities.

Equation \ref{Eofpsi} becomes
\begin{equation}
  \label{Eofpsie}
  \energy_S={ {1\over4}T_Bn_{B\infty}/n_{S\infty}
\left\{
    \exp\left({e\psi\over T_B}\right)-1
      -{e\psi\over T_B}{(n_{S\infty}+n_{B\infty})\over n_{B\infty}}\right\}^2
    \over
    \exp\left({e\psi\over T_B}\right)-1
      -{e\psi\over T_B}
    }.
\end{equation}
However, not all solutions of this equation give rise to solitons,
because a soliton must have a continuous variation of $\rho(\phi)$
over the range $0\le |\phi| \le |\psi|$. Consider the second derivative of
$\rho$:
\begin{equation}
{d^2\rho\over d\phi^2}= {3\over 4}{n_{S\infty}q_S^3\over
  \energy_S^2(1-q_S\phi/\energy_S)^{5/2}}+{n_{B\infty}q_B^3\exp(-q_B\phi/T_B)\over
  T_B^2}.
\end{equation}
It is always negative for electron acoustic solitions when $n_S$ is
positive\footnote{A negative density stream of finite spread might be
  used to represent a depression within the Maxwellian distribution
  function. Allowing such an electron stream invalidates this proof
  that $\psi$ cannot be positive. So does allowing an ion stream with
  significant potential response, but arguably that should anyway be
  called an ``ion acoustic'', not an ``electron acoustic'' soliton.}
because $q_S=q_B=-1$. Consequently if $d\rho/d\phi|_0<0$ (as it must
be for positive $\psi$) then at increasingly positive values of
$\phi$, $d\rho/d\phi$ decreases, instead of increasing. It therefore
does not become positive to allow $\rho$ continuously to become
positive and $V(\phi)$ to return to zero as is clear in figure
\ref{nvfsphi}(b). This demonstrates that there can be no
positive-potential purely electron acoustic solitions for any electron
distribution constructed from unreflected streams of positive density
($n_S$) and Boltzmann-like trapped density contribution. This remains
true if the streams have finite pressure. Figure \ref{evsphi}(b) shows
the resulting relationship between $\energy_S$ and $\psi$ for a range
of $n_{B\infty}/n_{S\infty}$.

\subsection{Soliton -- Hole Comparison}

The illustrative assumptions about the fluid behavior can of course be
generalized by introducing finite pressure (that is velocity spread,
or temperature) of the stream, or introducing multiple streams of
different velocities or temperatures giving rise to more complicated
charge functions $\rho(\phi)$ (see for example
\citealp{Lakhina2018}). However, notice that for any set of prescribed
external velocity distributions $f_{e}$,$f_{i}$, represented here by
$\energy_S$, $q_Sn_{S\infty}$, $q_Bn_{B\infty}$, $n_{C}$, $T_B$, our
simple fluid solitons have a single solution for amplitude
$\psi$. This discrete solution is a major characteristic of all fluid
solitons, which is not shared by electron or ion holes. Holes can have
continuous ranges of amplitudes and speeds depending on the degree of trapped
particle reduction, even for fixed passing particle distributions. The
discrete solutions for solitons arise because of the assumption of a
fixed form for the trapped particle distribution. In our simple case
we assumed effectively that it follows the Maxwellian dependence
$f(\energy)\propto \exp(-\energy/T_0)$ where $T_0$ is the background (i.e.\
untrapped) particle temperature. This assumption is mathematically
convenient but poorly justified physically because it implies an
enhancement of the distribution function on trapped orbits.

A more physically plausible assumption might be that the trapped
$f(v)$ is a flat plateau (for negative $\energy$) at its value where
$\energy=0$. That is what would arise for a potential structure that
gradually grew up from zero in a time long compared with the typical
transit time of the attracted particles\citep{Gurevich1968} because
newly trapped particles always sample the external distribution at
zero energy. Such a flat trapped distribution gives a mathematically
more complicated density function, but still gives discrete solutions
for $\psi$ versus the untrapped distribution's parameters. In our
illustrative case those parameters were $\energy_S$ and
$n_{B\infty}/n_{S\infty}$ but treated within a plasma frame in which
the distributions are fixed, they can be considered to dictate the
soliton's speed relative to the background plasma .

Electron and ion hole equilibria can still be found using the
pseudo-potential differential equation approach, but they do not
assume some fixed relationship between the trapped ($\energy<0$)
distribution $f(\energy)$ and the background untrapped ($\energy>0$)
$f(\energy)$. Instead they suppose that the trapped distribution is a
variable function that controls the hole amplitude $\psi$ and shape
$\phi(z)$ through its contribution to the charge density. Consequently
there is generally no fixed relationship between amplitude $\psi$ and hole
velocity $v_h$.

A characteristic of the flat plateau trapped distribution is that, unlike an
unshifted Maxwellian, its derivative is discontinuous at $\energy=0$.
The mathematical result, which is effectively inherent to holes, but
not always to solitons, is that the expansion of the density for small
$\phi$ possesses half-integer powers. As an important example, the
total density of a ``Schamel distribution'' electron hole having
trapped dependence $f(\energy)= f(0)\exp(-\beta \energy)$, for an
attracted species with a (single) shifted Maxwellian untrapped
distribution has the expansion
\begin{equation}\label{shallown}
n_e=1-{1\over2}Z_r'(U)\phi - {4\over 3} b \phi^{3/2}+{1\over
  16}Z'''_r(U) \phi^2+O(\phi^{5/2}),
\end{equation}
where $U=v_s/\sqrt{2}$ measures the shift velocity $v_s$ (scaled to
$\sqrt{T/m}$) of the Maxwellian,
$b=\exp(-U^2)(1-\beta-2U^2)/\sqrt{\pi}$, and $Z_r$ is the real part of
the plasma dispersion function\footnote{The $\phi^2$ coefficient was
  stated incorrectly in \citep{Hutchinson2017} and the present form
  corrects that error, and conforms with prior literature.}. The
$\phi^{3/2}$ term is the first non-zero half-integer power. Derivative
discontinuity is absent when $\beta=1$ and $U=0$ (unshifted
Maxwellian), but the flat trapped distribution has $\beta=0$, and so
has $b\not=0$ even for $U=0$. The Schamel distribution has the
attractive features of continuous $f(v)$ and non-singular $f_t'$
at $\energy=0$.\footnote{Recent papers
  \citep{Schamel2015,Schamel2018,Schamel2020} introduce
  generalizations of the Maxwellian trapped distribution shape that
  have discontinuities and slope singularities at
  $\energy=0$.}

It can be shown
\citep{Schamel1972,Schamel1973} that for small $\phi$
the density dependence (\ref{shallown}) corresponds to a nonlinear
wave equation governing the potential
\begin{equation}\label{mkdv}
 {\partial\phi\over\partial t}+(1+b\phi^{1/2}+\phi){\partial
   \phi\over\partial z}+{1\over 2}{\partial^3\phi\over\partial z^3}=0.
\end{equation}
When $b=0$ (and the 1 term is removed by a coordinate transformation)
this is the Korteweg-de Vries (KdV) equation \citep{Korteweg1895}
which is the standard of much soliton analysis, historically including
shallow water waves and the simplified ion acoustic soliton. That
soliton has shape (see \citealp{Sagdeev1966} eq.\ 40)
$\phi=\psi\,\sech^2([2\pi\psi/3]^{1/2} z)$.  The full equation
(\ref{mkdv}) with $b\not=0$ is sometimes called the modified KdV
equation.  It gives rise instead (ignoring the $\phi$ term in the
bracket) to the potential shape \citep{Schamel1986}
$\phi=\psi\,\sech^4([b\sqrt{\psi}/15]^{1/2}z)$. This solution shape is
a popular example for kinetic electron holes. Actually,
the soliton literature historically spoke of ``the Modified KdV
equation'' as having nonlinearity factor (i.e.\ the bracketed term in
eq.\ \ref{mkdv}) being $\phi^2$, which gives rise to solitary
potentials $\phi\propto \sech(x/\ell)$. A general nonlinearity
factor $\phi^n$ has also been considered by \citet{Dodd1982}.  So
\ref{mkdv} should really be called ``a'' modified KdV equation.

The ion acoustic soliton's amplitude is fixed by the (repelled
distribution's) stream energy in the hole frame, $\energy_S$,
giving $\psi=(3/2)[1-1/(2\energy_S)]$, and its small-amplitude width
is proportional to $\psi^{-1/2}$. By contrast, because the electron hole's
trapped distribution ($b$) can vary, its amplitude can vary even with
fixed untrapped distribution(s).

What is more, since nothing forces the trapped distribution to have
the dependence $f(\energy)= f(0)\exp(-\beta \energy)$, we are at
liberty to choose almost any different shape (not just $\sech^4$) for an
electron hole, and still avoid derivative singularity at
$\energy\to 0$ provided the distant potential decay has the correct
shielding length $\lambda$.

\textbf{In summary:} The key difference between solitons and
kinetic electron and ion holes is that solitons have particle
densities in the presence of the potential structure that are entirely
determined by the background particle distributions, while holes allow
the trapped particle distribution to vary independent of the
background. Solitons are usually treated using fluid approximations,
which often makes their mathematics more tractable, but that is not
essential. The central charge density sustaining a soliton generally
arises from streaming repelled particles being slowed and thereby
densified by the potential, rather than from the deficit of trapped
attracted particles in a hole. The soliton's speed relative to the
background stream(s) crucially controls the densification. The absence
of independent variability of the trapped density causes solitons to
have a discrete amplitude $\psi$ for given background distributions in
the soliton's frame. Changing soliton velocity $v_s$, in a frame of
reference in which the background distributions are fixed, changes the
background it sees in its own reference frame. Therefore a unique
functional dependence between $\psi$ and $v_s$ arises. Generally
$|\psi|$ increases with $|v_s|$ when $v_s$ is measured relative to a
narrow particle stream. Also, the soliton width generally has a
discrete inverse relationship approximately proportional to
$\psi^{-1/2}$. Holes have no such discrete relationships, and can have
a continuous range of amplitudes in a fixed background at fixed
velocity, as well as continuous ranges of velocities and widths.

\section{Historical Background}
\label{Historical}

\begin{figure}[ht]
  \includegraphics[width=\hsize]{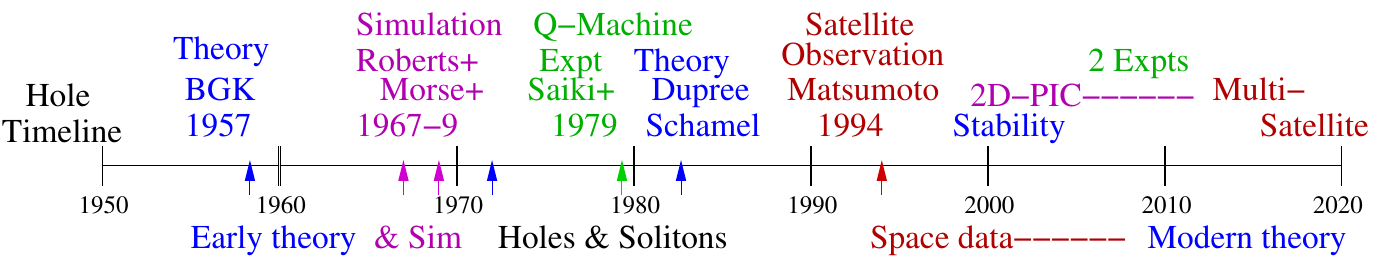}
  \caption{Phase-space hole research. Colors indicate theory (blue),
    simulation (magenta), laboratory experiment (green), and space
    satellite observation (red). A few key milestones are indicated
    with arrows. \label{timeline}}
\end{figure}

This section surveys the historical development of kinetic
electrostatic structures from their beginnings to approximately 2016.
Figure \ref{timeline} attempts to summarize in a single timeline that
past history, which we will now seek to fill out with more detailed 
explanation and citation.

\subsection{Initial theory and simulation 1957-1970}
\begin{figure}[ht]
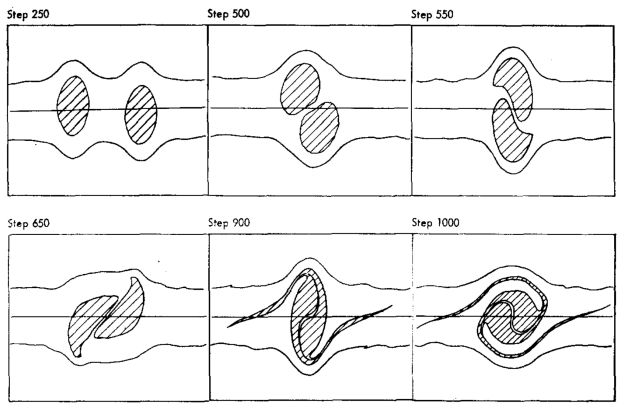

  \center
  \partwidth{0.6}{BNRholemerge}
  \caption{Early Vlasov simulation of hole merging in phase-space
    ($z,v$). Reprinted from \citep{Berk1970} with the permission of
    AIP publishing.
    The phase-space density is zero in the shaded region and unity
    elsewhere. 
    \label{BNRholemerge}}
\end{figure}

The work of \citet{Bernstein1957} can be considered the
historic foundation of the field of kinetic electrostatic plasma
structures. But in those days, even before experiments had persuaded
doubting plasma physicists that Landau damping was real, there were
many other nonlinear plasma wave aspects being pursued ahead of the
rather abstract possibility of solitary BGK structures. The first
subsequent paper to demonstrate that they were naturally formed was a
very early one-dimensional continuum Vlasov simulation by
\citet{Roberts1967}, of an electron two stream instability
growing, trapping electrons, forming a train of electron holes, and
undergoing a pairwise hole merging event to create bigger holes. More
detail of the trapping process for a single hole was given the same
year in \citet{Berk1967}, and more detail of the hole merging in
\citet*{Berk1970}. Fig.\ \ref{BNRholemerge} illustrates the merging
process.

\citet{Morse1969} observed the same sorts of phenomena, but with a
particle in cell code, and with the important addition of two and
three dimensional simulations that, unlike the one dimensional case,
did \emph{not} observe hole formation, thus calling into question the
applicability of the one-dimensional approximation to unmagnetized
plasmas. Analytic BGK theory of a ``shock'' that now would be called a
double layer, was also published by \citet{Montgomery1969}.

\subsection{Experiment, Simulation, and Analytic Consolidation 1970-1993}
The year 1972 saw the publication of a seminal paper by
\citet{Schamel1972}, in which he introduced the idea of supposing that
the velocity distribution of trapped particles be represented by a
Maxwellian $f(v)\propto \exp(-\beta v^2/2)$ of (possibly) negative
inverse temperature $\beta$. This ansatz, used with the
pseudo-potential solution method, is frequently called the Schamel
hole. The work addresses non-linear periodic waves of arbitrary
wavelength, including infinite, which represents the solitary case,
and introduces the notion of a Nonlinear Dispersion Relation
(NDR). For solitary potentials with cold ions, the NDR expresses
the fact that for fixed $\beta$ there is a unique relationship between
hole speed relative to the ions and the peak hole potential $\psi$.
For modest $|\beta|\lesssim 1$ this is a small modification of the
relationship for ion acoustic solitons given earlier by
\citet*{Vedenov1961a}.  The paper also shows that trapping
effects when $\beta\not=1$ can be represented approximately by a
single ``modified Korteweg-de Vries'' partial differential equation
with an extra non-linear term equal to $\sqrt{\phi}$ times a
coefficient $b=(1-\beta)/\sqrt{\pi}$ proportional to the difference of
electron distribution in the trapped region from the external
Maxwellian. This equation gives rise to wave solutions called
``snoidal'' or ``cnoidal'' with narrower peaks than troughs, and in
the infinite wavelength limit to solitons of the shape
$\phi(z)=\psi\, \sech^4(z/\lambda)$, rather than the more familiar
$\psi\, \sech^2(z/\lambda)$ form arising from the standard soliton
KdV equation.

In 1972 also, \citet{Davidson1972} published a helpful wide-ranging
book about nonlinear plasma theory, which briefly addresses solitons
and particle trapping but not significantly our current emphasis,
holes. \citet{Sakanaka1972}  reported continuum Vlasov
simulations of ion-beam plasma interactions giving rise to a shock and
simultaneously an ion ``vortex'' or ion hole, though not a solitary
hole. During the ensuing years of the 1970s decade, several important
theoretical papers concerning double layers appeared, culminating in a
review by \citet{Block1978}, which summarizes the double layer
field. But a critical result giving the first unambiguous laboratory
observations of an electron hole appeared in two 1979 papers:
\citet*{Saeki1979}, and \citet*{Lynov1979}. They discuss the same
experimental data, obtained in a long magnetized plasma filled
wave-guide (Q-machine), but the second gives expanded modelling in the
form of PIC simulations; and \citet*{Lynov1980}  expand it
further. Under some circumstances the experimental data and simulation
showed merging of two electron holes into one.

Presumably inspired by this experiment, \citet{Schamel1979}  immediately
published a theoretical analysis of the electron hole and
Gould-Trivelpiece soliton that were observed in it, accounting for the
important finite transverse extent imposed by the waveguide. The
effect is to require an extra term
$-{\partial^2 r\phi \over r\partial r^2}=k_\perp^2 \phi(0,z)$ with
$k_\perp$ fixed by the waveguide, in addition to
$-{\partial^2\phi\over \partial z^2}$, in Poisson's equation. Ignoring
any ion response, which is a good approximation for the experiment,
and approximating the hole amplitude $\psi$ as small, it is shown that
for an electron hole to exist, the trapped inverse temperature $\beta$
must be sufficiently negative; the potential form is then
$\phi=\psi\,\sech^4(z/\ell)$ with $\ell=(15/b\sqrt\psi)^{1/2}$.
The alert reader will observe that $b$ and $\sqrt\psi$ must therefore
be inversely proportional to one another in order to satisfy the
constraint that the asymptotic potential form is
$\phi\propto\exp(-z/\lambda)$, where $\lambda$ is the plasma screening
length (the Debye length modified by $k_\perp$, independent of hole
properties). For a hole moving relative to the Maxwellian background
electrons at speed $v_h$, the value of $b$ is
$b=(1-\beta-v_h^2)\exp(-v_h^2/2)/\sqrt\pi$. Small $\psi$ (the limit
under discussion using a Taylor expansion of the densities) then
implies large positive $b$ which requires large negative $\beta$,
regardless of moderate values of $v_h^2$. It is unfortunate therefore
that Schamel's paper
assumes $\beta$ is fixed and of limited size (-3), and concludes
incorrectly that in the limit of vanishing amplitude ($\psi\to 0$) and
perpendicular wave-number ($k_\perp\to 0$) the hole must move at speed
$v_h^2=1.71(T_e/m_e)$. This conclusion ignores the normal situation
that in the limit, $b\sqrt\psi$ is finite for an electron hole, not
zero; and a self-consistent $\beta\propto 1/\sqrt\psi$ exists to
permit any $v_h$ such that $\exp(-v_h^2/2)$ is not vanishingly
small. Nevertheless, this experiment and the analysis and simulation
it inspired remain important milestones in the physics of kinetic
electrostatic structures.

Another important theory paper in 1979 was
\citet*{Schwarzmeier1979}. In it the instability of adjacent peaks in
an initially periodic non-linear BGK wave was simulated and
analytically calculated. The results agree rather well; and represent
the first stage of the merging of adjacent electron holes at the
potential peaks into larger amplitude and eventually solitary holes.
This process had been seen in lower resolution in the simulations of
\citep{Roberts1967} and \citep{Morse1969}. But the work of
\citeauthor{Schwarzmeier1979}, together with its refinement by
\citet{Ling1981}, is notable in being the first (and for a long time
the only persuasive) application of the mathematical methods of
\citet*{Lewis1979}  to the problem of determining the general unstable
eigenmode of a BGK equilibrium.

Analytic \emph{ion} hole equilibria based on the Schamel approach were
addressed by \citet{Schamel1980}. The ion hole equations are
highly similar to those for the electron hole, and a paper with
additional details treating both followed soon after:
\citet{Bujarbarua1981} ; then \citet{Schamel1982a}  covered
much the same ground, with additional explanatory diagrams and
mathematical details, and added discussion of double layers.

In 1982 \citet{Schamel1982} also addressed the stability of a solitary
electron hole using the general mathematical apparatus of
\citet{Lewis1979}. His truncated series expansion of the Vlasov
solutions, using a so-called fluid approximation was of questionable
accuracy, but also he assumed erroneously that the most dangerous
(unstable) mode was a symmetric potential perturbation, which
subsequent simulations and theory have shown to be incorrect. 

\citet*{Pecseli1981} in 1981 reported the ``first experimental
observations'' of an ion hole. In effect it was of the type simulated by
\citet{Sakanaka1972}. More extended interpretation of this research in
a beam-plasma device is presented in \citet*{Pecseli1984},
together with PIC simulations.

The paper of \citep{Dupree1982} arose primarily from attempts
to understand the PIC simulations of his students \citet*{Berman1982}
which showed (for a mass ratio $m_i/m_e=4$) nonlinear growth of
current-driven electrostatic ion instabilities \emph{below} the linear
growth threshold. He had long been interested in the phase-space
density granulations that occur in non-linear plasma turbulence, which
he called ``clumps'': locally coherent positive or negative
phase-space density perturbations. But in \citep{Dupree1982} he begins
by criticizing his prior clump theory and pointing out that holes
(density depressions) have the ability to trap particles
self-consistently by their fields, which density enhancements do not.
The paper argues that Maxwellian (Schamel) holes are states of maximal
entropy, and shows how to calculate the effective mass, charge, and
energy of a hole, illustrating it with analytic calculations for a
``step function'' trapped distribution $f(v)$, which proves to be not
much different from a Maxwellian hole. It was followed by
\citet{Dupree1983}  which addressed in detail the evolution of a
one-dimensional hole when it grows in amplitude or accelerates,
behaving as an (almost) rigid body that obeys Newton's second law. The
mathematical detail in its 255 equations makes the paper rather
forbidding. But it systematically derives the mass, momentum, and
energy changes arising from acceleration and growth, under shallow
hole approximations, and shows some foundational concepts concerning
what might be termed a hole's kinematics (i) the effective mass of a
hole is negative, (ii) if a hole moves into a phase-space region with
higher background $f(v)$, then its depth (and consequently its
potential amplitude) increases because the trapped $f(v)$ remains
constant. (iii) electron reflection from an ion hole in a current
carrying plasma can cause an acceleration into a higher $f_i(v)$ and
hence such growth.

\citet*{Temerin1982}  is the first reported observation in space of
double layers and ``solitary waves'' (bipolar excursions of parallel
electric field). Its data was compared with simulation and analytic
theory by \citet*{Hudson1983}.

\begin{figure}[ht]
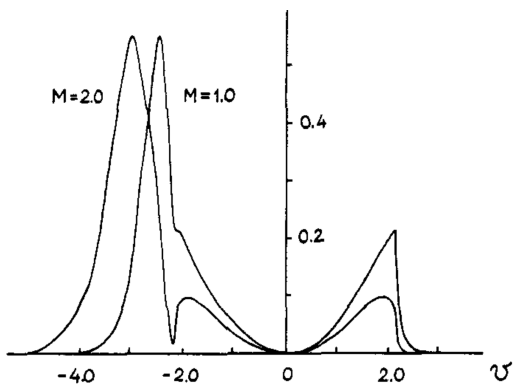
\center
  \partwidth{0.5}{Turikovfv}
  \caption{The distribution function $f(v)$ of Turikov's fast, deep
    holes. From \citep{Turikov1984} copyright IOP
    Publishing. Reproduced with permission. All rights
    reserved. Amplitude $\psi=5$. M is his notation for normalized
    hole velocity $v_h$.\label{Turikovfv}}
\end{figure}

\citet{Turikov1984} introduced a different type of
theoretical electron hole, whose equilibrium can apparently move at
any speed $v_h$ (even considerably greater than thermal speed)
relative to the passing electrons. He used an integral equation method
with prescribed potential shape, whose length $\delta$ was regarded as
adjustable; but he prescribed also that the trapped distribution $f(z,v)$
is exactly zero at the hole center ($z=0$, $v=0$). There is then a
monotonic increasing relationship between spatial width $\delta$ and
speed $v_h$. Allowing $\delta$ to vary in this way does not give rise
to Debye shielding asymptotic potential decay, and therefore induces a
slope singularity (which is negative for $|v_h|\gtrsim1.3$):
$df/dv\to-\infty$ at $\energy\to0$. The trapped $f(v)$ is then
non-monotonic, consisting of a peak at slightly negative $\energy$,
but a lower value (even $\sim 0$) at the trapping boundary
$\energy=0$, as illustrated in Fig.\ (\ref{Turikovfv}).
Although this theoretical structure has a hollow phase-space vortex of
trapped particles, it is not a hole in the background
phase-space. Instead, the net positive charge at the hole spatial center
is produced by the acceleration of the (predominant) passing electrons,
reducing their density as they move through the attracting potential;
while the negative charge required in the hole wings is provided by
the trapped peak in $f$.  The lower value at its edge $\energy=0$,
matching to the passing distribution, is permitted by the slope
singularity. Turikov's simulations show that these equilibria are
unstable for $|v_h|\gtrsim 2$ on a timescale $\sim10/\omega_{pe}$, so
they are then extremely transitory.

In 1986 \citet{Schamel1986} presented a review recapitulating the
electron hole experiments and his own theory of electron and ion
holes.  Dupree's negative hole mass is mentioned; and remarks made on
the methods and difficulty of stability analysis of holes. Double
layers are also substantially reviewed, together with experiments and
simulations.

\begin{figure}[ht]
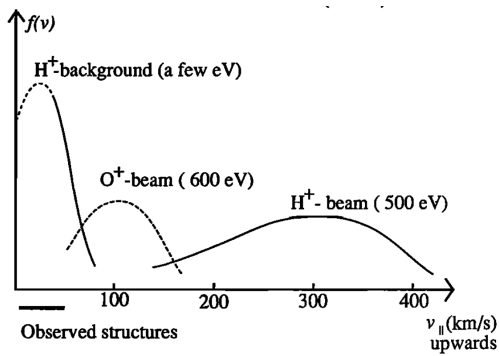

  \center
  \partwidth{0.5}{viking}
  \caption{The schematic model of the plasma environment in which slow
    ion holes (velocities indicated by the horizontal bar) were
    observed by Viking. From \citep{Koskinen1990}.\label{viking}}
\end{figure}
The Viking satellite launched in 1986 clearly identified negative
potential solitary structures (which were most likely ion holes) in
the magnetosphere as reported by \citet*{Bostrom1988}. Probes 80m
apart measured density or electric field, seeing up to $\sim$50\%
density drops that propagated along the magnetic field from one probe
to the other, hence determining the velocity. Simultaneously, bipolar
electric field excursions were observed, consistent with passing
through the gradients of a solitary potential valley: the hole. More
detailed interpretation, \citet*{Koskinen1990}, of the plasma
environment of the polar regions in which they were observed, which
had a substantial energetic upward-propagating mostly proton beam,
indicated that they were propagating at approximately the speed of the
bulk ions (not the beam), as illustrated in Fig.\ (\ref{viking}).

The work of \citet*{Ghizzo1988}  addressed the stability of a
periodic train of electron holes to hole merging. Their linearized
analysis assumed an eigenstructure not unlike that of
\citet{Schwarzmeier1979}, and they also performed high-resolution
Vlasov simulations, observing the initial ``macroparticle'' behavior
of individual holes, but concluding that the eventual state after long
enough time was virtually always a single large merged hole.

\citet{Collantes1988a}  analyze the linear stability of a single
electron hole, first for wavelengths short compared with the hole
extent, by using $f(v)$ at the hole center in a supposed uniform infinite
plasma analysis.  They find that there is no unstable mode with short enough
wavelength to satisfy this approximation, unless the hole is of the
unusual high velocity $v_h$ type introduced by
\citet{Turikov1984}. Then second, they use the methods of
\citet{Lewis1979}, to identify a set of perturbation eigenfunctions of
extent comparable to the hole. For a $\sech^4(\alpha z)$ equilibrium
they select a symmetric $\sech^3(\alpha z)$ perturbation shape. For
the Vlasov solution they use an expansion like \citet{Schamel1982},
which is justified only for short transit time particles, but they
carry it to higher order derivatives. They find instability with
growth rate of order unity. Previous and subsequent simulations show
no such instabilities in single isolated holes of typical length. This
disagreement is not explained but is probably attributable to
unsatisfactory approximations.

\citet{Raadu1988}  and the review of \citet{Raadu1989} 
represent the state then of double layer theory; and (in the latter)
simulation, experiment, and applications in space and astrophysical
plasmas. Double layer research saw continuous theoretical and
observational development thereafter.  But remarkably, the subsequent
period saw very little new development of the theory of electron and
ion holes, until a new impetus arose.

\subsection{Ubiquitous holes in space, multidimensional simulation, stability, 1994-2008}

Arguably the high-sample-rate (up to 12kHz\citep{Matsumoto1994a})
observations from the Geotail satellite published by
\citet*{Matsumoto1994}  opened a new era by identifying ``Broadband
Electrostatic Noise'' (BEN), with the presence of solitary holes.
Simulations by \citet{Omura1994,Omura1996} published alongside the
observations suggested electron holes generated by electron streaming
as being the most likely explanation.  Demonstrating the solitary
nature of BEN, which had been observed for decades in many missions
using spectral instruments, notably in the ``Plasma Sheet Boundary Layer''
PSBL of the Earth's magnetotail, indicated that holes occur elsewhere
than in the polar regions. A wide ranging review in 2000 by
\citet*{Lakhina2000} gives helpful explanations of the near-Earth
plasma regions mentioned in this section, and discusses where
broadband fluctuations, including electron holes, occur. It also
surveys many theoretical ideas concerning the associated
mechanisms.

Geotail instruments also identified holes in the Earth's bow shock
\citep{Matsumoto1997}. Solitary holes in the data from the Galileo
satellite in the magnetotail when crossing a magnetic flux rope,
were observed by 
\citet*{Mottez1997}. The Polar satellite observed large amplitude
electric fields in the PSBL at intermediate distances, reported by
\citet*{Cattell1998}, including electron holes analyzed by
\citet*{Franz1998}. The auroral region plasma observations from the
FAST satellite,
\begin{figure}[ht]
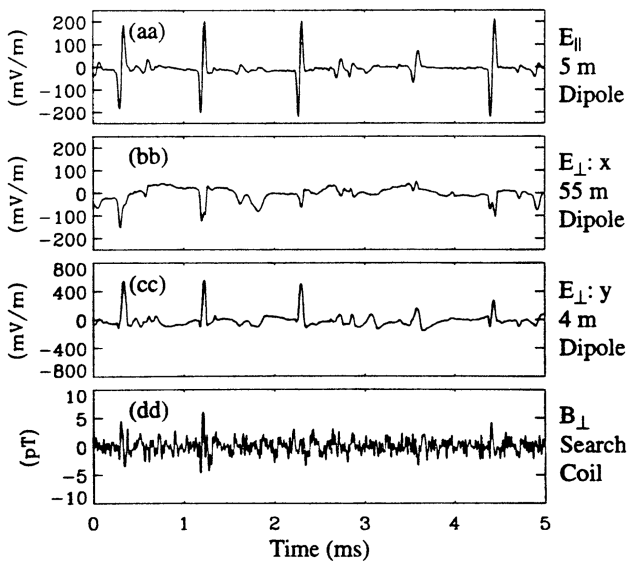

  \center
  \partwidth{0.6}{Ergunholes}
  \caption{Electric and magnetic field pulses observed from electron
    holes. Reprinted with permission from \citep{Ergun1998} copyright
    (1998) by the American Physical Society.\label{Ergunholes}}
\end{figure}
\citet{Ergun1998a,Ergun1998}, showed holes whose finite
transverse extent was indicated by unipolar transverse electric field
$E_\perp$ and magnetic field $B_\perp$ pulses, at the same time as
bipolar parallel electric $E_\parallel$ pulses, as illustrated in
Fig.\ \ref{Ergunholes}.
Further details are in \citet{Ergun1999}.

Pulsed electric fields indicating electron holes were seen in the
(120kHz!) data of the WIND satellite at the Earth's bow shock, by
\citet*{Bale1998}, and near the L1 Lagrange point (the free solar
wind, albeit with unusual potential shapes,) as reported by
\citet*{Mangeney1999}. (However more recent bow shock
observations\cite{Wang2020,Kamaletdinov2022}, to be discussed later,
call into question the presumed positive polarity of the structures'
potentials, and implicate instead ion holes.)  Thus, in just a few
years it had become clear that with the right instruments ---
rapidly-sampled electric probes --- electrostatic solitary structures
could be seen on occasions almost everywhere one looks in space.

Meanwhile, \citet{Saeki1998}  ran 1-D PIC simulations that included the
response of ions with $m_i=100m_e$, $T_e=40T_i$. They started with an
artificially grown hole, stationary with respect to the particle
distributions, having peak potential $\psi=1.2$. With immobile ions
thereafter the hole persists quiescent but with small symmetric
oscillations for at least 160 $(\omega_{pe}^{-1}$). However, turning
on the ion response, the hole immediately elongates and by time 40 has
split into two holes propagating in opposite directions, ending up
approximately 100 ($\lambda_{De}$) apart at time 160. These separating
holes are accompanied by an ion compressional pulse to which the
electron response seems coupled. These are called coupled hole
solitons (CHS), they are trailed by ion acoustic wave wakes. A
theoretical equilibrium analysis of the resulting CHS determines a
maximum hole amplitude $\psi$ for given speed and hole phase-space
area, with which the simulation observations appear to agree. Some
large initial potentials split up into more than two CHS structures.

A surge of simulation was triggered by the satellite measurements. And
the growing power of then-current high-performance computers enabled
multidimensional simulations with respectable resolution to be run for
long times extending deep into the non-linear stages of instabilities.

\citet*{Miyake1998a}  investigated bump-on-tail instabilities,
in a two-dimensional PIC simulation of limited extent, exploring
notably the effect of background magnetic field strength which we
express as the cyclotron frequency $\Omega$ (normalized to
$\omega_{pe}$) on electron holes. They found that at $\Omega=0.2$
holes formed but quite quickly decayed away while at $\Omega=1$, they
were sustained much longer, to times approaching
900$\omega_{pe}^{-1}$. At intermediate $\Omega=0.4$ they showed the
decay rate was sensitive to discrete particle simulation
noise. Writing the bounce frequency of electrons in the holes
$\omega_b$ (which depends on hole depth), these three strengths
correspond to $\Omega/\omega_b=$ 1, 5, and 2 respectively.

\citet*{Oppenheim1999} and \citep{Goldman1999}  performed 2-D
two-stream instability simulations with much larger domains, so the long time
behavior of truly isolated holes could be discerned. With strong
magnetic field $\Omega=5$, well defined holes extended in the
transverse direction were formed by time 448, but eventually (by time
1920) they were broken up by rather slow growing instabilities,
accompanied by long parallel wavelength streaks of small transverse
extent ($k_\parallel\ll k_\perp$) that were identified as lying on the
``Whistler'' branch of the cold plasma dispersion relation. Greater
clarity was obtained by starting the simulation in a 1-D mode
suppressing transverse potential variation; then after a time of
``many thousands'', when 1-D holes had developed, switching to 2-D
mode to observe the transverse instability grow from negligible
amplitude, to break up the holes in the next few thousand
($\omega_{pe}^{-1}$). Exploration of magnetic field strength showed
little difference for $\Omega>5$ but at $\Omega=0$ holes do not form,
and for $\Omega<1$ they form but quite quickly break up by transverse
instability, yet without generating whistler waves. 

The relationship between hole amplitude ($\psi$) and hole parallel
length was investigated by \citet*{Muschietti1999}. They
derived the minimum possible hole length $\delta$ for a given $\psi$
of holes with assumed Gaussian potential shapes. It is constrained by
the non-negativity of the trapped phase-space density, as explained in
\citep{Hutchinson2017}, leading to $\delta_{min}\sim \psi^{1/4}$. They
compared FAST observations with the theoretical dependence and found
good agreement within (substantial) uncertainty. These facts are
evidence against an alternative theory that the phenomena might be ion
acoustic solitons, because solitons have the opposite dependence of
length on amplitude. They also reported the beginnings of
two-dimensional simulation of pre-formed electron holes.

\citet*{Muschietti2000}  pursued further the 2-D simulation of
the transverse instability using prepared analytic electron hole equilibria
starting uniform in the transverse direction and observing the growth
of their kinks. The magnetic field strength and perpendicular
electron temperature were varied while keeping $\psi$ and the hole
length fixed. This fixes the bounce frequency for deeply trapped
electrons, $\omega_b$. They found that when $\Omega/\omega_b>1$ kink damping,
rather than growth, occurred. This stability criterion is in
qualitative agreement with rigorous theory to be discussed
later. \citeauthor{Muschietti2000} speculated that the mechanism of
the instability was ``electron focusing'' on the concave side of the
kink. That idea was disproved much more recently when the true
mechanism was discovered.

An observational statistical analysis of lasting importance was
that of Polar satellite data by \citet*{Franz2000}. Over a
range of magnetic field strengths $0.04<\Omega<4$, they found a
systematic relationship involving the ratio of the peak perpendicular
and parallel electric fields $E_\perp/E_\parallel$ for each hole. The
statistical average over 1003 holes was found to be well approximated
by the equation

\begin{equation}
  \label{Franz}
  \left\langle E_\perp^2\over
    E_\parallel^2\right\rangle^{-1/2}
  =\left(1+{\omega_{pe}^2\over \Omega_e^2}\right)^{1/2}.
\end{equation}
Simple mindedly, for a single potential structure of peak height
$\psi$ with perpendicular and parallel scale sizes $L_\perp$ and
$L_\parallel$ the order of magnitude of the perpendicular and parallel
electric fields are proportional to $\psi/L$. So the authors proposed
that $E_\parallel/E_\perp \simeq L_\perp/L_\parallel$ and that this
field ratio was therefore a measure of the aspect ratio, that is the
oblateness in 3-D, of the holes. This experimental scaling is widely
referred to in subsequent work. It amounts to the intuitively
plausible requirement that the transverse dimensions of an electron
hole cannot be smaller than roughly the electron
gyro-radius. \citeauthor{Franz2000} also offered a speculative
theoretical basis for the scaling, drawing on gyrokinetic
equations. We shall say more about the experimental measure and the
theory later.  \citet{Bale2002}  observed electron holes in data
from WIND, during bow shock crossings, displaying positive correlation
of length with amplitude and satisfying the width/amplitude constraint
$\delta>\delta_{min}\propto \psi^{1/4}$. Their electric fields were
predominantly in the parallel direction, indicating substantially
oblate holes, as expected from eq.\ (\ref{Franz}) since $\Omega$ was
small $\sim1/80$.

Addressing the related questions of transverse instability and
transverse hole extent, \citet*{Miyake2000}  performed 2-D PIC
simulations of initially symmetric two-Maxwellian-stream electron
distributions, at $\Omega=1$. A comparison was made between
simulations with immobile ions and mobile ions of mass $m_i=100m_e$
having ion temperature high enough to damp ion acoustic waves. Using
immobile ions, the unstable electron waves quickly organize themselves
by merging into a single hole extended across the entire transverse
domain. With mobile ions, by contrast, the initial organization is
into a few unaligned individual potential peaks of aspect ratio
approximately unity, which then align themselves into a hole of
greater transverse extent but with imperfections or gaps. Subsequently
the mobile ion case shows signatures of perpendicularly propagating
``Lower Hybrid'' waves that cause a slow decay of the hole amplitude.

Fully 3-D PIC simulations were performed by \citet*{Singh2001},
having 36 electrons per Debye cube $\lambda_{De}^3$: impressive for
the day, but subject to significant particle discreteness noise.  The
plasma was initialized with Maxwellian electrons plus a beam of
density 10\%, velocity $v_b=4$ (relative to electron Maxwellian), and
temperature $T_b<T_e$. The magnetic field strength $\Omega$ was
varied.  A qualitatively rather similar picture emerged of electron
holes forming, breaking up and re-forming in the first few hundred
$\omega_{pe}^{-1}$. For $\Omega>2$, longer lived holes were
observed. The authors used a mass ratio $m_i/m_e=1836$, and later in
time, they observed broken-up holes of approximately spherical shape
associated with shorter wavelength oblique waves they called ``Lower
Hybrid'', as well as longer parallel wavelength ``whistlers'' at
high $\Omega$.  Further discussion of these simulations is in
\citet{Singh2003}.

\citet{Schamel2000}  recapitulates electron and ion hole theory
based on the pseudopotential approach, but includes treatment of
periodic as well as solitary potentials. The treatment is unified
mathematically; and the small amplitude limit and relation to Van
Kampen modes is addressed.

Revisiting mathematical kinetic analysis of the transverse instability
(without reference to the much earlier \citep{Schamel1982}),
\citet*{Vetoulis2001}  attempted to address the coupling of an electron
hole to a wave via bounce resonance. However, their expansion of the
potential perturbation in a Fourier series of sinusoidal waves led
them to approximate the hole influence as a perturbation on the wave,
rather than supposing (more plausibly) that the wave was a small
perturbation on the hole oscillation. They concluded that
$\partial f/\partial \energy >0$ at bounce resonance was necessary for
instability. Although this is the familiar criterion for sinusoidal
wave growth in a uniform plasma, it turns out, as we shall see later,
that it does not apply to holes.

\begin{figure}[ht]
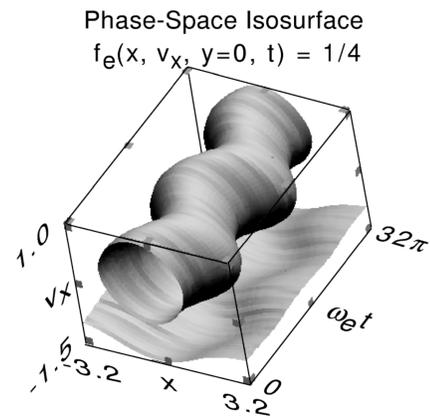
\center
  \partwidth{0.4}{NewmanKink}
  \caption{Vibrating electron hole kink. The bounding surfaces of the
    non-zero-$f(v)$ waterbag are
    shown. Reprinted with permission from \citep{Newman2001a}
    copyright (2001) by the American Physical society.\label{NewmanKink}}
\end{figure}

\citet*{Newman2001a}  simulated in two dimensions a waterbag electron
hole at high $\Omega>\omega_{p}$. They imposed an initial uniform kink
displacement in the parallel direction $\xi$, with transverse
dependence $\propto \exp(ik_yy)$ as a model of the transverse
instability's eigenstructure. The hole experiences a vibration, illustrated
in Fig.\ \ref{NewmanKink} showing the time evolution of the phase space
surface at fixed $y$. In view of the $\exp i(k_yy-\omega t)$
dependence this figure could also be construed visually as the
kink's $y$-dependence with the $t$-axis corresponding to $(k_y/\omega)y$.
An analytic treatment derives the predicted real part of the frequency
of the oscillation, which agrees very well with the simulation. They
proposed a coupling between the hole and the wave as the mechanism to
explain the growth of oscillations and accompanying whistler waves
that were observed in a larger domain able to accommodate them. This
work, though highly simplified, identified several important features of
the transverse instability.

A somewhat similar approach, accounting only for parallel electron
motion, was taken by \citet*{Berthomier2002}, in which the
resonant energy transfer from a sinusoidal whistler wave to the
trapped particles of an electron hole and the energy transfer by
Landau damping to passing electrons both contribute terms proportional
to $+df/d\energy|_{resonant}$ to the growth rate.  Again, this
approach effectively ignores the motion of the hole, and makes
coupling to whistler waves the primary mechanism. Instability with growth
rate of order a percent of the wave frequency, was found over most of
the domain $0\le k_\perp\lesssim 1$ and
$0.1\lesssim \omega\lesssim 1$. This mechanism is further explored for
three-dimensional holes by \citet{Berthomier2008}.

\citet{Jovanovic2002}  is a more comprehensive assault on the
linearized transverse instability in a magnetized hole, solving
Vlasov's equation by integration over prior time accounting for the
full helical motion of the electrons in the unperturbed potential.
Expansion in perpendicular cyclotron harmonics naturally arises.
Different approximations of sometimes questionable applicability are
adopted for handling different resonances and non-resonant and passing
particles analytically. It concludes that bounce resonance ``yields
the destabilization of the hole''. At low frequency
($\omega\simeq \omega_b$), large growth rates and hole breakup might
occur, while at high frequency when $\Omega\pm\omega\sim \omega_b$
``the instability manifests in the emission of linear waves''. No
purely growing instabilities ($\Re(\omega)=0$) are found. These
predictions are now known not to be in agreement with simulation.

\begin{figure}
  \partwidth{0.65}{Rothdata}\hskip-2em(a)\
  \vbox{\ifx\twocoltrue\undefined\hsize=0.33\hsize\else\hsize=0.6\hsize\fi
    \includegraphics[width=\hsize, height=1.4\hsize]{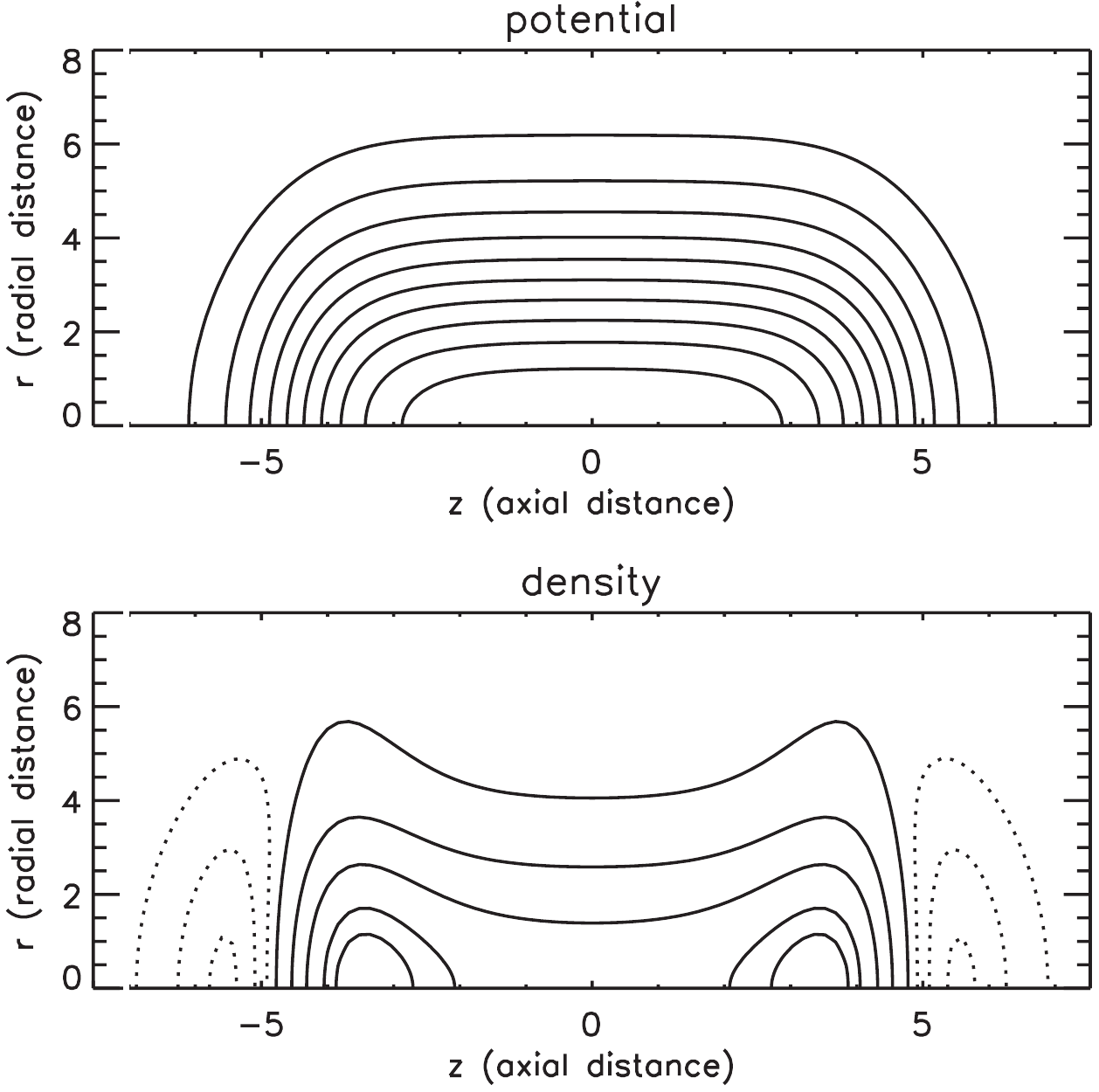}\hskip-1.5em(b)\par\smallskip
  }
  \caption{(a) Perpendicular and parallel electric field as a function of
    time, observed by the FAST satellite (time relative to 1996/12/29
    19:36:24), indicating holes of modest transverse extent but
    substantially elongated in the parallel direction (pink boxes). In
    the blue box, multiple such holes seem to be
    interacting. (b) Model contours of an elongated hole potential
    and electron density. After \citep{Roth2002}. \label{Rothdata}}
\end{figure}

\citet*{Roth2002}  presents FAST satellite data that shows
electron holes of limited transverse extent, but strongly elongated in
the parallel direction. This shape is indicated by separated positive and
negative parallel electric field spikes, between which there is
unipolar substantial perpendicular electric field. See Fig.\
\ref{Rothdata}(a).
A model of these elongated holes [Fig.\ \ref{Rothdata}(b)], shows that
they require non-monotonic trapped velocity distributions. And the
authors analyze the interaction of multiple such holes, solving single
particle orbits in prescribed fields. They also present a
multidimensional PIC treatment of interacting holes, in which the
smaller hole is absorbed (``cannibalized'') by the larger.

Axisymmetric three-dimensional electron hole equilibria, but without
parallel elongation, were analyzed by \citet{Chen2002}, and
revisited by \citet*{Chen2004}. They used the integral equation
method and derived analytic forms for the required trapped
distribution with assumed potential shapes Gaussian in the parallel
and perpendicular directions. They derived lower limits on the
parallel and perpendicular extent as a function of hole amplitude and
ion to electron temperature ratio. The limits arise from the
requirement that the distribution must be
non-negative. \citet*{Chen2005}  develop the treatment more
fully, and show favorable comparison with Polar spacecraft
observations. \citet*{Franz2005}  present those observations in
more detail and argue that only electron holes (not electron-acoustic
solitons) are consistent with them.

In \citet*{Krasovsky2004}, theoretically analyzing
multidimensional electron holes, it is argued that, whatever the
spatial form, substantial velocity anisotropy of the particle
distribution is essential for holes to exist. This argument was given
more rigorous form by \citet{Ng2005}  who qualify it by showing
that if the background distribution has dependence on angular momentum
(conserved in a spherically symmetric hole) then in
principle spherically symmetric solutions exist. The background
distributions at infinity are however pathological (see
\citep{Hutchinson2017} endnote 68), so the practical conclusion
remains the same.

\citet*{Friedland2004}  report an unusual type of electron hole in a
pure electron plasma trap. The holes are directly excited by applying
a resonant wave perturbation whose frequency is chirped on a
time-scale slow compared with the electron bounce/transit time. The
velocity resonance then moves from the outer wing of the background
electron distribution into its bulk. As it does so, it drags a
``bucket'' of weakly populated trapped particle orbits into a much
more strongly populated region, forming an electron hole. Theory,
simulation, and laboratory measurements are reported to be consistent.

Meanwhile simulations were growing in size, speed, and sophistication,
and illustrating a great variety of complex behavior.
\citet*{Goldman2003}  performed a one dimensional simulation
with equal temperature ions and electrons ($m_i/m_e=400$) in a long
$640\lambda_{De}$ box with open (rather than periodic) boundaries,
starting with electron drift (relative to ions) of $v_d=v_{te}$, but
initializing with a local charge-neutral density depression. A double
layer rapidly forms; and accelerates a beam of electrons into the
background, giving rise to two-stream instability, electron hole
formation, and merging. In the opposite direction, on a longer
timescale, an ion beam gives rise to trains of ion and electron
holes. Similar trains of ion holes had been observed by
\citet*{Boerve2001} in 2.5 dimensional hybrid PIC simulations of ion
beam-plasma interactions with Boltzmann electrons. The ion holes were
not fully solitary structures in either study, but were clearly ion
phase-space vortices with central phase-space depletion.

\citet*{Drake2003}  performed three-dimensional PIC simulations
($m_i/m_e=100$) of magnetic reconnection with a guide field
($\Omega=2.5\omega_{pe}$). They observed large current densities,
well beyond the Buneman stability limit, parallel electron beams, 
and consequent electron holes along the x-line and magnetic
separatrix. The holes and their electron scattering were implicated in
anomalously high effective resistivity.

\citet*{Eliasson2004}  simulate electrons and ions of large mass
ratio ($m_i/m_e=29500$) using a (one-dimensional) Fourier Vlasov
code. They start with a Schamel type pure electron hole (uniform ion
density) at zero velocity with respect to ions, and observe that after
dwelling stationary for a time $\sim 120\omega_{pe}^{-1}$, the
electron hole suddenly moves away with a velocity $\sim 0.55v_{te}$,
leaving behind a stationary negative potential peak associated with
the ions that have been repelled during the dwell period. This is the
clearest early demonstration of ``self-acceleration'' of a stationary
electron hole by repulsion from an ion density depression. They also
observe merging of two holes of the same type when they collide.
\citet*{Eliasson2004a}  consider ion holes with and without
trapped Langmuir waves providing extra ponderomotive force on the
ions, and simulate collisions of counter-propagating ion holes. 

A different way to generate electron holes was simulated by
\citet*{Califano2005}, who applied a large amplitude forcing
field to one end of their one-dimensional Vlasov-Maxwell (not just
electrostatic) code. The Langmuir wave packet generated was chosen to
have frequency $\omega_0=1.1$ and by ``wave breaking'' was observed to
induce electron holes that traveled much faster than the group
velocity of the packet.

A more extensive two-dimensional electromagnetic PIC study of warm two
stream electron distributions was undertaken by \citet*{Umeda2006},
but with immobile ions. They specified the initial thermal velocity
spread of the beams, $v_t$, normalized to the beams' drift velocity
$\pm v_d$, varying the ratio $v_t/v_d=0.25-0.5$, and the magnetic
field strength $\Omega=0.5-10(\omega_{pe})$.  In the nonlinear state,
the passing electron temperature $T_e$, hole potential $\psi$, and
bounce frequency $\omega_b$ were estimated from the simulation.
\begin{figure}[ht]
  \centerline{\vbox{\hsize=25em
  \hbox{$\omega_{pe}t$\qquad\qquad $\Omega=1$\qquad\qquad
    $\Omega=10$\qquad\qquad $\Omega=0.5$\qquad}\noindent
  \includegraphics[width=\hsize]{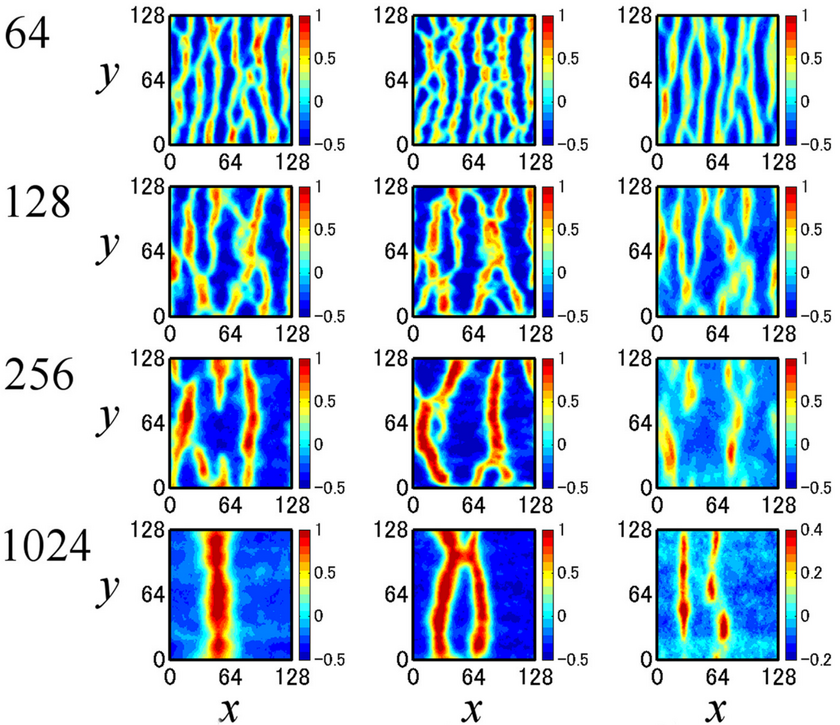}
  }}
\caption{Hole evolution in warm two-stream instability PIC 2-D
  simulations for three magnetic field strengths $\Omega=1,10,0.5$ at
  four times $t=64,128,256,1024$, for the low-potential-energy case
  $v_t/v_d=0.5$. Whistler waves are not significantly excited. The
  parallel direction is $x$. From \citep{Umeda2006}.  \label{UmedaSim}}
\end{figure}
The crucial new finding, illustrated in Fig.\ \ref{UmedaSim}, was that
when $v_t/v_d=0.5$, giving hole potentials $\psi/T_e\lesssim 1$, the
final state was a long-lasting (up to $t$=10000) one-dimensional hole
(uniform in the transverse direction) when $\Omega\gtrsim 1$, while
for $\Omega=0.5$ the hole was broken up transversely with only slowly
decaying $\psi$ (reaching zero at time 5000), in approximate agreement
with the magnetic field strength criterion $\Omega\gtrsim \omega_b$
for transverse stability. By contrast, for $v_t/v_d=0.35$ (colder
streams) the holes are higher in amplitude ($\psi/T_e$ up to 4), and
give rise to whistler waves that break up the holes transversely by
$t$=1000 even for $\Omega = 2$, as had been observed for example by
\citet{Oppenheim1999} with similar beam parameters. They conclude that
the whistler breakup effects do not occur when ``the potential energy
of the holes is smaller than the electron thermal energy''. Later
simulations with much longer domains and including ion dynamics by
\citet{Umeda2008}, indicate somewhat enhanced susceptibility to
low frequency wave excitation which causes the holes to decay faster:
on a timescale of a few thousand.  \citet*{Lu2008}  revisit
two-dimensional two-stream instability including ions $m_i/m_e=1836$
traveling at the velocity of one of the electron streams, but only
with $v_t/v_d=0.25,0.22$: the high potential energy case. They observe
the whistler waves, and with immobile ions ($m_i/m_e=\infty$) the wave
amplitude is much greater.  Artificially low $m_i/m_e=100$ permits
lower hybrid waves to be excited quite early in the simulations.

\citet{Newman2007}  performed one- and two-dimensional continuum
Vlasov simulations of a double layer, showing how the resulting
two-stream electron distribution on the higher-potential side gives
rise to electron holes propagating away from the layer and merging as
they go. The two-dimensional hole shapes are visually similar to those
in Figure \ref{UmedaSim}, and the effects of different degrees of
magnetization were also explored. The double-layer acceleration
results in continuous production of holes in this open-boundary
simulation.

In analytic theory, \citet*{Goldman2007}  performed analysis
aimed at trying to deduce information about the passing particle
velocity distribution shape based on the observed spatial length of
the hole's potential, for small amplitude $\sech^4$ holes.

In further observational analysis, \citet{Pickett2004} made a search,
at Earth-distance 4.5-6.5$R_E$, for solitary structures observed by
more than one of the satellites of the Cluster mission. Unfortunately
no persuasive correlated examples were discovered. This is perhaps not
entirely surprising since distances of at least 80km separated the
individual craft, but it places an upper bound on the lengths over
which the electron (or ion) holes preserve their shape. Later,
pursuing correlated holes, \citet{Pickett2008} reported a single
example of a solitary structure seeming to appear at two satellites.
The satellites were at Earth-distance 11.8$R_e$ and spaced by 30km
parallel and 40km perpendicular to the magnetic field. The
observations' time difference was 22ms; so if they really were of the
same hole, one deduces parallel speed of 1300km/s, and a very flat
oblate shape.

The
electron hole Cluster data analyzed statistically by
\citet{Cattell2005}  indicated correlation between the presence
of holes and relatively narrow electron beams. The beams were
predicted by reconnection simulations: an important indication that
electron holes are part of the complicated processes involved in
reconnection.  It was found in Cluster data by \citet{Pickett2005}
 that solitary waves occur throughout the Earth's magnetosheath
from Bow Shock to Magnetopause. They infer that holes are being
generated throughout that region, perhaps as a result of two-stream
instability.

\subsection{Lab and Space holes, renewed theory, 2008-}
\label{labspace}

The first new laboratory observations of electron holes since 1979
arose in reconnection experiments and were reported by
\citet*{Fox2008}  and \citet{Fox2012}. The holes were detected
by small (60$\mu$m dia) electric probes sampled at up to 5GHz. Two
probes spaced 4.6mm apart along the magnetic field observed a delay of
order 1.2ns in the arrival time, which enabled the hole parallel speed
and direction (in the positive electron acceleration direction) to be
deduced. The holes were determined to be roughly spherical and showed
negligible arrival time delay when the probes were spaced instead in
the perpendicular direction. Unusual features included long hole
lengths of $\sim 60\lambda_{De}$ and fast speeds up to
$2v_{te}$. Possible local enhancement of electron temperature might
decrease these normalized estimates, but in any case they are not
theoretically impossible in substantially distorted electron
distributions.

In 2010, laboratory electron holes (as well as wave packets and
irregular fluctuations) were observed by \citet*{Lefebvre2010}. They
were generated by an electron beam injected into LAPD, a large
magnetized basic plasma device, and were detected with small arrays of
electric probes of diameter 10$\mu$m, sampled at speeds well above the
local plasma frequency (0.3-0.7GHz). The observed coherent propagation
from one probe to another provided speed and size measurements. The
mechanism of hole formation was rather uncertain, because they were
observed in an electron velocity distribution having beam particles of
energy $\sim 60$eV, greatly broadened by earlier instabilities. The
cold background plasma with $T_e\sim 0.2$eV made up the remaining
$\sim75$\% of the density. Hole amplitudes of $\sim0.1-0.25$V were
observed, extending upward to 0.75V at high magnetic field
$\Omega/\omega_{pe}\simeq 4$. 

\citet*{Andersson2009}  observed a new phenomenon, parallel magnetic
field enhancement inside electron holes, during a ``bursty bulk flow''
event measured by a THEMIS satellite at $\sim10R_E$. Prior
perpendicular magnetic perturbations (e.g.\
\citep{Ergun1998a,Ergun1998,Ergun1999}) had been attributed to Lorentz
transformation of perpendicular electric fields in moving holes. But
parallel $B$-fields cannot arise by that mechanism. Instead
$B_\parallel$ perturbations, always enhancements, can arise from
electron $\E\wedge\B$ drift current in holes of limited transverse
extent. The hole velocities ($\sim 10^8$m/s c.f.\
$v_{te}\sim 0.4\times10^8$m/s) were deduced from the relative
magnitudes of perpendicular $\B$ and $\E$ taking $B_\perp$ to arise by
Lorentz transformation. The resulting hole lengths were typically
$L_\parallel\simeq 20-30\lambda_{De}$, somewhat greater than the
deduced $L_\perp$. In these hot plasmas, $T_e\sim 8$keV, potentials
were $\sim 0.5T_e$. The authors note that long, fast ($v_h>v_{te}$)
holes like these were characteristic of the laboratory reconnection
observations of \citet{Fox2008}.

\citet{Tao2011}  followed up with detailed theory of
electromagnetic electron holes. They addressed the key question of
whether it is justifiable to include the electron $\E\wedge\B$ drift,
but ignore the corresponding ion drift, which in a steady uniform
plasma is exactly the same and would cancel the current. Using two
dimensional test particle orbit calculations they determine the ratio
of the actual current density to the presumed $\E\wedge\B$ drift
current as a function of the ratio of transit time to cyclotron period
$\delta t \Omega/2\pi$. The (plotted) current ratio is approximately unity
above $\delta t \Omega/2\pi\simeq 0.4$, but declines steeply below,
becoming $<0.15$ at $\delta t \Omega/2\pi<0.15$. Electrons in the
THEMIS observations have $\delta t \Omega/2\pi\simeq 1$, validating
their current; ions have cyclotron frequencies $\sim 1000$ times smaller,
making their current negligible.  The authors also derive a more
accurate estimate of the hole velocity $v_h$ by including the drift
current effects, and use it to analyze in more detail the same data as
\citet{Andersson2009}. They find $\sim20$\% lower length and speed
estimates; and invoke the two-dimensional electromagnetic simulations of
\citet{Du2011} in support of their model.

\begin{figure}[ht]
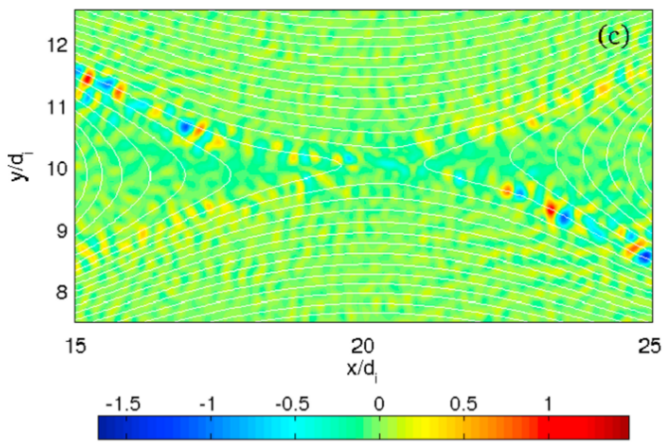
\center
  \partwidth{0.65}{LapentaRecons}
  \caption{Parallel electric field contours in a reconnecting
    magnetic field. Electrons are strongly accelerated inward along
    the top left and bottom right separatrix, giving rise to electron
    holes that appear as blue and red bipolar $E_\parallel$
    regions. After \citep{Lapenta2011}.\label{LapentaRecon}}
\end{figure}
Massive two-dimensional PIC simulations of reconnection at physical
mass ratio, in boxes large enough to avoid strong boundary condition
influence, were performed by \citet{Lapenta2010,Lapenta2011}.
It was found that ``bipolar structures'' in the parallel electric
field (electron holes, see Fig.\ \ref{LapentaRecon}) were generated
for all values of the guide field, and not just when the guide field
is strong as was the case in \citep{Drake2003}. This new result
accorded better with the Cluster observations \citep{Cattell2005}. A
notable reconnection event encountered by Cluster and analyzed by
\citet*{Viberg2013}  identified electrostatic solitary
structures ``localized to the separatrix regions'', broadly consistent
with the simulations.  \citet*{Goldman2014}  followed up with
simulation and analysis that shows the electron holes emitting
quasi-parallel ($k_\perp\ll k_\parallel$) whistlers by a Cerenkov
process. The whistlers grow till they cause the holes to decay, and
can then persist further than the holes.

\citet*{Lesur2014}  perform one-dimensional Vlasov and PIC
simulations of high resolution showing persuasively that the supposed
sub-critical ion acoustic instability growth obtained by
\citet{Berman1982}, which \citet{Dupree1982} developed his ion hole
growth mechanism to explain, does not really happen. Lesur was able to
reproduce Berman's result by using the small number of PIC particles
he used; but using much greater numbers, and hence reduced noise, or
using the continuum Vlasov code, the effect was suppressed.
Dupree's hole growth phenomenon remains a possible real effect, but it
does not produce subcritical Buneman instability and hole appearance. 

\citet*{Malaspina2013}  made a statistical study of electron
holes in the free solar wind, and discovered that 47\% of them occur
within 6000km of a current sheet, whereas only 17\% randomly chosen
times are within the same range. Holes that were even closer tended to
have higher amplitudes. They conclude that mechanisms producing
electron holes are active near current sheets (where reconnection is
likely occurring) in the solar wind. And \citet*{Malaspina2014,Malaspina2015} 
observed in data from the Van Allen Probes double-layers generating electron
holes all associated with ``dipolarizations'' of the magnetic field
and other plasma boundaries within the magnetosphere.

\citet*{Vasko2015}  report magnetic perturbations associated
with electron holes measured by the Van Allen Probes in the outer
radiation belt. The key novelty is that, unlike the prior observations
\citep{Andersson2009} and theory \citep{Tao2011} of $\E\wedge\B$
electron current, the parallel field $B_\parallel$ is \emph{depressed}
rather than being enhanced within the holes. They propose that this
sign reversal might be explained by diamagnetic current, associated
with a trapped electron population that is extremely hot (e.g.\
$T_\perp\simeq$ 0.3 or 6keV), anisotropic ($T_\perp/T_\parallel>2$),
and dense ($n_t\sim 65$\% of unperturbed), even though the cold
background is only a few eV. If this interpretation is correct, these
would be extremely exotic electron hole parameters. \citet*{Mozer2015}
reported a month later on ``Time Domain Structures'' (TDS) observed by
Van Allen Probes in the outer radiation belt. These are also of
various intense and exotic types, arising shortly after plasma
injections into the belt. They appear to be responsible for production
of anisotropic velocity distributions strongly enhanced at small pitch
angles (i.e.\ \emph{parallel} to $B$).

A previously under-appreciated environment in which electron holes are
to be expected is in the wake of unmagnetized bodies such as the Moon,
as plasma such as the solar wind, flows past. Solitary electrostatic
structures were observed by satellite Galileo in the wake of Jupiter's
moon Europa, \citet{Kurth2001} ; and near to the (Earth's) Moon,
\citet{Hashimoto2010}. \citet*{Pickett2015}  surveyed ESW
observations by Cassini in the vicinity of Saturn, and noted holes
clustered in the wake of its moon Enceladus, although impacts from its
dust plume somewhat polluted the measurements.  Moreover the Moon wake
had long been modelled analytically
(\citep{Gurevich1969,Gurevich1975}), and by simulations
(\citep{Farrell1998a,Birch2001a}) at low mass ratio ($m_i/m_e=20$,
inadequate to separate the species) that hinted at hole formation. In
2015 however, physical mass ratio simulations by
\citet*{Haakonsen2015}, and analysis by \citet*{Hutchinson2015} showed
theoretically how electron holes are spawned and grow in the wake. The
underlying instability mechanism is the inflow along transverse
magnetic field lines of electrons and ions into the density-depleted,
negatively charged, wake. In combination with the steady downstream
convection of the fieldlines and particles, a dimple (or
``notch'') arises in the electron velocity distribution within the wake,
found from the solution to the Vlasov equation
\citep{Hutchinson2012}.  The dimple is Penrose unstable and nonlinearly
forms small holes. Those holes grow as they are convected downstream
into the rising wake density by a mechanism analogous to that of Dupree
\citep{Dupree1983}. The simulations observe this hole amplitude
growth, especially for holes that are prevented by ion interactions
from being repelled out of the wake. The grown holes disrupt the ions
even before the ion distribution itself becomes unstable.

\begin{figure}[ht]
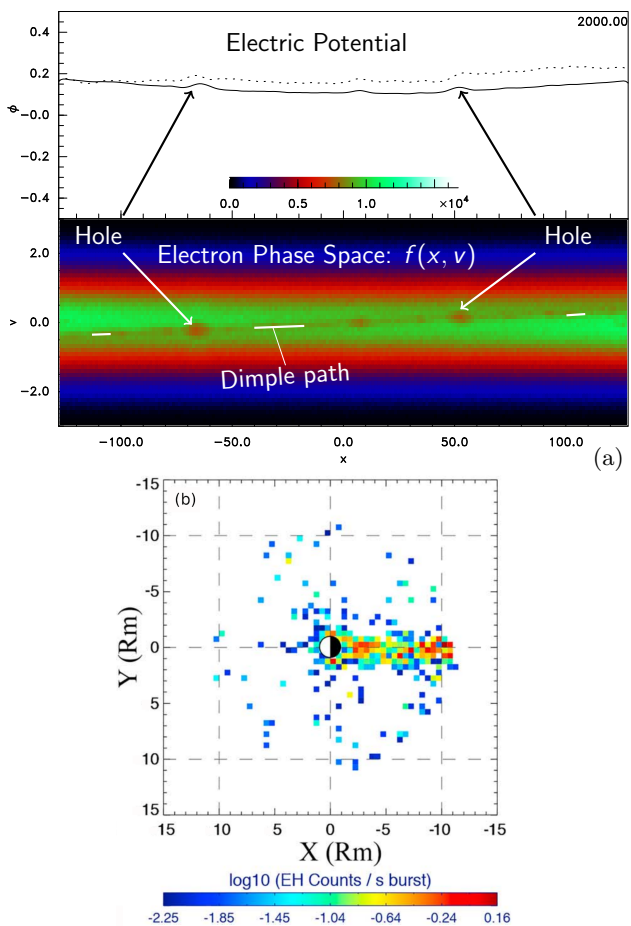

  \partwidth{0.6}{denriseholes}\hskip-1.5em(a)
  \partwidth{0.38}{wakeholes2}
  \caption{(a) Simulation of electron holes induced in a wake. (b)
    Observations of hole occurence near the moon. The wake streams
    away from the sun toward negative X. From \citet{Hutchinson2018b}.
    \label{denriseholes}}
\end{figure}

In view of the substantial accumulated data of the ARTEMIS mission's
two satellites orbiting the (Earth's) Moon out to about 10 Moon radii,
\citet{Hutchinson2018b} undertook to compare the theoretical
predictions of wake electron hole production with observations inside
the Moon wake when it is in the solar wind. Analytic solution of the
Vlasov equation characterizes the counter-streaming electron (and ion)
distributions filling in the wake along the magnetic field. They are
linearly unstable. PIC simulations were also performed that model the
rising density of the wake, as illustrated in figure
\ref{denriseholes}(a).
The lower frame shows color contours in phase space of a fully
developed nonlinear state in which the dimple appears as a depressed
phase-space density $f_e(x,v)$ diagonal line in which are embedded
electron holes nonlinearly emerging from its instability. Those holes
are observed to propagate out of the (open boundary) simulation along
the track of the dimple, which is also the phase space orbit of
electrons. The potential across the modelled wake is depressed in the
center, which is responsible for the hole acceleration.

It proved not to be possible to determine in the satellite data the
hole velocities, because of sampling speed limitations, but hole
existence was reliably determined. The statistical data is shown as a
two-dimensional histogram projected onto the ecliptic plane in figure
\ref{denriseholes}(b).  The Moon wake streams away from the sun toward
negative X, and in it are by far the majority of the observed electron
holes. The 161 single holes observed outside the wake in the 10 $R_m$
disk traversed are typical of elsewhere in the solar wind where they
are often associated with current sheets. These are only $\sim4$\% of
the total observed, the rest are clustered along the wake. The
detailed profile of the hole observations within the wake shows that
the hole density is somewhat hollow, peaking off the wake axis. It is
concluded that holes are formed within the wake, and propagate
outward, confirming the theoretical predictions. But must also break
up on a length scale less than the wake width, since few holes are
observed to propagate outside the wake.

Since the Moon's orbit around the earth takes it from the solar wind
across the bow shock and into the magnetosphere, the ARTEMIS data also
documents how electron hole occurrence varies correspondingly.
\begin{figure}[ht]
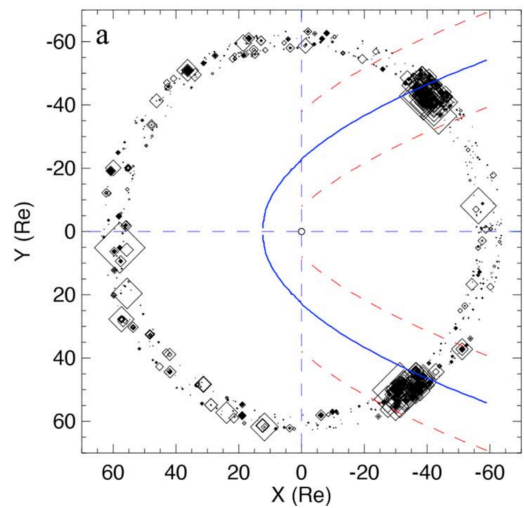
\center
  \partwidth{0.5}{moonorbit}
  \caption{Occurrence of electron holes near the moon as it moves
    from the solar wind across the Earth's bow shock (blue parabola) into the
    magnetosphere. Symbol area denotes potential amplitude.
    From \citet{Malaspina2019}\label{moonorbit}}
\end{figure}
\citet{Malaspina2019}  report the Artemis data of figure
\ref{moonorbit}, showing all the holes detected. As we already know
from figure \ref{denriseholes} almost all of them are in the Moon's wake
when it lies in the solar wind, to the left of approximately the blue
parabolic curve. However, the statistics show that only a very small
fraction of those that occur when the Moon is in the Magnetosphere, to
the right of the curve, are observed to lie in the wake. Instead they
are essentially isotropically distributed around the Moon. Since there is
little systematic steady plasma flow there and no appreciable wake, a different
mechanism is their cause. Also note the intense electron hole
occurrence in the vicinity of the bow shock and magnetopause (between
the two red dashed lines). It is concluded that at least four
different processes cause the holes observed in the data. When the
moon is in the solar wind the holes are caused by (i) its wake, and
(ii) current sheets in the solar wind. The magnetospheric mechanisms
can be grouped as (iii) magnetotail processes and (iv) processes at
the interface of magnetosphere and solar wind. 

\citet*{Norgren2015a}  were successful in finding
multi-spacecraft detection of electron holes in Cluster data obtained
seven years earlier. The holes were observed in the PSBL where the
Debye length was $\lambda_{De}\sim 2$km; and the hole parallel lengths
were typically $2-4\lambda_{De}$. The two spacecraft were separated by
$\sim30$km parallel and $\sim20$km perpendicular to the magnetic
field. Time delays of 50-70ms for the holes to pass between them were
observed; and identification of the corresponding holes was made
persuasive by the preserved shapes of trains of multiple holes of
different amplitudes.  The hole amplitudes were $\sim0.1T_e$ and
speeds in the ion frame were approximately 2\% of the electron thermal
speed. That would place their velocity within the ion distribution
spread if $T_i\simeq T_e$. These are then \emph{slow} electron
holes. Their generation mechanism could not be definitively
identified. However, \citet*{Norgren2015}  analyze different
electric field data from Cluster together with measured velocity
distributions to explore the sorts of instabilities that might give
rise to the observed electron holes. They use two Maxwellian electron
components and one ion component. Although their fits to the observed
$f_e(v)$ do not result in instability, they show that a narrowing of
electron beam spread can give instability, and suggest that effect
might represent the distribution prior to the effects of instability
spreading the beam. \citet*{Graham2016}  present more systematic
statistical analysis of hole parameters as a function of background
plasma parameters, for the magnetopause data from Cluster. In summary,
they find a wide range of hole speeds (relative to ions) from
$\sim v_{ti}$ up to $\sim v_{te}$. There is strong linear correlation
of hole length with $\lambda_{De}$: the peak to peak ($E_\parallel$)
length is $\simeq 9\lambda_{De}$. Maximum potentials are small
$\sim 0.01T_e$, and easily satisfy the existence requirement
(\citet{Chen2004}) of non-negative trapped distribution. An extensive
review of the Cluster mission observations of solitary electrostatic
structures has been published by \citet{Pickett2021}.

Throughout this prior history of electron hole research a remarkable
missing element was direct experimental measurement of the trapped
phase-space deficit responsible for sustaining the potential. The
reasons are that laboratory experiments have not had velocity
distribution measurements (at all) and the impressive satellite
measurements of the velocity distributions have not had sufficient
time (and hence space) resolution to distinguish spatially between
inside and outside the hole.
The MMS particle distribution instruments give fast single energy,
angularly resolved, measurements every 0.195ms, which is shorter than
the hole transit time. The energy is scanned sequentially in time to
produce the entire distribution.  \citet*{Mozer2018} used MMS data
from four satellites transiting 20 similar holes to reconstruct a
statistically ``superposed epoch'' approximation to the distribution
function with sufficient resolution, for the first time.
\begin{figure}[ht]
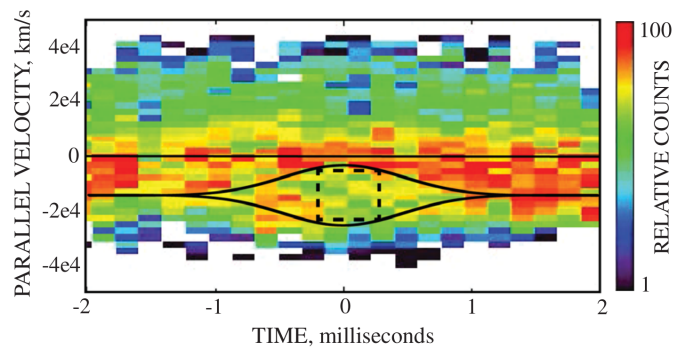

  \center
  \partwidth{0.65}{mozerfofv}
  \caption{Reconstructed ``superposed epoch'' phase space contour plot of
    the velocity distribution in an electron hole set observed in the
    magnetosphere. Reprinted with permission from \citet{Mozer2018}
    copyright the American Physical Society (2018).  \label{mozerfofv}}
\end{figure}
The result
is shown in figure \ref{mozerfofv} in which the horizontal (time) axis
corresponds to parallel position ($z$). It is therefore equivalent to
a phase-space contour plot of $f(x,v)$. The trapping region lies
within the square box, in which there is a detectable decrease of
phase-space density relative to the surrounding region. Although the
statistical significance is somewhat disappointing, this result
required a tour de force of analysis.

The MMS Electron Drift Instrument (EDI) data for a different train of
holes was analyzed by \citet{Norgren2022}. It measures the difference
in electron flux between two relatively narrow collimated directions
at comparable positive and negative velocities, obtaining 1000 samples
per second: (just) sufficient to resolve the satellite's passage
through the holes. This measurement's value is that it can be used to
analyze single holes and demonstrate the flux decrease expected from
the reduced trapped phase-space density. The correlation between hole
presence and flux decrease is very good; although the (one
dimensional) model predictions of the flux decrease are roughly a
factor of two less than those observed, perhaps because of
multi-dimensional effects. Nevertheless this data contributes
importantly to confirming trapped electron deficit experimentally.

\section{Time Dependence and Hole Kinematics}
\label{Kinematics}

We now move to a discussion, organized topically, of progress since
approximately 2016. The perspective adopted is primarily a theoretical
one, aimed at reporting and clarifying the understanding of different
physical mechanisms responsible for electron and ion hole
behavior. There have been many important contributions too from
satellite observation and these will be brought out within the
relevant topics. The present section concentrates on ways in which holes
respond to the background plasma, limiting the discussion to
essentially one-dimensional effects.

\subsection{Landau Damping and Resonant Interactions}
\label{LandauD}

For a steady self-consistent electron hole equilibrium, obviously
collisions eventually become important and cause it to decay on the
collision timescale. Simulations usually have some equivalent degree
of dissipation arising from particle discreteness or higher order
effective viscosity for example. Our treatment continues to neglect
collisions and ignore the dissipation timescale.

However, the question sometimes arises whether soliton or hole
equilibria experience Landau damping when there is a slope on one of
the background distribution functions at the velocity of the potential
structure. After all, Landau damping is an important collisionless
dissipation mechanism. The one-dimensional answer is that in most
situations electron hole equilibria do \emph{not} experience Landau damping;
but something else can happen.

One way to understand this answer
qualitatively is to realize that a BGK soliton or hole is like (and
sometimes actually is \citet{Carril2023}) the final nonlinear stage of
a coherent kinetic wave, in which Landau damping (or growth) has
stopped in a perfectly collisionless plasma because of non-linear
particle trapping (see e.g. \citet{Swanson1989a} chapter 8, and early
theory such as \citet{ONeil1965}).

Long ago, it was shown that nonlinear steady undamped periodic or
solitary waves can be constructed arbitrarily close to uniform
equilibria, by choice of the trapped distribution.
\citet{Holloway1991,Buchanan1993,Korn1996}. Later observations by
\citet{Montgomery2001} of a signature of electron acoustic wave
contributions to stimulated Raman scattering, which was unexpected on
the basis of strong linearized Landau damping, motivated the further
question as to how undamped periodic waves can grow, driven (by
modulated ponderomotive forces) out of an initially uniform plasma
background. \citet{Krapchev1980} had addressed the ``adiabatic''
evolution under the conservation of phase-space action, finding a
dispersion relation of what is sometimes called the ``thumb'' shape,
which corresponds to the hypothetical ``undamped wave'' of
\citet{Stix1962}. The dispersion curve has, in addition to the
Langmuir wave, a lower frequency branch that today would be called an
electron acoustic wave.

Simulations driven by an applied sinusoidal field, to model the SRS
situation of \citeauthor{Montgomery2001}, found non-linear waves that
persisted approximately undamped well after the driver was turned
off. They were dubbed Kinetic Electrostatic Electron Nonlinear or KEEN
waves, and are comprehensively reported by \citet{Johnston2009}, and
recapitulated more recently in \citet{Bertrand2019} chapter 5.7-8.

Waves of the KEEN type have been shown sometimes to experience what is
called the Negative Mass Instability (NMI)
\cite{Dodin2013,Hara2015}. It is a type of Trapped Particle
Instability \cite{Kruer1969} arising from the positive derivative of
the trapped velocity distribution with energy (action), and consists
of bunching of trapped electrons locally on their phase space orbits
within the trapped vortex. \citet{Brunner2004} proposed that NMI is
responsible for the burstiness of SRS.  \cite{Dodin2014} suggest that
the NMI plays a vital role in the \emph{formation} of persistent
waves. But countless simulations show fully formed steady
electron holes with no NMI.

\citet{Valentini2012}, surveyed the applicability of Landau damping and
discussed the suppression of Landau damping to give undamped waves.
But a related published comment \citet{Schamel2013} disputed how to
understand the important phenomena. Controversy and misunderstanding
still persist.

A different perspective is to recognize that Landau damping
arises inherently from the initial conditions, which is why Landau
used a Laplace transform (not a Fourier transform) in his original
derivation\citep{Landau1946}. Those initial conditions are that there
is an initial wave potential in which the initial particle
distribution is perhaps linearly but not non-linearly consistent with
it. The (initial) Landau damping occurs in the process of acquiring
the required non-linear consistency, and when or if that is acquired
the damping stops. But hole or soliton equilibrium conditions are
already non-linearly self consistent. So the Landau linearized
analysis, leading to damping at a rate proportional to the resonant
$|df/dv|$, does not apply. In summary then, linear Landau damping does
not apply to electron and ion hole equilibria. 

An important alternative phenomenon that can happen, however, is that
the solitary potential structure can accelerate. One way this occurs
is as a consequence of imbalanced reflection of the repelled species
(when a non-zero slope exists on the background distribution). This
phenomenon is (like Landau damping) caused by resonant particles
(moving at or near the structure velocity). There are also other ways
a hole can experience an instability that feeds on its own structure
and particle energy, regardless of repelled species
reflection. Moreover any transverse variation of the potential can
give rise to particle orbits in which parallel energy is not conserved
even in a steady rest frame, and as a result particles can trap or
detrap from the structure, and/or experience stochastic
orbits. Furthermore if there are background field fluctuations, caused
by other waves or moving electrostatic structures, these also can
cause stochastic diffusion of the particle orbits in phase space. In
general, orbit perturbations are liable to cause orbit detrapping and,
for holes at least, gradual decay of the density deficit that is
responsible for sustaining the potential.

In succeeding sections we shall see that much of the recent
theoretical progress in electron and ion hole physics consists of
developing a rigorous understanding of the effects of acceleration,
instability, and stochastic orbits, which are the important
collisionless mechanisms causing holes to (grow or) decay, and thereby
determining their lifetimes.

\subsection{Kinematics of Holes as Composite Bodies}
\label{composite}

The wake simulations of \citet{Haakonsen2015} showed electron holes
spawned by the dimple instability arising from electron parallel
cross-wake influx, usually then propagating out of the wake along the
phase-space orbits of electrons. However, a few holes did not
propagate out, instead remaining within it and being convected into
downstream increasing density regions, all the while growing to large
amplitude and eventually disrupting even the ion streams. The pressing
theoretical question was why these few behaved differently in their
response to the background parallel electric field environment. This
turns out to be answered by detailed consideration of the kinematics
of the electron holes, that is, of their momentum and energy
conservation. \citeauthor{Dupree1983}, long before, had pioneered
these considerations, discovering some important results. However, his
treatment omitted a vital component of the momentum conservation of
the repelled species (ions for electron holes), by making a
small-potential expansion that omitted the lowest-order non-zero term.
\citet{Hutchinson2016} and \citep{Zhou2016}  therefore derived
the momentum conservation expressions for electrons and ions
associated with an electron hole either accelerating or growing in
amplitude, without small-potential approximations and showed that they
agreed with particle in cell simulations, as will now be outlined.

The key to electrostatic hole momentum conservation is the recognition
that the electric field itself of a one-dimensional solitary potential
structure exerts zero total force on the plasma. This is so because
the Maxwell stress of the ($z$-component) electric field $E^2/2$ is
zero at distant positions $z_1$ and $z_2$ on either side of the hole
($z_1<0<z_2$). A simple demonstration is to write the force density
exerted as $\rho E={dE\over dz}E={d\over dz}E^2/2$, and integrate from
$z_1$ to $z_2$, to obtain the total force
$-\int_{z_1}^{z_2}\rho d\phi= [E^2/2]_{z_1}^{z_2}$ which is zero for
$|z_{1,2}|\gg 1$. Notice that the so-called quasi-potential,
introduced earlier, $\int\rho d\phi$, is in fact minus the Maxwell
stress (as noted by \citep{Andrews1971a}). Zero total field force
implies that the total particle momentum must be conserved. In a
\emph{steady} potential structure, no net particle momentum changes
occur. Passing particles flow out of the region $z_1<z<z_2$ carrying
the same momentum they brought in; trapped particles bounce within the
hole and no average momentum change occurs. However, when there is
some time rate of change of the potential, even within the initial
rest frame of the hole, then orbit energy is no longer conserved,
particle ``energization'' effects occur, and more importantly for our
present discussion, momentum exchange occurs which we will call
``jetting'' because the hole acts somewhat like a jet engine. When a
hole accelerates, individual passing particles no longer carry out the
same momentum they carried in, and trapped particles experience
non-zero average acceleration, yet the total particle momentum must
still be conserved.

The derivation of the momentum change of particles of a particular
energy proceeds using the approximation that the transit time
$\delta t$ of the orbit across the hole is short compared with the
timescale of potential change. For a hole of speed $U$ and
acceleration $\dot U$, the particle acceleration time is of order
$v/\dot U$, and the approximation is that $\delta t \dot U/v$ is
small.  The total time derivative of passing particle momentum is then found to be\citep{Hutchinson2016}
\begin{equation}
  \label{passingmom}
  \dot P_p = m\dot U\int_{z_1}^{z_2}\int
  \left[-2+3{v_1\over v}-\left(v_1\over v\right)^3\right]f_1(v_1)dv_1 dz,
\end{equation}
where $v_1$ is the external velocity (at $z_1$) and $f_1$ the external
distribution, and $v=\sqrt{v_1^2-2q\phi(z)}$ is the velocity at $z$
where potential is $\phi$ (approximated by short transit time as
unaffected by $\dot U$). [We retain the mass $m$ in this expression so
as here to regard time as measured in the same units for electrons and
ions. In our standard dimensionless notation, one can consider this to
be a time scaling factor for $\dot U$, different for the different
species.] The momentum derivative of trapped particles is simply
the total trapped mass times the hole acceleration:
\begin{equation}
  \label{trapmom}
   \dot P_t=m\dot U\int_{z_1}^{z_2}\int_{-v_0}^{v_0} f_t(v) dv dz,
\end{equation}
where $v_0=\sqrt{2|\phi(z)|}$ is the velocity of a particle of zero
energy, which is the upper bound of trapped particles. These
expressions make no approximation with respect to hole potential being
small, but if it is small, then for the attracted species the
distribution can be expanded as the first two terms of its Taylor
expansion about $v_1=0$ as $f_1(v_1)=f_0+v_1f_0'$ and one finds
\begin{equation}
  \label{passingaprx}
  \begin{split}
  &\dot P_p \simeq m\dot U\int_{z_1}^{z_2} - 2f_0 v_0 dz,
  \pamper
  \dot P_p+\dot P_t \simeq m\dot U \int_{z_1}^{z_2}\int_{-v_0}^{v_0}(f_t-f_0)
  dv dz.  
  \end{split}
\end{equation}
These expressions agree with those of
\citeauthor{Dupree1983}. However, this approximation does not work for
the repelled species, instead giving a divergent integral. Physically this
problem arises because reflected particles are not properly treatable
by the short transit time approximation. If there are reflected
particles then (as Dupree noted) a net reflection force (momentum
transfer rate) independent of any acceleration arises
\begin{equation}
  \label{reflected}
  \dot P_r = m \int_{z_1}^{z_2}
  \int_{-\sqrt{2\psi}}^{\sqrt{2\psi}}-2 f_1(v_1)|v_1|v_1dv_1 dz,
\end{equation}
and dominates the repelled particle effects.  However, if the number
of reflected particles is negligible, then $\dot P_r\to 0$ and a
different small-$\phi$ approximation of equation \ref{passingmom}
applies, giving
\begin{equation}
  \label{passingrepel}
  \dot P_p= m\dot U\int_{z_1}^{z_2}\int-3\left(\phi\over mv_1^2 \right)^2
  f_1(v_1)dv_1dz.
\end{equation}
This is the important repelled particle force term, second order in
$\phi$, that Dupree's treatment omitted.

Reference \citep{Hutchinson2016} shows that if there is a uniform
background force accelerating the particles at a rate $\dot v_b$, then
in the above expressions one should replace $\dot U$ with
$\dot U-\dot v_b$. An electron hole for which ion interaction can be
ignored then conserves momentum when $\dot U=\dot v_b$. That is, it
accelerates at the same rate as an electron. This explains the
simulation observation of electron holes moving out of a wake along
the background electron phase-space trajectory. Situations where both
electron and ion forces apply require more extended analysis. One such
situation is the observed self-acceleration of initially stationary
electron holes by interaction with ions. It is shown theoretically
that the ion jetting force (\ref{passingrepel}) becomes small compared
with electron (\ref{passingaprx}) when the hole speed relative to ions
is greater than approximately
$\left(m_i/m_e\right)^{1/4}\sqrt{T_e/m_i}=6.5c_s$ (dimensional units,
hydrogen), and that holes initialized gently at zero speed self
accelerate to a calculable speed of this order of magnitude, which
agrees remarkably well with simulation \citep{Zhou2016}.

\begin{figure}[ht]
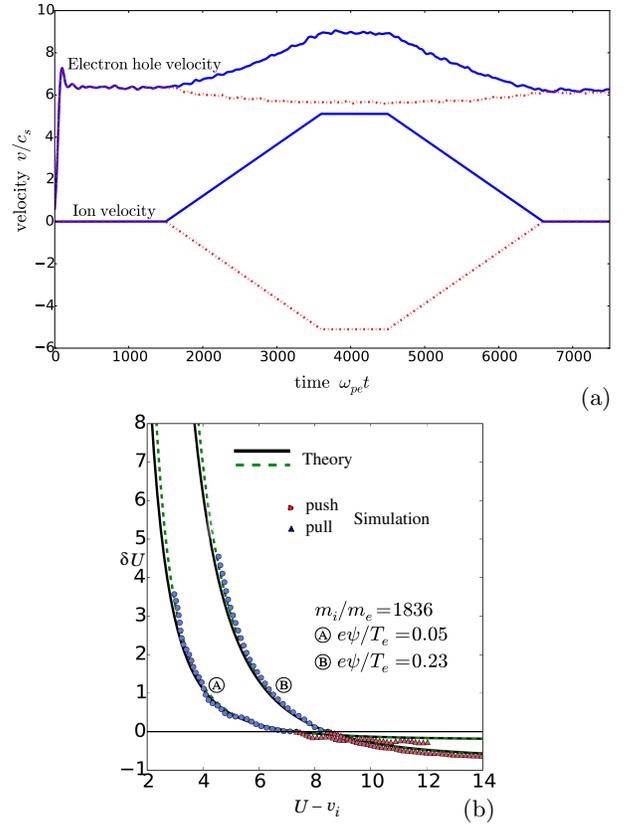

  \partwidth{0.59}{hole_pushing}\hskip-1.5em(a)\ \ 
  \partwidth{0.365}{pushpull}\hskip-1.5em(b)
  \caption{(a) Artificially pushing ion velocity toward hole velocity
    (blue), or pulling it away (red) \label{pushing} in PIC
    simulation. (b) Agreement of simulation results with analytic
    theory. Reprinted from \citep{Zhou2016}, with the permission of
  AIP Publishing.}
\end{figure}

A more specific simulation using a specially-developed hole-tracking
PIC code implemented an artificial background force to accelerate or
decelerate controllably the ions (but not the electrons). The process
is illustrated in figure \ref{pushing}(a).
The hole is initialized with a slight positive velocity and quickly
speeds up to just over $6c_s$ as expected from self-acceleration. Then
at time 1500 the cold ions are artificially accelerated (blue lines)
till 3500 and then at 4500 decelerated back to zero speed. The hole
velocity responds by accelerating by not quite as much, and then
decelerating back to its $\sim 6c_s$ value. A second simulation (red
dashed lines) instead decelerates and then accelerates the ions. The
hole changes its velocity much less, but still reversibly. In figure
\ref{pushing}(b) is shown the remarkable quantitative agreement of the
analytic theory and simulation, for the hole velocity change during this
process $\delta U$, versus the closeness of approach of the ion velocity 
$U-v_i$. Two different hole amplitudes are shown. The theory line is
(in $c_s$ velocity units)
\begin{equation}
  \label{pushtheory}
  \delta U = (M^4_{ie}/3)\left([U-v_i]^{-3}-[U_0-v_{i0}]^{-3}\right),
\end{equation}
where $U_0-v_{i0}$ is the initial velocity difference when
$\delta U=0$, and $M^4_{ie}/3\simeq (m_i/m_e)\psi/T_e$ for shallow
holes.

The inverse third power of the velocity difference shows that the
strength of the ion interaction becomes great if the ion stream
approaches the hole velocity. And it turns out to be easily strong enough
in the wake simulations that, when the hole velocity lies between the
velocities of two ion streams, it is essentially trapped by them and
cannot be accelerated past by background electric field. In general,
electron holes cannot smoothly overtake a narrow ion stream. And this
is the explanation for the few holes in the wake simulation that are
trapped and remain in the wake long enough to grow to large
amplitudes.  Their velocity is trapped between two ion streams. The
other holes that propagate out along the electron phase space
trajectories have untrapped velocities.

All these comparisons show that electron holes do indeed behave like
composite entities whose momentum conservation properties we can
calculate, giving quantitative predictions of their motion. 

\subsection{Oscillatory Instability of Electron Hole Velocity}

\begin{figure}[ht]\center
  \includegraphics[width=0.9\hsize]{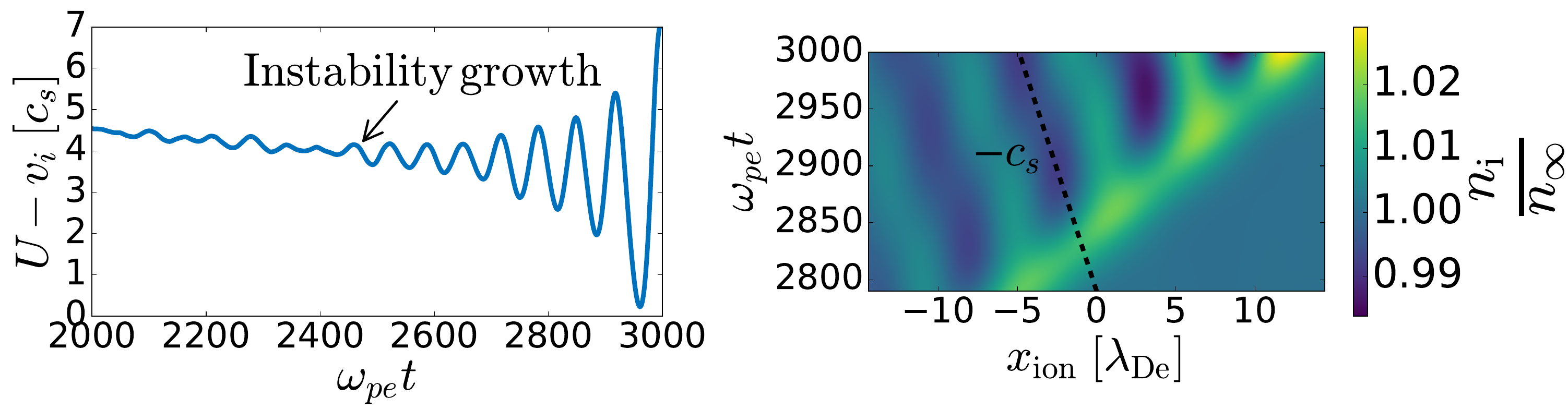}\hskip-0.52\hsize(a)
  \hskip0.38\hsize(b)
  \caption{(a) Threshold and growth of the velocity instability. (b)
    Ion density contours versus position in the ion frame
    ($x_{\rm ion}$) during the instability showing trailing ion
    acoustic oscillations. From \citep{Zhou2017}
    reprinted with permission from Cambridge University Press.
    \label{velinstab}}

\end{figure}

In the course of the simulations described in the previous section, a
new phenomenon was observed: oscillatory instability of the hole speed
when the ion stream velocity approaches the hole speed. It was
comprehensively analyzed in a follow up paper by \citet{Zhou2017}
.  Figure \ref{velinstab}(a) illustrates the approach of the ion
velocity ($v_i$) toward the hole ($U$) caused by artificial pushing
(time 2000 to $\sim 2400$) and the subsequent growth of the
instability, during which trailing ion density oscillations are
visible \ref{velinstab}(b).

The kinematic analysis of this oscillation proceeds as before for
electrons, because they remain well approximated by the
short-transit-time assumption, giving the same $\dot P_e$. But ions do
not satisfy this approximation. Instead, their perturbed response must be
approximated as an integral along unperturbed orbits (for small
oscillatory perturbation) of the energization
$\delta \energy = \int \dot U v dt$, where now $\dot U$ is
proportional to $\exp(-i\omega t)$ with $\omega$ the complex
frequency. After extremely heavy algebra, a long expression is found
for $\dot P_i$, which reduces for $\psi\ll U^2$ (shallow holes) and
writing the mean (unperturbed) hole velocity $U$ in the ion frame
($v_i=0$), to yield a form
\begin{equation}
  \label{pipe}
  {\dot P_i\over\dot P_e} = -{F(\omega/U)\over G(U)},
\end{equation}
where $F$ is a complex function that depends only on its argument
$\omega/U$ and spatial integrals over the normalized shape (mostly
width) of the hole potential $\tilde \phi(z)\equiv \phi(z)/\psi$;
while $G$ is a real function that is effectively
$(m_e/m_i)(U^2/\psi)^2 \dot P_e$, and has dependence only on $U$ and
$\tilde\phi$ integrals, not $\omega$. The dispersion relation is
momentum conservation: $\dot P_i/\dot P_e=-1$, and a Nyquist analysis
shows that unstable (positive) growth rate $\gamma=\Im(\omega)$ occurs
for $U$ less than a critical value $U_c$, at which the real frequency
is $\omega_c$. The critical speed for a small amplitude hole shape
$\psi\,\sech^4(z/4\lambda)$ is $U_c\sim (m_i/m_e)^{1/4}\psi^{1/2}$, but
in Figure
\begin{figure}[ht]
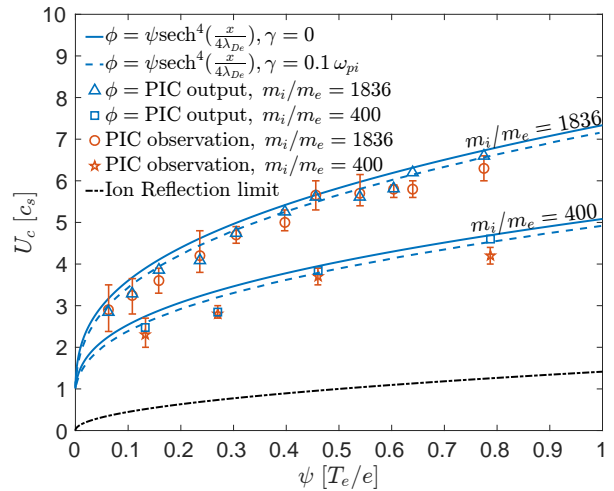
\center
  \partwidth{0.6}{instability_boundary}
  \caption{Critical hole speed at which instability occurs. Lines:
    theory for $\sech^4$ shape potential. Points: comparison of theory
    (triangle, square) versus observed onset (circle, star) for the
    simulation's actual potential shape. From \citep{Zhou2017},
      reprinted with permission from Cambridge University Press. 
    \label{instbound}}
\end{figure}
\ref{instbound} the calculations are carried out numerically. The
lines show the instability boundary and (dashed) the place just below
it where $\gamma=0.1\omega_{pi}$, for an exact $\sech^4$ shape versus
hole amplitude $\psi$. Two different mass ratios are shown. The points
are PIC simulation comparisons, showing in blue the analytic $U_c$ for
the actual (slightly variable) hole shape in the simulation, and in
red the observed instability onset. The agreement is excellent. The
agreement between theoretical and observed real part of the frequency
(not shown) is equally good.

\subsection{Hole-Soliton Coupling and Slow Electron Holes}
\label{SlowEHoles}

As described earlier, some simulations by \citet{Saeki1998} generated
a solitary structure called a Coupled Hole Soliton (CHS).  The way
this phenomenon works is essentially through the momentum interaction
of electron holes coupling to ions. In an ion acoustic soliton, a
positive (electron attracting) potential is generated by densification
of streaming repelled ions. This positive potential attracts an
electron hole, and can effectively trap it so that it is attached to
the soliton through its electric field. In \citet{Zhou2018} a series
of controlled one-dimensional numerical experiments were conducted to
explore the behavior of CHS states. A CHS is launched by a carefully
prepared initial state of the code, in which the ions local to the
structure are given a velocity offset and the local electron
distribution is given a Gaussian dimple (a local depression of
$f_e(v)$) at an appropriate velocity. Without the dimple, a simple ion
acoustic soliton tends to be launched; but with it, a CHS results. The
soliton moves away in a direction determined by the local ion
velocity offset, leaving behind a train of ion-acoustic oscillations,
but carrying with it an electron hole that is clearly visible in the
electron phase space and density.

The hole's presence is observed to assist the formation of the
soliton, and once formed somewhat enhances its speed relative to the
standard soliton dispersion relation (equation \ref{Eofpsi}). A
frequently cited aspect of KdV soliton behavior is that colliding
solitons appear to pass through one another retaining their amplitude
and speed. In contrast, two colliding electron holes, if their speeds
are not greatly different, have long been observed to combine with one
another to form a deeper electron hole (see e.g. Figure
\ref{BNRholemerge}). In this study, colliding CHSs display a kind of
hybrid behavior.  During the collision, while the ion density
perturbations of the holes overlap, the electron holes combine with
one another within the potential peak, moving at electron orbit speeds, much
faster than the ions. However, the ion compressions appear to continue
with almost their original velocities, and as they begin to separate,
they appear to tear apart the large combined electron hole so that one
still sees two CHSs propagating away from the collision, although not
with exactly the same amplitudes as before.

\begin{figure}[ht]\center
  \includegraphics[width=0.8\hsize]{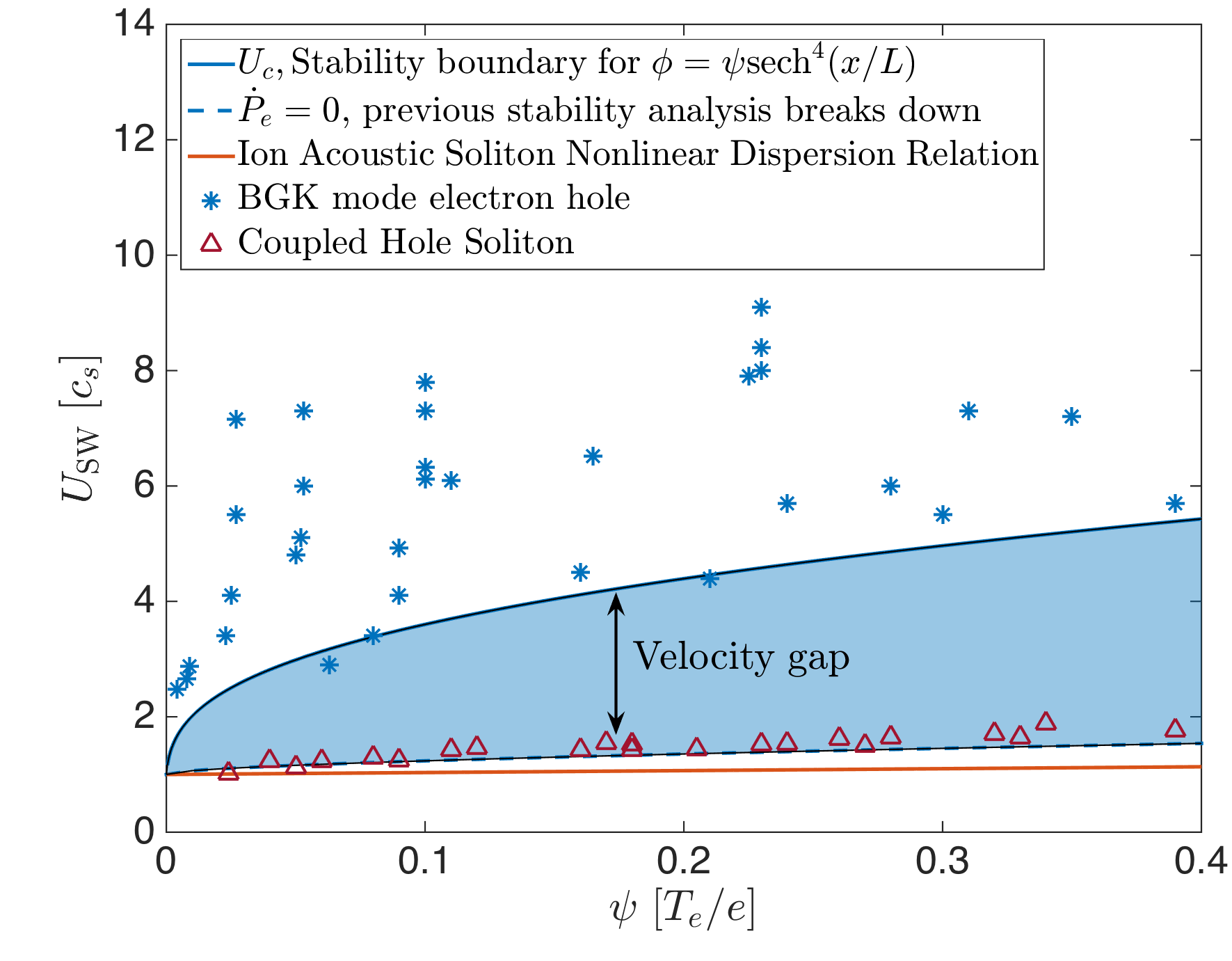}
  \caption{The theoretical unstable gap (shaded region) between
    (simulated) free electron holes (blue points) and Coupled
    Hole-Solitons (red points). Reprinted from
    \citep{Zhou2018} with the permission of AIP publishing. 
    \label{thegap}}
  
\end{figure}
Extensive exploration of CHS existence in the simulations (Figure
\ref{thegap}) showed that there is a speed gap between the CHS state,
which is a little above the ion acoustic soliton speed, and the free
electron hole state, which is several times $c_s$. This is
attributable to the oscillatory instability which occurs in the gap
and generally causes the electron hole either to accelerate out of
the gap, becoming a free BGK electron hole with little ion
interaction, or to be captured into a CHS state. Discussion is given
of transition out of the CHS state by what the paper calls ``Ion
Landau Damping''. I now believe that the amplitude decay is mislabeled
by this name, for the reasons given in section \ref{LandauD}. But in
any case, whatever the decay mechanism, it is observed that if a CHS
decays in amplitude, thereby moving into the speed gap, its position
begins to oscillate within the potential well created by the ion
density perturbation of the soliton. And the oscillation can grow
sufficiently large that the hole escapes the soliton influence and
moves off at speed above the gap.

The opposite process of transition from a free electron hole to a CHS
was also demonstrated, using two symmetric, oppositely propagating ion
streams ($v_i=\pm4 c_s$), and causing the hole amplitude to grow by
growing the plasma density. The open 100$\lambda_D$ long domain was
injected with increasing fluxes of electrons and ions, which forms an
increasing, somewhat concave, density profile. The initially
stationary small hole $\psi\simeq 0.1$ in unity density grows in
potential as the background density grows until, at time 1200
($\omega_{pe}^{-1}$), the increase is stopped and the density flattens
at 3.2, with (by time 1800) a local ion density enhancement and ion
acoustic waves being generated by the hole. The hole settles, the
waves disappear, and by time 3330 a CHS persists with $\psi\simeq 3$,
and a local density peak of $\sim 4$: well above the 3.2 background,
showing the strong ion response.  This sequence occurs only when the
density growth rate is sufficiently fast, otherwise the oscillatory
instability occurring during the transition across the unstable gap
disrupts the hole and parts of it are accelerated to high speed by ion
interaction and lost.

A final reported set of numerical experiments in this paper concerned the
Buneman instability, arising from strong relative ion-electron drift.
The wave train that it generates in the nonlinear stage, when the ions
are cold $T_e=20T_i$, $\bar v_e=45c_s$, consists of CHS-like ion
density enhancements coupled with electron phase-space holes. By
contrast, with equal temperatures $T_e=T_i$, $\bar v_e=70c_s$ the
instability generates waves that appear to consist of trains of pure
electron holes, traveling at 4-5$c_s$ (relative to the ions) above the
oscillatory velocity instability gap.

In an attempt to detect the instability gap in nature,
\citeauthor{Zhou2018} plotted satellite data from \citet{Graham2016}
on a figure like \ref{thegap}. The comparison was unfortunately
inconclusive, mostly because the hole amplitudes were so small that
the gap was as narrow as the apparent uncertainty in hole speed.
Further observational analysis of MMS magnetopause data by
\citet{Steinvall2019} did reveal some velocities along the red region
of \ref{thegap} that might have been coupled hole-solitons, and a few
holes of high amplitude above the velocity gap, that was suggestive of
fast holes, as well as many of low amplitude around and below $U=1$.

A later major statistical study of hole speeds in MMS data,
\citet{Lotekar2020}, analyzed (during fast-flow events in the
magnetotail) more than 8000 holes, of which 2400 were observed on at
least three spacecraft.\begin{figure}[ht]\center
  \includegraphics[width=0.44\hsize]{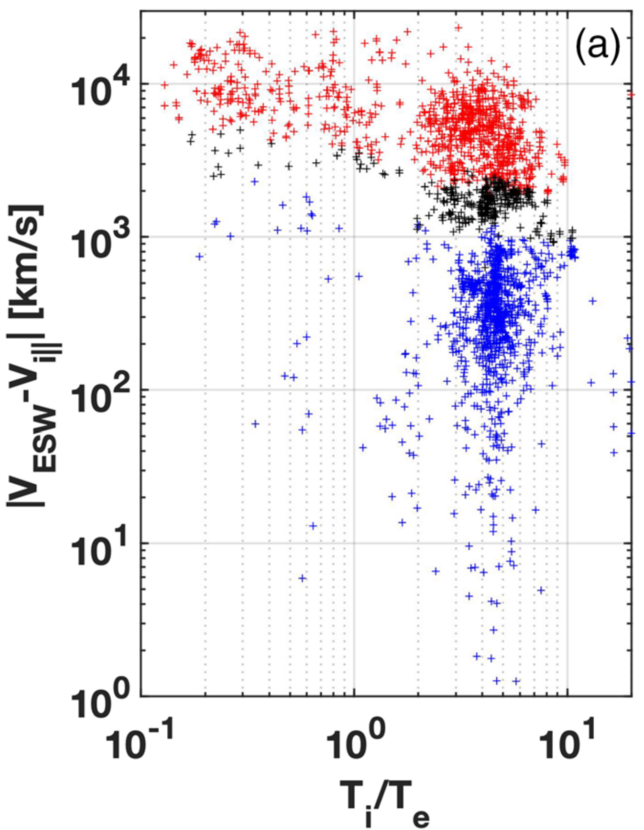}
  \includegraphics[width=0.474\hsize]{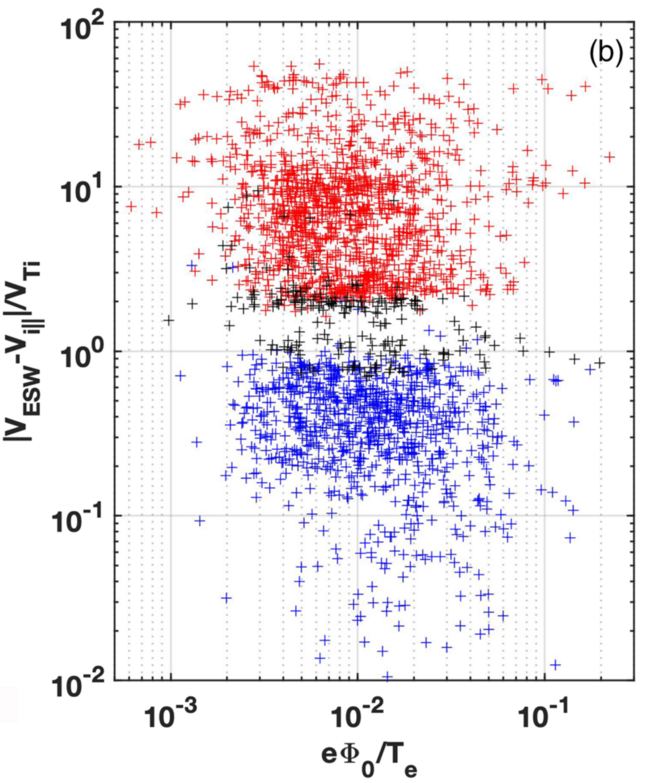}
  \caption{Hole speeds observed by MMS (a) versus background
    temperature ratio, (b) normalized to ion thermal speed versus hole
    amplitude, showing a suggestive gap just above the ion thermal
    speed. From \citet{Lotekar2020}.\label{Lotekar}}
\end{figure}
Such observations allow much more precise determination of hole
speed. The parallel separation of the craft was a few up to about
10km, and by careful cross-correlation events were selected that were
holes propagating from one craft to others. Velocities relative to the
craft down to 100km/s could be determined. These were all electron
holes, having positive potential polarity (only about 20 were excluded
from analysis because of being negative). Distributions of hole
length, amplitude, duration, and occurrence provide important evidence
concerning their generation and stability, that will be mentioned in
other contexts, but here we focus on the velocity statistics shown
in Figure \ref{Lotekar}.

The colors separate slow holes (blue) with speed (relative to ions)
$v_h<0.05$ of the electron thermal speed, fast $v_h>0.1$ (red), and
intermediate (black). We observe that when $T_i<T_e$ almost all holes
are fast, but there are many holes in plasmas with $T_i\sim 3T_e$ that
are slow, having speeds that are less than $v_{ti}$.  There appears to
be an almost empty gap in the intermediate speed range, which is
suggestive of the theoretical forbidden range where the oscillatory
instability occurs. However, there is no sign of a restriction of the
slow hole velocities to the soliton velocity, as observed in the CHS
simulations. So probably these slow holes (with $T_i>T_e$ unlike the
simulations) are not coupled hole soliton structures.

The question then arises as to what permits them to be so slow. As we
have noted earlier, holes simulated starting at slow speeds, within
the ion distribution, are usually observed to self-accelerate by
interactions with the ions, ending up at speeds generally
significantly exceeding the ion acoustic speed. Yet the slow holes of
\citet{Lotekar2020} have not experienced this effect. How do they
exist? The theoretical answer is given by \citet{Hutchinson2021c}. One
way to explain it intuitively is in terms of the ion density change
that accumulates in the hole. For a single-humped background ion
distribution such as a Maxwellian, in the interior of which the hole
velocity resides, the density perturbation caused by a positive
potential structure (electron hole) is negative: ion charge density is
decreased. However, in the opposite extreme, if the ion distribution
is composed of the sum of (say two) Maxwellians with different mean
velocities, then if they are sufficiently spaced apart, a hole between
them will instead cause an increase of the ion charge. In this second
situation, the electron hole controlling the potential experiences an
\emph{attraction} to the ion density perturbation, rather than a
repulsion. This can prevent self-acceleration. The alternative
intuitive explanation, which forms the basis of a more thorough
calculation of the velocity stability threshold, is to recognize that
slow holes by definition cause ion reflection. And, although the
reflection force averages to zero in equilibrium, including for
example a symmetric ion distribution, if a hole accelerates away from
that equilibrium, then the ion force becomes non-zero. The sign of the
force and its resulting acceleration is such as to amplify the
velocity perturbation, yielding unstable growth of the motion, unless
the background ion velocity distribution is double humped and the
equilibrium lies near its local minimum. Thus the theoretical answer
to how slow electron holes can exist is that they require a local
minimum in the ion velocity distribution to avoid unstable acceleration.
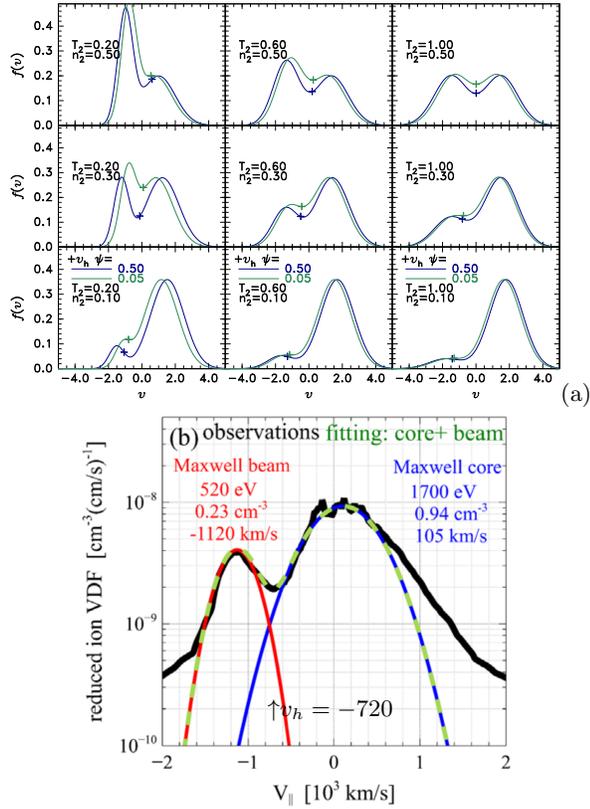
\begin{figure}[ht]
  \partwidth{0.53}{shapes}(a)
  \begin{tikzpicture}
    \node[anchor=south west,inner sep=0] (image) at (0,0) {
      \partwidth{0.44}{iondistribK}
    };
    \begin{scope}[x={(image.south east)},y={(image.north west)}]
     \draw[black] (.41,.2) node [anchor=south west] {$\uparrow$$v_{h}=-720$};
    \end{scope}
  \end{tikzpicture}
  \caption{(a) Various two-Maxwellian ion distribution shapes that
    permit electron holes marginally stable against
    self-acceleration. Reprinted with permission from
    \citep{Hutchinson2021c} copyright the American Physical Society
    (2021). (b) Example measured parallel distribution function (black
    line) with the hole velocity indicated by the arrow. Reprinted
    with permission from \citep{Kamaletdinov2021} copyright the
    American Physical society (2021). \label{shapes}}

\end{figure}
A range of different model ion distributions for which electron holes
are theoretically marginally stable against the self-acceleration
instability is shown in Figure \ref{shapes}(a). The equilibrium hole
velocity is marked on each curve with a cross.

Almost simultaneously with this theory, \citet{Kamaletdinov2021}
 performed an in-depth study of the measured ion distribution
functions during which slow electron holes were observed by MMS. An
example is shown in Figure \ref{shapes}(b). The observed hole velocity
lies right in the local minimum of the double-humped ion distribution,
as the theory says is necessary for a stable equilibrium. Moreover,
the authors obtained the ion distributions for all the approximately
1000 slow electron holes identified by \citeauthor{Lotekar2020}, and
found that the ion velocity distributions essentially all had this
sort of double-humped character, and the hole velocity lay in the
distribution's local minimum. Thus these well-characterized slow
electron holes show remarkable agreement with the theoretical
requirements for slow holes, and the statistics of this comparison are
very extensive.

Some simulations of slow electron holes in double-humped ion
distributions that the quasi-static theory of \citep{Hutchinson2021c}
indicates should be stable, showed a different phenomenon of
``overstability'' treated theoretically by
\citet{Hutchinson2022}. That is, they exhibited growing velocity and
position \emph{oscillations}. When the oscillation grows to sufficient
amplitude in the simulations, the electron hole escapes the local
influence of the attractive ion density rise, no longer a slow hole.
This phenomenon is not represented within the quasi-static theory, and
instability calculations require a full scale numerical solution of
the linearized Vlasov equation for the ions, whose transit time is
comparable to the oscillation period, and which therefore introduce a
phase-shift dependent on $\omega$ into the response equations.
\begin{figure}[ht]
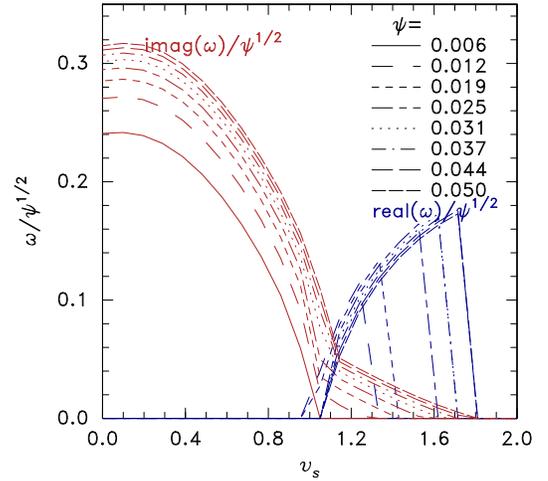
\center
  \partwidth{0.5}{omegaofv1}
  \caption{Real and imaginary parts of the frequency versus the
    velocity shifts ($\pm v_s$) of two equal Maxwellian ion
    components, for a range of electron hole amplitudes
    $\psi$. From \citep{Hutchinson2022}.\label{omegaofv1}}
\end{figure}
More details will be discussed of the solution
techniques employed when we come to treat transverse stability. But
the comprehensive one-dimensional slow hole instability theory results
are summarized in Figure \ref{omegaofv1}.
When the shifts $\pm v_s$ (in ion thermal units $\sqrt{T_i/m_i}$, $T_i=T_e$) of the two symmetric Maxwellian ion
distribution components is small, there is pure growth of the hole
velocity given by a purely imaginary $\omega$. It is approximately
$\Im(\omega)=0.3\psi^{1/2}$. As $v_s$ is increased, giving rise to
flattening of the distribution and eventually a local minimum, the
growth rate decreases, and for small $\psi\lesssim 0.01$ just stabilizes at
$v_s\simeq 1.05$. However, holes of greater amplitude instead become
overstable, with finite real frequency, whose magnitude grows during
deepening of the local $f_i$ minimum. The growth rate for a specific
amplitude continues to decrease until the instability eventually
ceases. Thus the quasi-static analysis needs to be complemented by
this fuller eigenfrequency determination to give a complete picture of
the theoretical one-dimensional stability of slow electron holes. They
need a rather deeper local minimum for stability than is indicated by
the quasi-static approximation, or else they must have a very small
potential amplitude $\psi$.

\section{Asymmetric Electron and Ion Holes}
\label{asymmetry}

\subsection{Theory of hole asymmetry and acceleration}
In the previous section we have described how the kinematics of holes
governs their behavior as composite entities capable of free motion
and acceleration. The observed quantitative agreement between
simulation and analytic calculation justifies the approximation that any
hole shape variation, internal vibrations or distortions, is a small
effect on the overall momentum balance.

A further assumption implicit in almost all the treatments discussed
is that the hole potential profile is symmetric in space about the
hole center. When the charge density is indeed a function only of
potential, this symmetry is inherent in the governing equations,
because trapped densities are themselves symmetric (in equilibrium at
least), and so even asymmetric passing velocity distributions do not
immediately induce any spatial asymmetry.

However, asymmetric kinetic potential structures do exist. Double
layers are the best known example, where the potential difference
across the structure is the dominant factor, and is often assumed to
be monotonic. Asymmetries in electron and ion holes can also
arise from the same mechanism as in double layers: particle
reflection. In the context of holes, when the potential variation is
dominated by a local positive or negative peak, and the potential difference
between the two sides of the hole is small, the dominant reflection
effects arise from the repelled species: ions at an electron hole, or
electrons at an ion hole. Such reflections break the symmetry when the
background velocity distribution of the reflected particles is
asymmetric in the hole frame, making the charge density a function of
\emph{both} the potential and the sign of the position ($z$) relative
to the potential peak ($z=0$). Reflected particles contribute only to
the density on the side of the peak from which they approach the
hole. Obviously this is a dominant effect for a double layer, but it
may be a more moderate factor for an electron or ion hole.
Figure \ref{symholes}(ii) illustrates schematically the
sort of situation we have in mind, where the ion hole asymmetry is
presumed to arise from a small shift of the electron distribution's
velocity center.
\begin{figure}[ht]
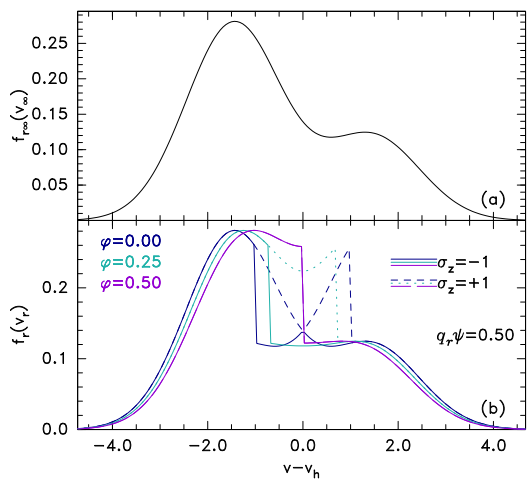
\center
  \partwidth{0.5}{ionreflect}
  \caption{Example of asymmetric repelled species velocity
    distributions in the presence of reflection. (a) Background
    (incoming) distribution. (b) Distributions at three different
    potentials on either side ($\sigma_z$) of the peak. \label{ionreflect}}
\end{figure}
Figure \ref{ionreflect} illustrates actual repelled species velocity
distributions at different potentials (b) when the background
distribution (a) is asymmetric.  Discontinuities arise at velocity
corresponding to marginally reflected particles, and the densities are
different on the different sides ($\sigma_z$) of the hole.

Finding a theoretical solitary structure equilibrium in this situation
requires a considerably more elaborate process that we now
outline. The original papers\citep{Hutchinson2021c,Hutchinson2023}
give much more detail. Determining whether a particle velocity is
reflected requires knowing the height $\psi$ of the potential peak. It
is therefore easiest to use a variant of the BGK (integral equation)
approach. Knowing $\psi$ and the specified repelled species background
distribution $f_r(v)$ provides the repelled density
$n_r(\phi,\sigma_z)$.  Since the repelled density is different at
$z\to\pm\infty$, the attracted species must also have asymmetric
distant density to satisfy neutrality. This $n_a$ asymmetry can only
be induced by a potential difference
$\Delta\phi=\phi(+\infty)-\phi(-\infty)$. (The zero of potential is
taken at $[\phi(+\infty)+\phi(-\infty)]/2$.) Whatever the attracted
species distribution is, its distant density is a function only of
potential. Thus, $\Delta \phi$ can be found by solving the equations
$n_a(\pm\Delta\phi/2)=n_r(\pm\Delta\phi/2,\sigma_z=\pm1)$.  For
example, if the attracted species is Maxwellian of reference
temperature, then $n_a=n_0\exp(-q_a\phi)$, which, given $n_r$, yields
$q_a\phi=-\ln(n_r/n_0)$.

For any arbitrary electron hole velocity $v_h$ relative to (say) the ion frame,
there will then generally be a net ion reflection force on the hole.
The hole force balance can then only be satisfied by a net transfer of
momentum to the attracted species, and in the general case when
attracted species reflection does not exactly balance it, hole
acceleration will occur. In other words, the hole will in fact not be
in equilibrium in respect of hole motion. Only at a set of discrete
hole velocities, and corresponding $\Delta\phi$, will the ion
reflection force be exactly balanced by attracted species reflection
from the potential difference $\Delta\phi$ across the hole.  Those
discrete true equilibrium hole velocities can be found by iterative
search. They generally lie close to any local maxima or minima in
$f_r(v)$. The stability of those equilibria with respect to slow
perturbations of the hole velocity is then determined by the sign of
the derivative with respect to $v_h$ of the total force.

It is helpful to distinguish between what is called the ``extrinsic''
force or momentum transfer, which arises purely because of
$\Delta\phi$, and the ``intrinsic'' force, which is present regardless
of $\Delta\phi$ \citep{Hutchinson2023}. The intrinsic momentum
transfer to the attracted species is attributable purely to
acceleration,
\begin{equation}
  \dot P_{aint}= M_a\dot v_h \qquad {\rm where} \qquad M_a=\int \tilde
  n dz.
\end{equation}
Here $\tilde n$ is the hole attracted density arising just from the
depression of its phase-space density relative to the flat plateau,
and is therefore negative. $M_a$ is the effective hole mass, also
negative. The intrinsic reflection force on the repelled species is
the electric field force times its charge density integrated over the
hole
\begin{equation}
  \dot P_{rint}=F_{rint}=\int_{q_r\phi>|\Delta\phi|/2} -n_rq_r{d\phi\over dz} dz.
\end{equation}
The extrinsic force, combined for both species, can be expressed as
the Maxwell stress difference across the potential range
$-|\Delta\phi|/2<q_r\phi<+|\Delta\phi|/2$ on the lower-potential wing
of the hole. This can immediately be found from the distant asymptotic
form $\phi\propto\exp(-|z|/\lambda)$, where $\lambda$ is the
generalized Debye shielding length, giving
\begin{equation}
  \dot P_{ext}=F_{ext}=\Delta\phi|\Delta\phi|/2\lambda^2.
\end{equation}
We then have a kind of Newton's Second/Third law:
$-F_{rint}-F_{ext}=M_a\dot v_h$,
in which the hole's mass $M_a$ is negative. Equilibria lie at
velocities such that $F=F_{rint}+F_{ext}$ is zero. But those equilibria 
for which $dF/dv_h$ is positive are unstable, because any small
difference of $v_h$ from the equilibrium causes acceleration that
increases the difference.
The general result is that equilibria near a local maximum of $f_r(v)$
are unstable and those near a local minimum are stable. 
\begin{figure}[ht]
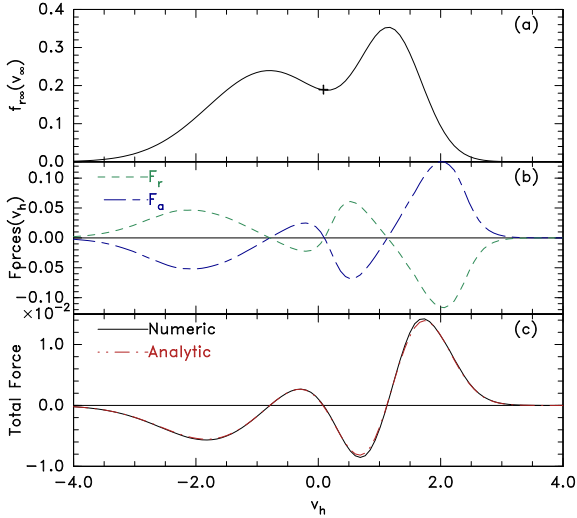
\center
  \partwidth{0.55}{fvsvhplot}
  \caption{Repelled and attracted species force balance as a function
    of hole velocity when $\psi=0.1$. Only the velocity with total
    force ($F=F_r+F_a$) zero and negative slope, marked with a cross,
    is a stable equilibrium. Reprinted with permission from
    \citep{Hutchinson2023}, copyright AIP publishing
    (2023).\label{fvsvhplot}}
\end{figure}
Figure \ref{fvsvhplot} shows the force variation with $v_h$ and the
fact that force-zeroes near peaks on the repelled species distribution are
unstable, and near (local) valley bottoms are stable to slow
acceleration.

It proves possible\citep{Hutchinson2023} by Taylor expanding the
repelled distribution about the hole velocity, for moderate or small
$\psi$, to derive an analytic approximation to the total force
\begin{equation}\label{asymforce}
F=  F_{rint}+ F_{ext} = -4\psi^2f_r'-{\textstyle{ 8\over 9}}\psi^3f_r'''
\end{equation}
where primes denote velocity derivatives. The resulting analytic line
in Figure \ref{fvsvhplot}(c) is in excellent agreement with the
numerical calculation. The corresponding analytic approximation for
the potential drop across the hole is
\begin{equation}
  \Delta\phi={ - 4\psi^2f_r'-{\textstyle{ 2\over 3}}\psi^3f_r'''\over
    1/\lambda^2 - f' +\psi f_r'''/12 }
={{\textstyle{ 2\over 9}}\psi^3f_r'''\over
    1/\lambda^2 - f' +\psi f_r'''/12 },
\end{equation}
where the second form sets $F$ (equation \ref{asymforce}) to zero so
it applies only at equilibrium.  In view of the $\psi^3$ factor, we
see that $\Delta\phi$ is thus very small unless the hole amplitude is
big $\psi\sim 1$. In equilibrium, small amplitude holes are thus
predicted to have potentials very close to symmetric. However, holes
that are accelerating, for which equation \ref{asymforce} is not zero,
can have more substantial asymmetries.

Holes with velocities near an equilibrium that is unstable because of
particle reflection experience exponential growth of their velocity
$\Delta v_h$ relative to the equilibrium velocity. The simplest
example is a hole with velocity at the peak of the repelled particle
distribution. In figure \ref{fvsvhplot} that would be the
zero-crossings of $F$ at $v_h=-0.8$ and $1.2$. The growth rate
$\gamma$ of $\Delta v_h$ there depends on the second derivative
$f_r''$. A single Maxwellian peak has
$f_r''=-(T_a/T_r)^{3/2}/\sqrt{2\pi}$ and gives a growth rate
$\gamma=-(4/\sqrt{2\pi})\psi^2\theta^{-3/2}/M_a$ in attracted species
units (ion units for an ion hole), where $\theta=T_r/T_a$. The hole
mass for a $\sech^4$ shaped hole is to leading order
$M_a\simeq-(16/3)\psi\sqrt{1+1/\theta}$ which gives
\begin{equation}\label{reflectgrowth}
  \gamma\simeq {3\over 4\sqrt{2\pi}}{\psi\over \theta\sqrt{\theta+1}}.
\end{equation}
A large repelled species temperature (large $\theta$) thus decreases
the acceleration growth rate. But in the case of an ion hole it is
really the curvature of the electron distribution very close to its
peak that determines stability and growth. A perfectly flat $f_r$
($f_r''=0$) there zeroes the growth rate, but would be very difficult
to measure. In any case, this growth rate, and the associated
acceleration rate for finite $f_r'$ are slow enough that treating an
ion hole as a composite entity lasting many bounce periods of its
trapped ions is a consistent approximation. In other words, equation
\ref{reflectgrowth} which represents a quasi-static treatment of the
particle dynamics, is accurate only if the timescale of growth is long
compared with particle transit or bounce times. That is certainly true
of electrons in an ion hole, since the growth rate is then in absolute
units of $\omega_{pi}$; and when $\psi$ is not large it is valid for
ions too. However, for an electron hole it is a poor approximation for
ion dynamics since the characteristic ion timescale is
$\sqrt{m_i/m_e}$ longer than electron. Electron hole unstable
acceleration needs a more elaborate treatment and can give a growing
velocity oscillation in some cases (see \citet{Hutchinson2022}, 2022,
and figure \ref{omegaofv1} in Section \ref{SlowEHoles}).

All of the mathematical discussion so far applies to electron or ion
holes, when we use the different normalizations, with the caveat just
mentioned about different timescale approximations. In addition, the
different \emph{velocity} scales make the practical importance of
reflection greater for ion holes than electron holes.  Electron holes
can easily have velocities relative to ions that make the ion response
and especially the ion reflection negligible, whereas ion holes, whose
velocities are generally limited to a few times $\sqrt{T_i/m_i}$,
almost never travel fast enough for electron reflection to be
negligible. Using the designation ``slow'' to indicate a structure
whose velocity lies within the repelled distribution, we can say that
electron holes are often not slow, whereas ion holes are almost always
slow. This velocity scale difference also means that electron holes
accelerated by ion reflection can remain intact once their speed is
raised beyond the ion thermal spread; while ion holes, if they are
accelerated, disengage from the attracted \emph{ion} distribution long
before disengaging from the electron distribution. The result is most
plausibly collapse of the ion hole, since at $v_h\gg v_{ti}$ there is
no ion phase space density in which a depletion could
occur.\footnote{An alternative speculation is that conceivably a hole
  of the ``Turikov'' type (\citet{Turikov1984}), figure \ref{Turikovfv},
  might persist.}  Nevertheless, ion holes are unambiguously observed
in space as is discussed in Section \ref{IonHoleObs}.

Before that, though, we should note that when an electron hole is
accelerating it can develop potential asymmetry, even when there is
negligible ion response. Two possible mechanisms for causing this
acceleration and potential asymmetry are (i) mirror force arising from
a parallel gradient of magnetic field strength, which is the topic of
the analysis of \citet{Vasko2016a,Kuzichev2017}; and (ii) background
conservative (e.g.\ gravitational or background electrostatic) force
with resulting density gradients, which are the subject of simulations
by \citet{Vasko2017b}. These papers are motivated in part by the
possibility of particle energization by adiabatic heating of trapped
electrons, and by the possibility that substantial total potential
drop, sometimes inferred in space, can accumulate from a sequence of
small drops across successive electron holes.

The theory of \citet{Vasko2016a} concentrates on an adiabatic
mechanism of particle energization. The conservation of magnetic
moment $\mu$ for a particle moving up a slow positive magnetic field
strength gradient increases its perpendicular energy ($\propto
|B|$). In the absence of any other forces, its total energy is
conserved and the parallel kinetic energy decreases by an equal amount
until the orbit is reflected when $v_\parallel=0$. However, an
electron trapped in the electric field of a moving electron hole is
constrained to move at an average speed equal to that of the hole. It
then bounces somewhat asymmetrically in the electric potential well,
in such a way that any loss of average parallel kinetic energy is made
up by the average of the hole electric field force. Therefore it is
possible for its perpendicular and total kinetic energy to increase
without limit as long as $|B|$ increases (and $\mu$ is
conserved). This energization might then give rise to particles with
energy much larger that thermal, without requiring a large electron
hole potential. This paper calculates that some of the perpendicular
energy given trapped particles comes from parallel energy lost by
untrapped particles. It supposes that the difference between these two
transfers is small, and then addresses the evolution of the hole
potential amplitude, assuming that the hole velocity is fixed and
given. (Invoking the ``given field'' approximation \citet{ONeil1965}.)
A weakness of this approach is that it is likely that (the ignored) hole
acceleration or deceleration is a more important effect than hole growth,
and neither are properly treated in the ``given field''
approximation. \citet{Kuzichev2017} reports continuum Vlasov
simulations of the process in a specified magnetic field
gradient. They do indeed show hole deceleration when moving into an
increasing $|B|$. It proves to be considerably faster than would be
predicted by the average mirror-force acceleration of trapped
particles.  Further simulation results with similar setup are described
in \citet{Shustov2021} where braking (deceleration) of holes in
reconnection layers with increasing $|B|$ is modelled. Varying the
hole amplitude shows that smaller $\psi$ gives slower deceleration,
and smaller $\Delta\phi$, but the mirror force is the predominant
mechanism for all amplitudes.

The simulations of \citep{Vasko2017b}, video examples of which can be
viewed at
\url{https://www.youtube.com/playlist?list=PLVd7b9xQL_fP6wqp84NJf5gPmiHmpYrRA},
are similar except that they
apply a force arising from general potential $U$ (independent of perpendicular
energy). The background density is taken to vary spatially in
accordance with a Boltzmann factor. These also show hole acceleration
noticeably faster than would occur for single electrons in this
potential. Its cause is not explained, just said to be unsurprising in
view of energy exchange with passing electrons. A feature of all these
simulations is that the starting hole velocity is quite large
1.3($v_{te}$), so these holes are out on the wing of the electron
distribution, approaching the unusual type of \citet{Turikov1984}
(see figure \ref{Turikovfv}). 

Electron reflection from the potential drop that is observed to be
developed is one important possible cause of deviation of the observed
effective charge to mass ratio from unity.  Another, perhaps more
important, is that in the non-uniform density, hole growth is
expected, which also gives rise to acceleration, and this mechanism,
derived in \citep{Hutchinson2016} is ignored in the comparison by
\citep{Vasko2017b}. The observed amplitude variation in the
simulations is rather small. So a complete explanation is still
somewhat elusive.

\subsection{Recent ion hole observations}
\label{IonHoleObs}

The Earth's bow shock is a region displaying large amplitude, often
electrostatic, fluctuations. In it the incoming ions moving at near
the solar wind speed are slowed and densified over a relatively short
distance in which the magnetic field strength is also enhanced by a
factor of a few. It is known that a substantial fraction of the
inflowing ions are actually reflected within the shock and return back
upstream with approximately their incoming speed. This phenomenon can
easily give rise to a kind of two-stream instability of the ions
(actually ion acoustic instability), theoretically analyzed long ago
by \citet{Stringer1964}. Therefore ion instabilities have long been
proposed as a major cause of bow shock electrostatic fluctuations (see
e.g.\ \citet{Akimoto1985}). Until recently it has been extremely
difficult to detect the velocity and hence potential polarity of the
observed solitary bipolar-field electrostatic structures in the bow
shock. The assumption that they were electron holes, made for Wind
satellite measurements by \citep{Bale2002}, was called into question by
analysis of the Cluster data (e.g. \citet{Hobara2008}) showing
negative potential polarity. But Cluster did not have electric field
measurements along its spin axis; so full three-dimensional field
diagnosis was not available.

From approximately 2018
onward, however,  careful
interferometric analysis by
\citet{Vasko2018a,Wang2020,Vasko2020,Wang2021,Kamaletdinov2022} have shown the
solitary structures observed by MMS, with full 3D field measurements
during several bow shock crossings, to be predominantly \emph{negative}
potentials. That is, they are (most probably) ion holes.
\citet{Wang2021}  present the most comprehensive data analysis,
including more than 2000 bipolar field structures of negative
polarity (setting aside the $\sim100$ observed positive
ones). Because the solitary structures had lengths comparable to the
antenna separations, corrections to the time dependence of the
voltage-differences were essential to determining the spatial shape of
the field $\E$ and the structure's propagation velocity $v_h$ and
direction $\k$. After applying these corrections the directions of
$\E$ and $\k$ were found from single-craft signal correlations to
coincide within $20$ degrees, and the uncertainty in $v_h$ was $<30$\%
at 90\% confidence level. The statistical distribution of hole
velocities (in the $\E$ direction) relative to the mean plasma ion
velocity peaked close to zero and extended up to approximately 2 times
$\sqrt{(T_e+3T_i)/m_i}$, where temperatures are measured in the
upstream plasma flow.
\begin{figure}[ht]
  \begin{tikzpicture}
\node[anchor=south west,inner sep=0] (image) at (0,0) {
  \includegraphics[width=0.475\hsize]{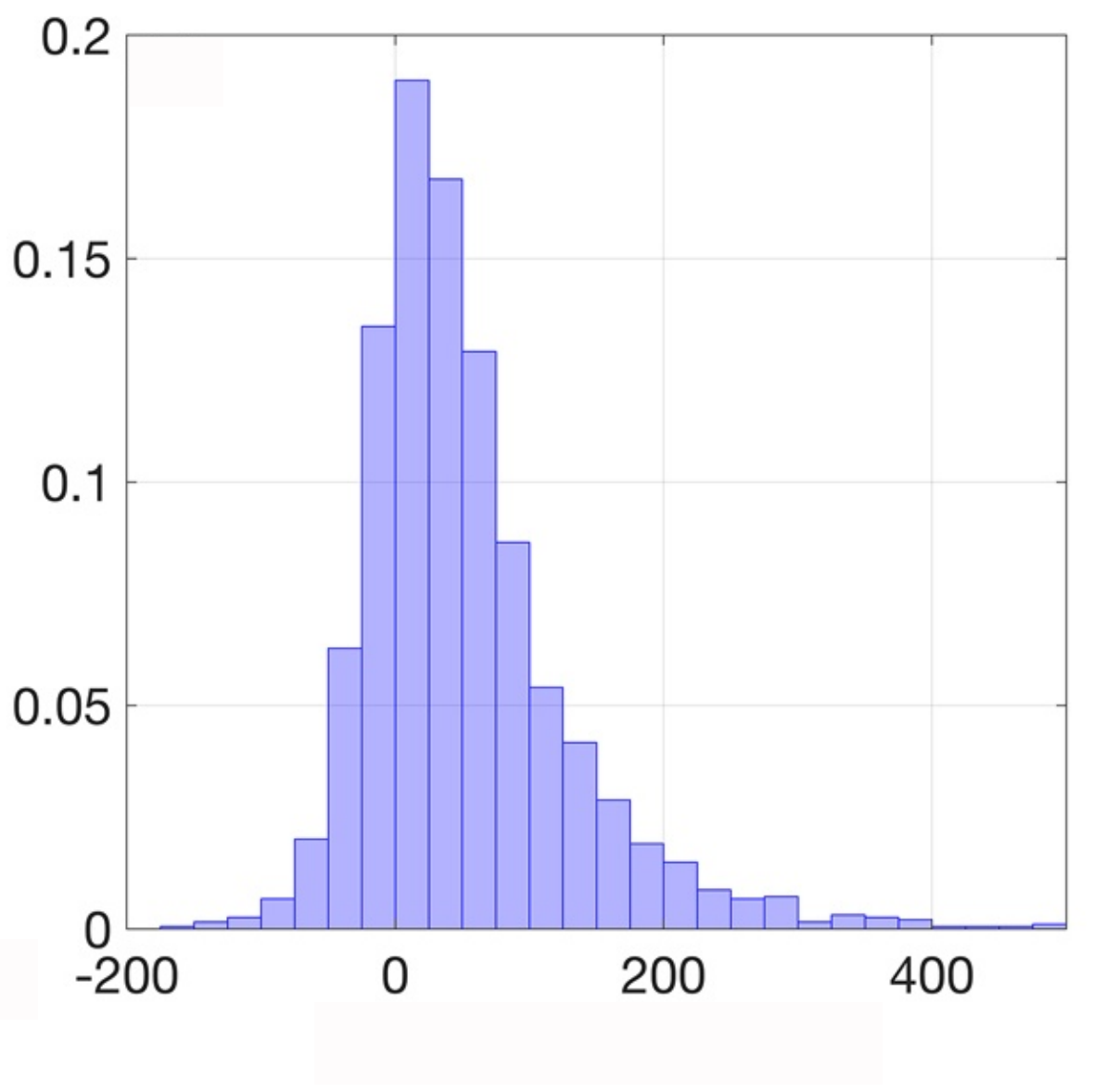} };
\begin{scope}[x={(image.south east)},y={(image.north west)}]
  \draw[black] (.1,.0) node [anchor=south west]
  {$v_{h}-\bar \v_i\cdot\hat\E\quad [km/s]$};
  \draw[black] (.12,.4) node [anchor=south west]
  {\rotatebox{90}{probability}};
\end{scope}
\end{tikzpicture}\hskip-1.5em(a)
  \begin{tikzpicture}
\node[anchor=south west,inner sep=0] (image) at (0,0) {
    \includegraphics[width=0.48\hsize]{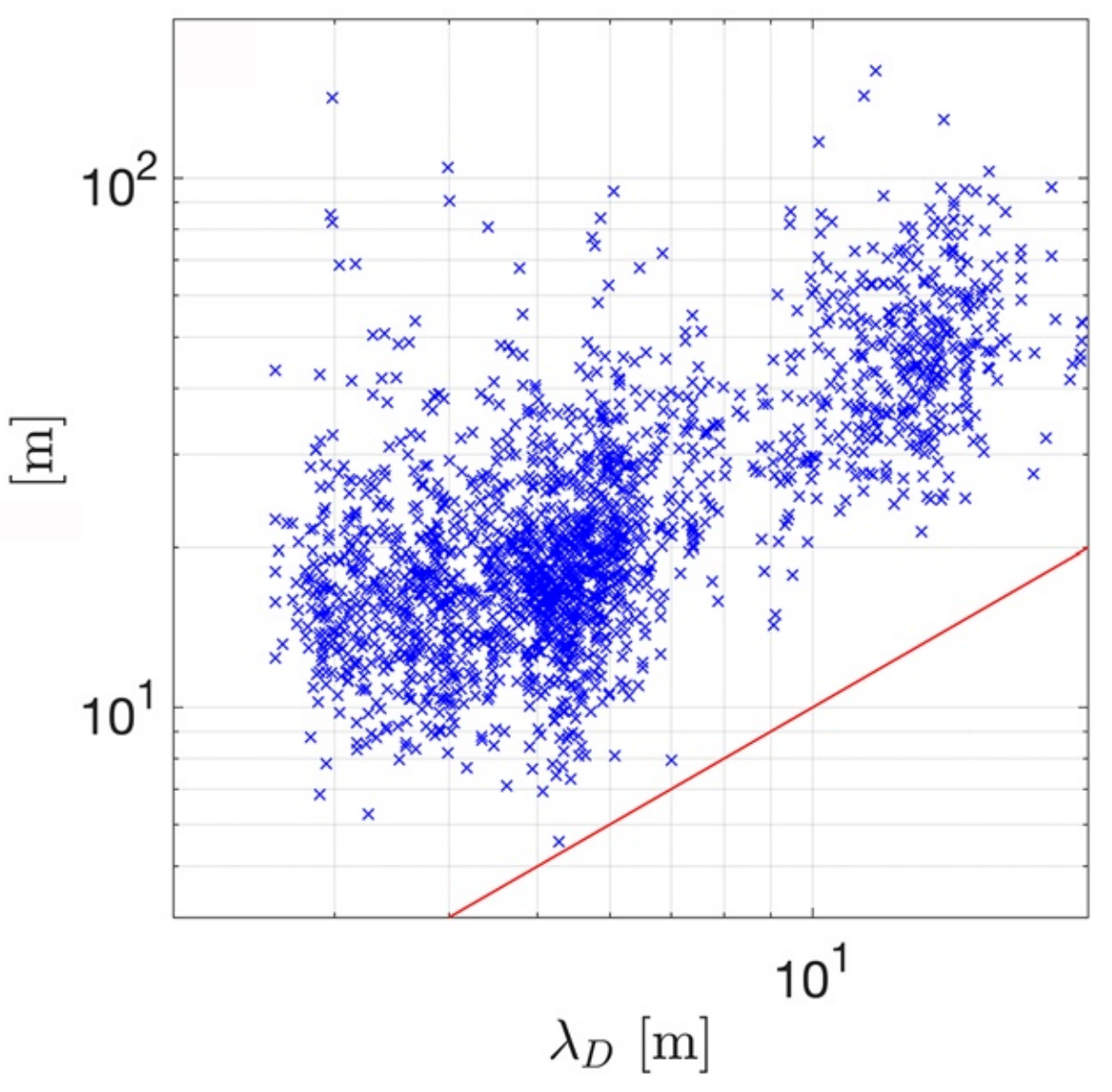} };
\begin{scope}[x={(image.south east)},y={(image.north west)}]
  \draw[black] (0.,.35) node [anchor=south west] {\rotatebox{90}{Length}};
\end{scope}
\end{tikzpicture}\hskip-1.5em(b)
\caption{Ion holes in the Earth's bow shock. (a) Histogram of observed ion
  hole velocities in the ion frame. (b) Hole lengths compared with
  Debye length $\lambda_D$. From \citep{Wang2021}\label{WangFigs}}
\end{figure}
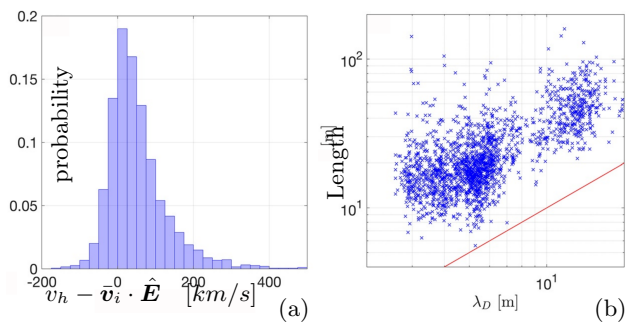
Figure \ref{WangFigs}(a) shows the absolute velocity distribution. 
The hole
length distribution peaks at approximately 3.5$\lambda_D$ with a range
from $\sim2$ to $\sim8$. Moreover the correlation of length with
$\lambda_D$ was unequivocal over nearly a factor of 10, as illustrated
in figure \ref{WangFigs}(b). Amplitudes
were mostly $\psi\lesssim 0.1 T_e$ but with a small proportion up to
$\sim0.3$.  Significantly, the principal direction $\k$ of hole
electric field and (hence) apparent propagation velocity $\v$, is mostly nearly
perpendicular to the shock normal, with little statistical difference
in the sign of the small normal velocity component. The magnetic field
is a somewhat preferred direction for hole propagation, though with little
difference in the sign of $\B.\v$, but there is also a substantial
fraction of the hole velocities spread almost uniformly across all
possible angles with respect to the magnetic field.

The interpretation offered by \citet{Wang2021} is based on the
reasonable assumption that forming ion holes requires at least several
bounce periods, with the bounce frequency (for intermediate trapping
depth) estimated as
$\omega_b=\sqrt\psi \lambda_{D}/L\simeq \sqrt\psi/3$, yielding
$\omega_b\sim \omega_{pi}/10$. The range 1 to 10 bounce periods in
typical bow shock plasmas then corresponds roughly to 10 to 100 ms as
the minimal hole lifetime and of order 1 to 10 km propagation
distance. Arguing on theoretical grounds that growth of an ion-ion
instability saturates when the bounce frequency is comparable to the
initial growth rate, and estimating the maximum growth rate
theoretically for cold ion components, yields a maximum saturation
amplitude (taken as $\psi$) proportional to $n_eL^2$. It is somewhat
greater than the observed hole amplitudes and has comparable slope
with respect to length $L$; so amplitude observations are at least
consistent with ion-ion instabilities being the underlying generation
mechanism. Moreover oblique propagation with respect to the relative
velocity is another theoretical characteristic of ion-ion
instabilities when the beam velocities are more than approximately
2$c_s$ apart, consistent with observations showing predominantly
tangential orientation of $\k$, assuming the ion beam velocity
separation is approximately in the shock normal direction.

Elsewhere, \citet{Wang2022}  have analyzed solitary structures
of negative potential polarity in the Earth's Plasma Sheet, including
the handful set aside by \citeauthor{Kamaletdinov2021} in their
electron hole study because of their negative polarity. The
structures, 153 in total, are simultaneously observed on all four MMS
spacecraft, providing accurate velocity diagnosis by inter-craft
correlation even though single-craft interferometry was not usable
because of spacecraft potential variations. They show evidence of
being strongly oblate (an order of magnitude wider than they are long)
since the single principal normal $\k$ is the same for all spacecraft
and the perpendicular electric fields are small. The angle between $\k$ and the
magnetic field is distributed approximately uniformly over its entire
range, 0 to 90 degrees. This observed non-alignment is attributed to
the unmagnetized ion parameters $\omega_{pi}\gg \Omega_i$ of this
plasma region.  The background ion velocity distributions in which the
ion holes occur are highly non-Maxwellian, but their second moment
corresponds to a temperature of approximately 5 keV, which should be
compared with the electron temperature of $\sim2$ keV. This
temperature ratio provides experimental demonstration that there is
really no requirement $T_e/T_i\gtrsim 3.5$ for ion hole existence, as
mentioned in section \ref{1dequil}. The hole speeds lie between 0.2
and approximately 3 times $\sqrt{T_e/m_i}$; this places them within the
ion distribution and disappearing as they move beyond its extent. Hole
amplitudes $|\psi|$ are from $\sim10^{-3}$ up to $\sim0.2 T_e$; and
there appear to be potential drops across them up to
$\sim{1\over 4}|\psi|$, which might suggest, based on the asymmetry
theory, that they are in the process of accelerating.

\section{Transverse Instability Theory}
\label{Stability}

The predominant theoretical development of electron and ion holes
discussed so far is of their one-dimensional equilibrium and momentum
balance. However, these phenomena occur in multidimensional space, and
there is plenty of observational evidence in the historical review
that holes are often multidimensional, of limited transverse
extent. Simulations of two stream instabilities in multiple dimensions
between approximately 1998 and 2001
\citep{Mottez1997,Miyake1998a,Goldman1999,Oppenheim1999,Muschietti2000,Oppenheim2001b,Singh2001,Lu2008}
observed instabilities breaking up the transverse extent of electron
holes.  And two-dimensional simulations established that electron
holes that start as quiescent one-dimensional entities (uniform in the
transverse direction) when the bounce frequency exceeds the cyclotron
frequency \citep{Muschietti1999,Muschietti2000,Wu2010}
$\omega_b\gtrsim\Omega$ are unstable to kinks with finite transverse
wavenumber $k$. Even when the magnetic field strength exceeds this
threshold, a much slower growing instability is observed in
some simulations, associated with long wavelength striations aligned with
the magnetic field that are identified as whistler waves.  Despite
various theoretical ideas and analyses attempting to explain the
threshold\citep{Vetoulis2001,Jovanovic2002} and the whistler-related
instability\citep{Newman2001,Berthomier2002,Roth2002}, none were
really persuasive about the underlying mechanisms. The original
speculative ``electron focussing'' mechanism of Muschietti tended to
be cited in the context of observations and simulations.

\subsection{Jetting as the instability mechanism}
\label{heuristic}

There matters lay until \citet{Hutchinson2017}  briefly reported
simulations that contradicted the focussing mechanism, and emphasized
that the transverse instability mechanism and threshold were still
open questions. This was a clearly unsatisfactory theoretical
situation, since the transverse instability is implicated in
determining the multidimensional structure of observed electron holes.
The transverse stability of pure electron holes (fast enough that ion
response is negligible) was therefore addressed.

In \citet{Hutchinson2018} the simulation evidence was reported in
detail together with a simple analytic explanation of the mechanism,
based on the kinematic momentum conservation, explained in one dimension
in the present paper's section \ref{Kinematics}. The
PIC simulations approximately confirm the long-standing criterion to
stabilize the fast growing instability, except for providing a
slightly higher numerical coefficient $\Omega \gtrsim
1.5\omega_b$. However, the electron focussing mechanism formerly
proposed to provide the mechanism was discredited since the electron
density enhancement was usually on the convex side of the kinked hole
not the concave. Instead this work proposed that the transverse
instability for negligible magnetic field strength is explained by
considering the constant transverse velocity $v_y$ of particles in the
presence of a kinked hole whose parallel displacement is
$z_h(y,t)=\xi\exp[i(ky-\omega t)]=\xi\exp[(ikv_y-\omega_i)t]$, when
$v_y$ is constant so $y=v_yt$, and $\omega=i\omega_i$ is pure
imaginary.  An electron with this transverse velocity therefore
experiences a hole whose acceleration (at the electron position) is
$\ddot z_h =-(kv_y-i\omega_i)^2 z_h$. Averaging this acceleration over
all $v_y$ gives
$\langle \ddot z_h\rangle= (-\langle k^2v_y^2\rangle+\omega_i^2)z_h
=(-k^2T_y/m_e+\omega_i^2)z_h$. In general, the momentum transfer to
electrons by jetting, $\dot P_e$, is proportional to the acceleration
$\langle\ddot z_h\rangle$. But since the momentum balance ignoring ion
response is $\dot P_e=0$, it does not much matter what the coefficient
of proportionality is, provided it does not significantly vary with
$v_y$. The instability growth rate must then satisfy $\langle \ddot
z_h\rangle=0$, yielding
\begin{equation}
  \label{trans}
  \gamma =\omega_i= \pm k\sqrt{T_y/m_e}.
\end{equation}
This treatment is valid for small enough $k$, but when $kv_y$
approaches the bounce or parallel transit frequency of electrons in
the hole, the coefficient of proportionality reverses its sign and
cancellation of the acceleration of different velocity particles
suppresses the instability, eventually reversing the velocity average
$\langle \dot P_e\rangle$. Because the transit time of the particles
responsible for jetting is roughly $L/\sqrt{\psi}$, where $L$ is the
hole length (typically $\sim 4$), the critical wavenumber at which the
instability is suppressed is approximately $k_c\simeq \sqrt{\psi}/L$
(using a thermal $v_y\simeq1$). The maximum growth rate occurs
somewhere below $k_c$: let us suppose half
($k_{max}\sim \sqrt{\psi}/8$); and the growth rate there will be
perhaps half of equation \ref{trans} because of the incipient
stabilization. That yields the estimate
$\gamma_{max}\simeq k_c/4\simeq\sqrt{\psi}/4L\simeq\sqrt{\psi}/16$.
\begin{figure}[ht]
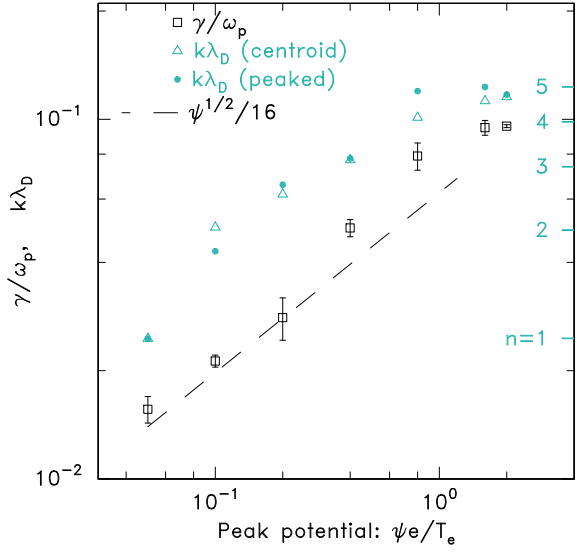
\center
  \partwidth{0.55}{depthscale}
  \caption{Comparison of growth rate and wave number observed in
    simulations with their theoretical
    estimates. Reprinted with permission from \citep{Hutchinson2018},
    copyright the American Physical Society (2018), showing agreement within
    uncertainties.\label{depthscale}}
\end{figure}
Figure \ref{depthscale} shows a comparison of these estimates with the
observed growth rate and wavenumber of the dominant transverse
instability observed in simulations. The agreement is better than
approximately 50\% (up to a hole amplitude of $\psi\sim1$), which is
within the probable uncertainty of the estimates and observations.
Thus, jetting is the mechanism of the transverse instability
when magnetic field is negligible. The effect of finite $\Omega$ is
certainly to weaken the instability, but a rigorous quantitative
calculation requires much more detailed mathematical treatment, to be
described next.

\subsection{Instability calculation including magnetic field}

A full mathematical calculation of the transverse instability based on
the jetting approach was undertaken by \citet{Hutchinson2018a}, of
which a summary description will be given here, omitting many
mathematical details. It is a treatment of the time-dependent
Vlasov-Poisson system linearized for small potential perturbations
relative to a steady one-dimensional nonuniform hole equilibrium. The
Vlasov solution for the electrons, which provides the perturbation to
the distribution $f_1$, is written as an integral over all prior time
of the first order acceleration term $\nabla \phi_1\cdot\nabla_vf_0$,
where $\phi_1=\hat\phi(z)\exp(ky-\omega t)$ is the potential
perturbation and $f_0$ is the unperturbed velocity
distribution. Linearization implies that the integration path can be
approximated as being along the \emph{unperturbed} orbit in phase
space, and the result is
\begin{equation}
  \label{eq:f1parts}
  \begin{split}
    f_{1}(t)& =q\phi_1(t){\partial f_0\over\partial \energy_\parallel}
    +
  q\int_{-\infty}^t\pamper\left(i(\omega-kv_y)
    {\partial f_0\over\partial \energy_\parallel}
    +ikv_y{\partial f_0\over \partial
      \energy_\perp}\right)\hp\, {\rm e}^{i(ky-\omega\tau)} d\tau.    
  \end{split}
\end{equation}
Here $\tau$ is the time prior to the instant under consideration $t$,
and the value of $\hp$ in the integral is at the prior orbit position
and time $y(\tau),z(\tau)$. The parallel and perpendicular energies
are $\energy_\parallel = q\phi+v_\parallel^2/2$ and
$\energy_\perp=v_\perp^2/2$ in normalized units.  The leading term is
called (somewhat misleadingly, but for historic reasons)
``adiabatic'', meaning that it is the perturbation that would occur
for an infinitesimally slow perturbation. No jetting is contributed by
this term. Jetting all comes from the integral term which is the
``non-adiabatic'' perturbation, and can be written
\begin{equation}
  \label{eq:tf0}\tilde f=
  iq_e\left((\omega-kv_y)
    {\partial f_0\over\partial \energy_\parallel}
    +kv_y{\partial f_0\over \partial
      \energy_\perp}\right)\Phi{\rm e}^{i(ky-\omega t)},
\end{equation}
where
\begin{equation}
  \label{eq:tp0}
  \Phi(z_t) \equiv \int_{-\infty}^t\hp(z(\tau)){\rm
    e}^{-i\omega'(\tau-t)}d\tau,
  \ {\rm and} \ \omega'=\omega-kv_y.
\end{equation}
When the magnetic field is ignorable, $v_y$ is simply a constant.
This Vlasov solution must in addition satisfy the Poisson equation
\begin{equation}
  \label{eq:Poisson}
  {d^2 \hp\over d z^2} -k^2\hp=
  -{q_e\over \epsilon_0} \int f_{\parallel1} dv_z=
  -{q_e\over \epsilon_0}\left( \hp {d n_0\over d\phi_0} 
  +\int \tf_{\parallel}dv_z\right).
\end{equation}

In order to proceed, one needs to know the spatial form or forms of
$\hp(z)$, that is, the eigenmode shape. Unlike the standard wave
treatments in a uniform background plasma, the eigenmodes for a
solitary hole equilibrium are not Fourier modes ($\exp(ik_zz)$), and
if one chose to express the eigenmode shapes in Fourier space they
would consist of an infinite sum of coupled Fourier contributions. In
principle \citep{Lewis1979} the linearized Vlasov-Poisson problem for
a solitary equilibrium is an integro-differential eigensystem for
which two sets of basis functions for the potential and for the
distribution perturbations ($\hp$ and $\tilde f$) are combined by
vectors of amplitude coefficients. The system becomes an infinite
matrix equation of which the eigenmodes are the required perturbations
whose corresponding eigenvalues $\omega$ need to be assessed for
stability. To form the matrix in its entirety would require a complete
set of evaluations of the prior Vlasov integral for the whole basis,
which would be overwhelming. Instead, we proceed by realizing that the
dominant eigenmode is a parallel shift of the original hole, whose
linearized form is $\hp=-\xi d\phi_0/dz$, where $\xi$ is the
displacement. Although this might be only an approximation to the
precise perturbation structure, the eigenvalue can be found to second order
accuracy in any deviation from the shift mode approximation by using
the ``Rayleigh Quotient'' treatment (see e.g.\ \citep{Parlett1974}). A more
familiar plasma example of such an approach is the use of the energy
principle together with an assumed approximate displacement to
estimate MHD instability growth rate for a magnetic confinement
configuration. The present application proceeds by substituting the
shift mode into the full Poisson equation, multiplying by $\hp$, and
integrating over all space arriving at
\begin{equation}
  \label{eq:quotient}
\begin{split}
  F_E\equiv&  \xi k^2\int \epsilon_0\left(d \phi_0\over d z\right)^2dz
  \pamper=-\int{d \phi_0\over d z} q_e \int
  \tf_{\parallel}dv_z dz \equiv{F_p+F_t}.
\end{split}  
\end{equation}
The value of $\omega$ that satisfies this equation is the
approximation to the eigenfrequency of the mode. Physically, the left
hand side is the net Maxwell stress force on the kinked hole while the
right hand side is the jetting force of passing and trapped
particles. So the equation is the momentum balance of the hole as a composite
entity.

The paper shows that when $k=0$ then, in the approximation of short
transit time such that ${\rm e}^{-i\omega(\tau-t)}$ in equation
\ref{eq:tp0} can be taken as unity, exactly the same expressions are
obtained for the jetting as were derived by the more elementary
treatment (eqs \ref{passingmom}, \ref{trapmom} in section
\ref{composite}). And in this approximation the growth rate of the
transverse instability with $k\not=0$  is essentially as given in equation
\ref{trans}, though with minor ($< 50$\%) coefficient corrections at
large amplitude $\psi\sim1$.

The past time integral needed to evaluate the forces \emph{without}
short transit time approximation can (after integration by parts) be
written
\begin{equation}
  \label{eq:Ltilde}
  \tL(\omega')\equiv  \int_{-\infty}^t(v_z-v_\infty)i{\rm
    e}^{-i\omega'(\tau-t)}d\tau,
\end{equation}
where $\omega'=\omega-kv_y$. For passing orbits $v_\infty$ is the
distant velocity, and for trapped orbits it is zero. Then the
contribution to the passing and trapped jetting forces per unit
perpendicular velocity per unit parallel velocity (which must be
integrated $d^2v_\perp dv_\parallel$ over the velocity distribution at
$z\to\infty$ for passing and at $z=0$ for trapped orbits to give the
total) is
\begin{equation}
  \label{eq:dFpL}
\begin{split}
  {dF_{p,t}\over d^2v_\perp dv_\parallel} =& 
  i\xi\left(\omega'
    {\partial f_0\over\partial W_\parallel} +(\omega-\omega'){\partial f_0\over \partial
      W_\perp}\right)
  \pampert\int_{p,t}-q{d\phi\over dz}\omega' \tL_{p,t}(z,\omega') {dz\over
    v_z} v_\parallel.
  \end{split}
\end{equation}
All the integrals are carried out numerically and the dispersion
relation $F_p+F_t=F_E$ solved to give the growth rate $\gamma$ as a
function of $k$. The Maxwell stress $F_E$ is mostly ignorable, and the
result is in remarkably good agreement with the heuristic estimates
explained in section \ref{heuristic}.

When a uniform magnetic field is present, it is shown that after
considerable algebra one can extend the treatment, accounting for the
helical orbits (so $v_y$ is sinusoidal in time), using the same
function $\Phi$ but expanding in integer cyclotron harmonics $m$ with
$\omega'=\omega+m\Omega$ instead of $\omega+kv_y$.  The harmonics'
coefficients for Maxwellian perpendicular distribution are
proportional to $I_m(\zeta_t)$, where $I_m$ is the modified Bessel
function and $\zeta_t=k\sqrt{T_\perp}/\Omega=kr_L$ is the finite thermal
gyro radius parameter. Thus
\begin{equation}\label{eq:f1magnetic}
  \begin{split}
  \tilde f_{\parallel } =
 \sum_{m=-\infty}^\infty &   i\left[(m\Omega+\omega)
  {\partial f_{\parallel0}\over \partial W_\parallel}
  +m\Omega {f_{\parallel0}\over T_\perp}\right]
  \pampert q_e\Phi_m {\rm e}^{-\zeta_t^2}I_m(\zeta_t^2),
\end{split}
\end{equation}
where 
$  \Phi_m(z)\equiv\int_{-\infty}^t \hp(z(\tau))
{\rm e}^{-i(m\Omega+\omega)(\tau-t)}d\tau$.
\begin{figure}[ht]
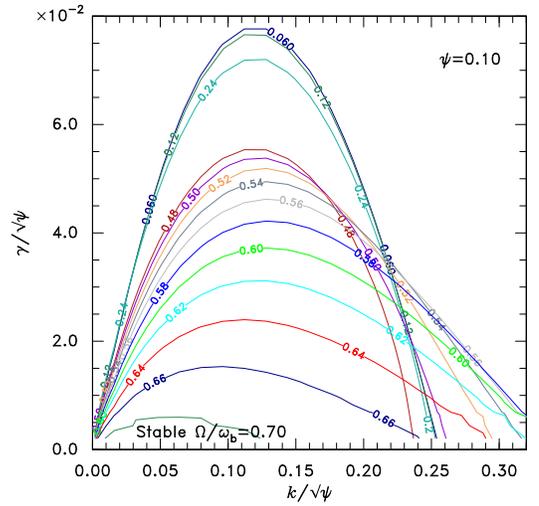
\center
  \partwidth{0.5}{transB}
  \caption{Growth rate $\gamma$ of the transverse instability versus
    wavenumber $k$, for a range of magnetic field strengths
    $\Omega/\omega_b$ (labeling the lines). Reprinted with permission
    from \citep{Hutchinson2018a} copyright Cambridge University
    Press.  \label{transB}}
\end{figure}
Evaluating $\tilde f_\parallel$, then substituting into $F=F_p+F_t-F_E$ whose
zeroes give equation \ref{eq:quotient} and plotting the zero contour
corresponding to equality on the plane $(k,\gamma)$, for imaginary
frequency $\omega=i\gamma$, gives the results of figure \ref{transB}. The
growth rate and perpendicular wave number are both proportional to
$\sqrt{\psi}$ except when $\psi$ is large, $\sim 1$. So they are
plotted normalized to $\sqrt{\psi}$ giving approximately a universal
curve regardless of amplitude $\psi$.

When the magnetic field is weak
$\Omega/\omega_b\lesssim 0.1$ it may be ignored and one gets the
unmagnetized result with $k_c=\sqrt{\psi}/4$, $k_{max}=\sqrt{\psi}/4$,
and $\gamma_{max}$ a little larger than $\sqrt{\psi}/16$. As the
magnetic field is increased it systematically decreases the peak
$\gamma_{max}$, and the (pure growth) instability disappears at
$\Omega/\omega_b=0.70$. The reason for the growth suppression is shown
quantitatively to be caused by the progressive sign reversal of
particle force contributions from orbits more deeply trapped. (All of
the quantitative results are for a $\sech^4$ potential, and ignore ion
response.)

This magnetic field threshold $\Omega/\omega_b=0.70$ for stability is
approximately half that observed in simulations. This discrepancy was
shown, in \citet{Hutchinson2019}, to be because there is an
additional type of instability, not considered in
\citep{Hutchinson2018a}, that has finite real part of frequency as
well as imaginary part. It is an oscillating instability, and occurs
slightly above this magnetic field strength threshold for purely growing
modes. The new results for this mode were obtained using the same
analysis but removing the restriction to zero real part of the frequency
$\omega_r$. They are also accompanied by detailed comparisons with PIC
simulation.
\begin{figure}[htp]
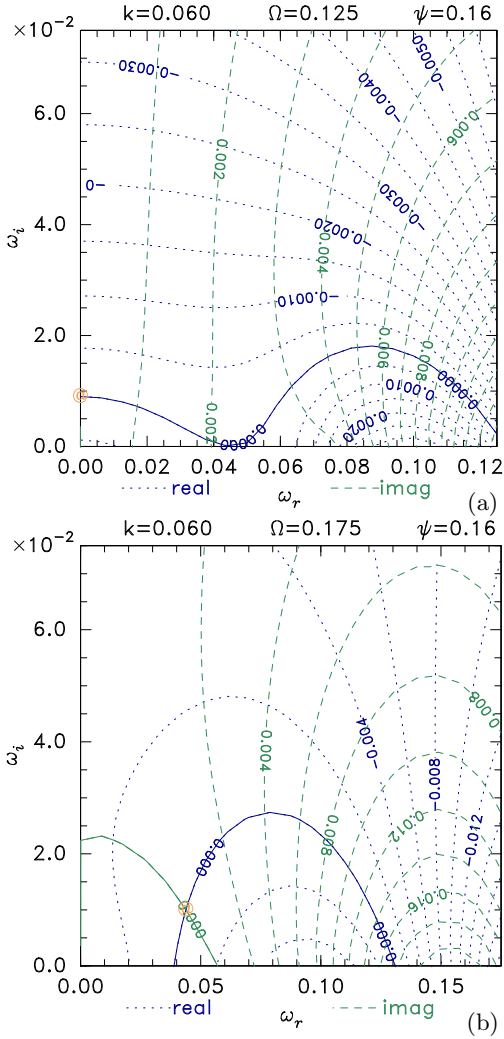

  \partwidth{0.48}{ocontr12}\hskip-1.5em(a)\hskip.5em 
  \partwidth{0.48}{ocontr17}\hskip-1.5em(b) 
  \caption{Contours of the real and imaginary part of the total force
    $F$, which are both zero at the eigenfrequency marked @. Reprinted
    with permission from \citep{Hutchinson2019}, copyright the
    American Physical Society.\label{ocontr}}
\end{figure}
Figure \ref{ocontr} shows calculated contours of the real and
imaginary parts of the total force $F$ that must both be zero at the
eigenfrequency, on the complex $\omega$ plane, for two cases. (a) has
magnetic field strength $\Omega=0.125$: below the pure-growth
stabilization threshold $\Omega=0.7\omega_b=0.7\sqrt{\psi}/2=0.14$
(for $\sech^4$ hole amplitude $\psi=0.16$). It shows an intersection
of the real $F$ zero contour (solid blue) with the imaginary $F$ zero
contour along the vertical axis, at a purely growing ($\omega_r=0$)
frequency $\omega_i\simeq 0.01$ (marked with @). By contrast, in (b)
at $\Omega=0.175$ above the 0.14 threshold, the real $F$ contour no
longer intersects the $\omega_r=0$ axis: no pure-growing instability
is present. But it does intersect a branch of the (solid green) imaginary
$F$ zero contour at position $\omega=(0.043,0.01)$ indicating oscillatory
growth.

\begin{figure}[htp]
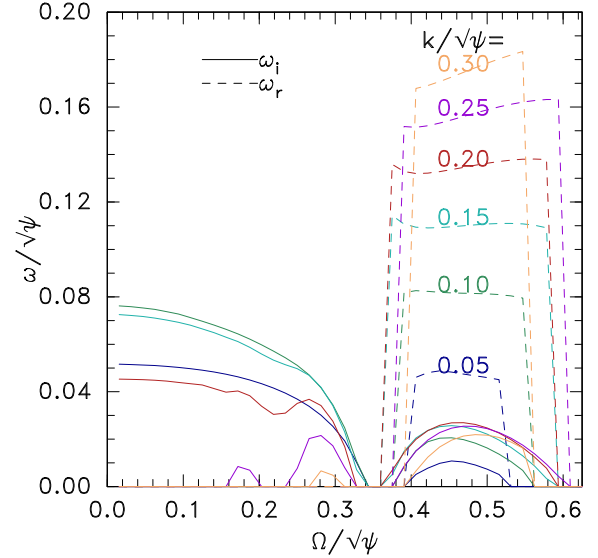
\center
  \partwidth{0.55}{ovbk16}
  \caption{Real and imaginary parts of the eigenfrequency versus
    magnetic field strength for a range of wave numbers. Reprinted
    with permission from \citep{Hutchinson2019}, copyright the
    American Physical Society.\label{ovbk}}
\end{figure}
Extensive numerical investigation of unstable eigenfrequencies is
summarized in figure \ref{ovbk}. As noted previously, for small and
moderate $\psi$, normalizing the frequency and wavenumber by
$\sqrt{\psi}$ yields essentially universal curves. Then the real
($\omega_r$ dashed line) and imaginary ($\omega_i$ solid line) parts
of the frequency are plotted versus $\Omega$ for the relevant range of
wavenumber $k$. When $\Omega/\sqrt{\psi}<0.35$ (equivalent to
$\Omega<0.7\omega_b$), pure growing instability occurs ($\omega_r=0$
and $\omega_i>0$) for $k$ in the middle of its range. That is the
prior result. However for $0.35<\Omega/\sqrt{\psi}\lesssim0.6$ a range
of oscillatory instabilities occurs whose $\omega_r$ is essentially
independent of $\Omega$ but determined by $k$. This is responsible
theoretically for the extended unstable range of $\Omega$.

The results of PIC simulations of initially one-dimensional $\sech^4$
holes are directly compared with these analytic results over a full
range of corresponding parameters. 

\begin{figure}[ht]
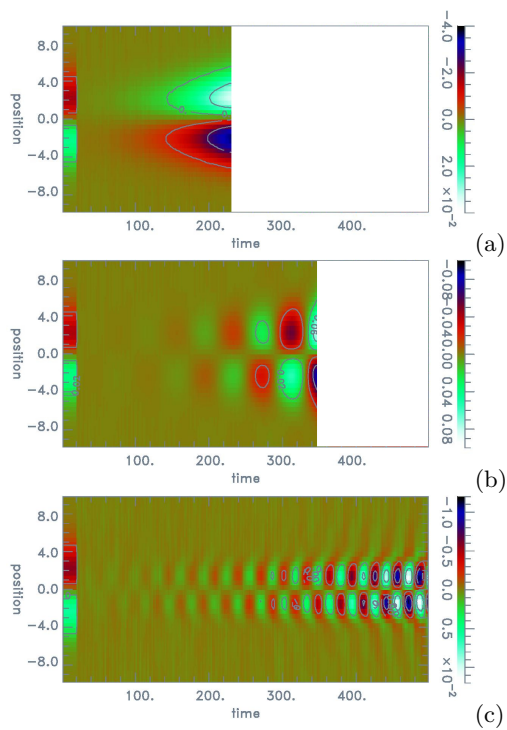

  \centerline{\vbox{\hsize=0.45\hsize
    \partwidth{1.}{B20cm1masked}(a)\par
    \partwidth{1.}{B25cm2masked}(b)\par
    \partwidth{1.}{B45cm6final}(c)\par
  }\hfil}
  \caption{Simulation instability perturbation parallel structure as a function
    of time and parallel position, as contours of $\hat \phi$.
    Hole amplitude is $\psi=0.49$.
    The three magnetic field strength cases are: (a) $\Omega=0.2$
    (b) $\Omega=0.25$ (c) $\Omega=0.45$. Reprinted with permission
    from \citep{Hutchinson2019}, copyright the American Physical Society. 
    \label{simmodes}}
\end{figure}
Figure \ref{simmodes} displays the perturbation mode structure
evolution for three magnetic fields. The color contours indicate the
PIC's perturbation $\tilde \phi$; time is along the horizontal axis;
and parallel position $z$ relative to the hole center is on the
vertical axis. (Only a local fraction of the total simulated $z$-length is
displayed). At the left ($t=0-20$) is shown an arbitrarily scaled
contour of $d\phi_0/dz$, the theoretical shift mode, for comparison
with the later growing simulated perturbation. The first case
\ref{simmodes}(a) has $\Omega/\omega_b=0.57$ below the threshold (0.7)
for pure growth stabilization. It shows pure growth with mode
structure highly similar to the shift mode. In figure
\ref{simmodes}(b), $\Omega/\omega_b=0.71$ just above the threshold for
oscillatory instability, the oscillations are clear, and have
comparable mode structure. Figure\ref{simmodes}(c),
$\Omega/\omega_b=1.29$ ($\Omega/\sqrt{\psi}=0.64$) is at the upper end
of the observed overstable region, and slightly above the analytic
limit of instability ($\Omega/\sqrt{\psi}=0.6$) shown in figure
\ref{ovbk}. Its mode structure is narrower than the shift mode,
showing that there is some deviation from the analytic mode
approximations, which explains the slightly extended unstable range.

With the exception of this extension and a more gradual cut off of the
simulations' growth to near zero at high $\Omega$, the detailed
comparison of 25 simulations shows very good agreement with analysis
of the real frequency (better than $\sim20$\%, despite some
limitations attributable to discrete $k$ enforced by limited $y$
simulation length); excellent agreement ($\lesssim 5$\%) with the
$\Omega$-thresholds for mode oscillation and growth; and good
agreement with growth rates.

\subsection{Strong magnetic field approximation}

The remarkable agreement with simulation of the analysis of magnetic
field stabilization of the transverse instability confirmed the
identification of the instability mechanism; but it left open some
important indications that, at even higher magnetic field, a new type
of instability with much longer transverse wave-length and much lower
frequencies persists. This strong-field region
instability probably corresponds to what was observed in the earlier
high-field simulations of
\citep{Goldman1999,Oppenheim1999,Newman2001a,Oppenheim2001b,Umeda2006,Lu2008,Wu2010},
usually in association with coupled long-parallel-wavelength
potential waves, identified as belonging to the ``whistler'' wave
branch. In figure \ref{simmodes}(c) (which is actually not in the high
field regime) one can see similar coupled streaks, but with higher
frequency, extending up and down to distances far beyond the localized
oscillating mode.

In \citet{Hutchinson2019a}  the high-field limit was addressed. It can
be analyzed in the approximation of purely one-dimensional motion of the
electrons, which corresponds to the general magnetized analysis when
only the $m=0$ cyclotron harmonic needs to be retained. Using the
general numerical Vlasov solution, the contours of force obtained are
shown in figure \ref{triangle}(a).
\begin{figure}
  \partwidth{0.48}{plot2psi09r}\hskip-1.5em (a)
  \partwidth{0.48}{triangle}\hskip-1.5em (b)
  \caption{Comparison of (a) the numerically calculated force
    contours, with (b) the imaginary part contours of the analytic
    approximation, showing good agreement. Reprinted with permission
    from \citep{Hutchinson2019a} copyright Cambridge University
    Press.\label{triangle}}
\end{figure}
The real and imaginary parts of the mode frequency $\omega_r$ and
$\omega_i$ are observed to scale such that if they are normalized as
$\hat \omega_r=\omega_r/\psi^{3/4}$ and $\hat\omega_i=\omega_i/\psi^{3/2}$
(not the same factor as before, or as each other), then contours that
are essentially universal are obtained at small $\psi$. Contours of
the real part of $F$ are to an excellent approximation vertical
lines. The horizontal positions of those lines are determined by the
value of $k$ through the real part of equation \ref{eq:quotient},
i.e.\ the real part balance between jetting and the transverse Maxwell
stress ($\propto \xi k^2\psi^2$). The imaginary part zero contour is
approximately triangular in shape and independent of $k$. For this
particular $k$, the real part zero contour intersects it at its
peak. This $k$ is the mode with the greatest $\omega_i$, the fastest
growing. 

A fully analytic linearized treatment of the stability of a $\sech^4$
potential shape hole is given in this work, which treats the shift
mode in the one-dimensional motion limit. It identifies three main
force terms contributing to the imaginary $F$. These are (i) the
imaginary trapped electron force arising from resonance between the
bounce frequency and the mode real frequency: $F_R$, (ii) the
intrinsically imaginary part of the passing particle force $F_p$, and
(iii) the non-resonant part of the electron force: $F$ (mostly
trapped) whose imaginary part arises only because of the imaginary
part of $\omega$ and the fact that $F\propto \omega^2$. These are
evaluated analytically using square-wave approximations of the trapped
position $z(t)$ (because the orbits dwell much of the time near their
turning points), and of the positional advance of the passing
particles $z-v_\infty(t-t_0)$ (because this advance mostly occurs near
the potential peak $z=0$). The total for small $\omega_i$ is
\begin{equation}
  \begin{split}
  \Im(F)=&\Im[F_R+F_{p}+ 2i{\omega_i\over\omega_r}F(\omega_r)]
\pamper=\xi[C_t\omega_r^5/\psi^{1/2}+C_p\omega_r^2\psi+C_i\omega_i\omega_r\psi],
\end{split}
\end{equation}
where the coefficients are obtained by the integration approximations
as $C_t\simeq 1000$, $C_p\simeq-7$, and $C_i\simeq 10$. The scaling
relations $\hat\omega_r=\omega_r/\psi^{3/4}$ and
$\hat\omega_i=\omega_i/\psi^{3/2}$ render the scaled expression
\begin{equation}
  0=\Im(F)=\xi\psi^{13/4}\hat\omega_r
  [C_t\hat\omega_r^4+C_p\hat\omega_r^2+C_i\hat\omega_i],
\end{equation}
showing why the frequency scalings have the form they do (and that the
scaling for force is $\hat F=F/\psi^{13/4}\hat\omega_r$ though this
scaling is not applied in the figure). Figure \ref{triangle}(b) shows
the analytic contours of $\Im(F)$. The comparison with the numerical
figure \ref{triangle}(a) is very satisfactory. It should be noted how
extremely small the growth rate $\omega_i$ is even for $\psi\sim 1$
and that it quickly becomes much smaller ($\propto\psi^{3/2}$) for
smaller $\psi$.
The absolute relationship between $k$ and $\omega_r$ for this shaped
hole is $k^2=(315/256) C_i\omega_r^2/\psi$, which is
$k/\psi^{1/4}=3.5\hat\omega_r$.

The overall agreement between numerical evaluations and analytic
approximations is strong confirmation of the correct identification of
the key components of the force that need to be balanced. The
identification shows that the resonant force has a \emph{stabilizing}
contribution, not, as has often been assumed in prior theories, a
destabilizing one. This counter-intuitive situation results from the
fact that the hole mass is negative, and the instability is a negative
energy mode \citep{Lashmore-Davies2005} (which grows as its energy becomes more negative).

The same paper \citep{Hutchinson2019a} reports comparisons of the
theory with new PIC simulations. These prove to be in good agreement
with the theory on the ratio $k/\omega_r$ of the unstable modes, but
only qualitative (of order factor of 2) agreement in growth rates,
with the simulations showing larger growth rates when instability
occurs. This is attributed to the fact that the maximum growth rate is
determined by the stabilizing resonant force at very low frequencies
corresponding to trapped particle orbits very close to $\energy=0$, at
which there is (in theory) a discontinuity in the distribution slope.
Any stochastic orbit wandering or numerical effective collisionality
in the simulation there is liable to reduce $df/dv$, and hence reduce
the resonant stabilization (i.e.\ make the hole more unstable).

It is observed, as in prior simulations, that small changes in the
computational domain size can suppress or enhance instability. This
implicates resonant whistler wave effects in finite domains as
influential in simulations (but not necessarily in nature), but it
does not show that coupling to whistler waves is the mechanism of
instability, as has sometimes been proposed. In contrast, what is
shown is that there is an instability for which coupling to whistler
waves, though sometimes relevant in simulations, is unnecessary to its
mechanism.

\citet{Umeda2006}  observed no instabilities in their high field
simulations with hole potential below $\psi=0.8$. Somewhat similarly,
in the present simulations no instabilities that grew to substantial
amplitude were observed even up to time 10,000 for initial hole
heights $\psi< 0.4$. There was some evidence of competition between
different transverse modes at quite low amplitude.

Summarizing the high magnetic field (one-dimensional motion)
transverse instability: there is theoretically a very slowly growing
oscillatory instability $\omega_i\sim 10^{-3}\psi^{3/2}$, that has
been observed in simulations only for deep holes $\psi>0.4$. Its
mechanism does not depend on coupling to whistler waves, but it is
observed to excite them (as do other oscillatory hole
instabilities). Its destabilization is by the imaginary component of
the passing particle jetting ($F_p$); the resonant trapped
particle force $F_R$ is a counterbalancing stabilizing effect. The
real frequency of the fastest growing mode (of transverse wavenumber
$k$) is $\omega_r\simeq 0.3 k\sqrt{\psi}\simeq 0.05\psi^{3/4}$.

\subsection{Ion response effect on transverse instability }

All the preceding discussion of this section is of electron holes in
which ion response is ignored, justified by supposing the holes fast
compared with the ion velocity distribution. However, in
\citet{Hutchinson2022} the transverse instability theory of slow
electron holes including ion response, was presented. It also
addressed the one-dimensional consequences for the dynamics of the
slow holes summarized for this review in subsection \ref{SlowEHoles}.

The many degrees of freedom of this more complicated situation have
not all been explored but the paper presents two representative figures
of the dependence of the complex frequency on ion distribution shape,
transverse wavenumber, and magnetic field strength. Broadly speaking,
ion distributions that are double humped with the hole velocity lying
in their local minimum, represented by two equal Maxwellians shifted
by $v_s=1.3$ (compare figure \ref{omegaofv1}) show the least effect of
ion response on the transverse instability, and indeed when amplitude
is small ($\psi=0.01$) the results are little changed by ion
response. At moderate amplitude ($\psi=0.09$), however, the effect is to
increase the growth rate of the intermediate regime
($0.35<\Omega/\sqrt{\psi}<0.6$) oscillatory instability by perhaps a
factor of two, and to cause the high field ($0.6<\Omega/\sqrt{\psi}<1$)
regime to have a growth rate $\omega_i/\sqrt{\psi}\sim 0.03$,
comparable to the intermediate, and much higher than the strong magnetic
field approximation for electron-only response.

For a single-humped ion distribution ($v_s=1$), the destabilization of
the high field regime is present even for low amplitude
($\psi=0.01$). Moreover the intermediate and high field regime
instabilities become purely growing ($\omega_r=0$) for some longer
transverse wavelengths, showing that the ion reflection force hole
acceleration mechanism, which for a single-humped distribution gives
pure growth, has begun to be more important than the electron
jetting. At $\psi=0.09$ these single-humped ion interactions dominate
all but the shortest transverse wavelengths, and the growth rates are
several times higher than when ion interactions are absent. Thus, ion
reflection forces for slow electron holes of moderate depth in single-humped
ion distributions strongly enhance the transverse instability and make
the fastest growing modes of longer transverse wavelength. There is
doubtless scope for further theoretical and computational exploration
of the transverse instability of slow electron holes.

\subsection{Generalized multi-mode transverse instability}

Although the mechanisms of transverse hole instability are
comprehensively identified in the prior subsections, there remain some
quantitative discrepancies in the comparisons with simulation that are
thought to arise from the fact that the actual linearized perturbation
potential structure is not exactly the shift mode that has been
assumed. An appropriate challenge is then to establish a theory that
predicts the modification of the eigenmode by deviations from the
shift mode, and how much that changes the instability eigenvalue
($\omega$). In effect, this is the challenge that the early work of
\citet{Lewis1979} analyzed and set forth an outline mathematical
program to solve. For non-linear wave states \citet{Schwarzmeier1979}
used this technique to solve for the unstable merging of adjacent
peaks of a periodic wave. Much later \citet{Siminos2011} used a Fourier
method to construct the predominant eigenmode for this one-dimensional
problem.

However, the very few previous attempts to implement
such a program for the electron hole transverse instability were
unsuccessful either because of incorrect identification of the
predominant form of the instability (e.g. addressing symmetric
perturbations instead of antisymmetric) or because their method of
integrating the Vlasov equation was inappropriate. Theoretical
coupling to whistler waves was previously investigated based on ad hoc
assumptions about the coupling strength that were not well
justified. Calculating this coupling rigorously is formally part of
the same mathematical program. The recent work of \citet{Chen2023}
 successfully carried out a rigorous analysis of this problem
for $\sech^4$ shaped electron hole equilibria, as will now be
explained.

The unstable linearized eigenmode $\hat\phi(z)$ of a solitary
potential structure is not exactly the shiftmode when $\omega$ is
finite. In principle it can be represented exactly by a weighted sum
of any complete (infinite) set of functions. The prescription of
\citep{Lewis1979,Lewis1982} is that the most convenient
functions are the eigenfunctions of the adiabatic Poisson operator. For an
equilibrium of the form $\phi_0(z)=\psi\,\sech^4(z/4)$, the adiabatic
operator is
\begin{equation}
  \label{poissonop}
  V_a        
  = \left[{d^2\over dz^2} - {\phi_0^{'''}\over \phi_0'}\right]
  = \left[{d^2\over dz^2} +{30\over 16}\sech^2(z/4)- 1\right],
\end{equation}
It can be shown that the eigenmodes of this operator that satisfy
$V_a\ket{u}=\lambda\ket{u}$ (using bra-ket notation) consist of just 5
discrete eigenmodes of limited $z$-extent, plus a continuum of
eigenmodes that extend to $|z|=\infty$ with sinusoidal behavior far
from the hole.
\begin{figure}[ht]
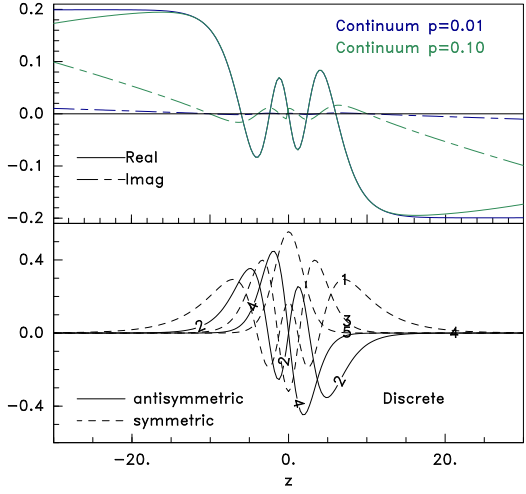
\center
  \partwidth{0.5}{eigenmodes}
  \caption{The eigenmodes of the adiabatic Poisson operator.
    Reprinted with permission from  \citep{Chen2023} copyright
    Cambridge University Press.\label{eigenmodes}}
\end{figure}
Figure \ref{eigenmodes} illustrates these eigenmodes. Of the discrete
modes numbered 1-5, three are symmetric (1,3,5), and two are
antisymmetric (2,4). Mode 4 is the shift mode, which, normalized to make
$\bra{u}\ket{u}=\int u^*u dz=1$, is
\begin{equation}
  \label{shifteigenmode}
  \ket{4} = -3\tanh(z/4)\sech^4(z/4)\sqrt{70}/16,
\end{equation}
proportional to $d\phi_0/dz$. The other modes, including the
continuum, similarly have analytic forms helpful for
evaluation. They will not be given explicitly here. The eigenvalues can
be written
\begin{equation}
  \label{eigenvalues}
  \lambda_u=u^2/16-1,
\end{equation}
where $u$ is the integer mode number for the discrete modes and for
the relevant antisymmetric continuum modes $u$ is equal to
$4i\sigma_zk_z=i\sigma_zp$, where $k_z$ is the positive parallel
wavenumber of the mode in the external sinusoidal region, and
$\sigma_z$ is the sign of $z$. The shift mode $\ket{4}$ eigenvalue
$\lambda_4$ is zero. That is why it is the dominant mode, and is
exactly the relevant mode in the limit of low frequency.

For an infinitesimally small $\omega$ these eigenmodes are exactly the
eigenmodes of the linearized Vlasov-Poisson problem, and are uncoupled
from one another. However, when $\omega$ is finite, the Vlasov-Poisson
problem acquires a non-zero non-adiabatic contribution that is written
as a non-adiabatic operator $\tilde V$, such that Poisson's equation
becomes
\begin{equation}
  \label{fullpois}
  (-k_\perp^2+V_a+\tilde{V})\ket{\hat\phi}=0,
\end{equation}
where the squared transverse wavenumber $k_\perp^2\equiv\lambda_\perp$
can be considered the full eigenvalue. The operator $\tilde V$ couples
the adiabatic eigenmodes together, but since they are a complete set,
any perturbation mode can be written as a weighted sum and integral of
eigenmodes:
$\ket{\hat\phi}=\sum_{j,discrete}a_u\ket{u}+\int_{p,continuum}a(p)dp$,
where $p=u/i\sigma_z$ is real and positive.
The sums and integrals are limited to the antisymmetric forms when
considering the dominant transverse instability. No coupling arises
between symmetric and antisymmetric modes. Substituting into Poisson's
equation, left-multiplying by any adiabatic eigenmode $\bra{s}|$, and
using the orthogonality of the eigenmodes,
we get
\begin{equation}
  \label{eigengen}
  0=(\lambda_s-\lambda_\perp)\bra{s}\ket{s}a_s+\sum_j\bra{s}|\tilde{V}\ket{j}a_j
+\int\bra{s}|\tilde{V}\ket{p}a(p)dp.
\end{equation}
This is a matrix equation $\M\a=0$ in which the off-diagonal matrix
elements arise from the non-adiabatic couplings
$\bra{s}|\tilde V\ket{j,p}$, and which, if $\omega$ and hence
$\tilde V$ is specified, requires a specific value of $\lambda_\perp$
to make $|\M|=0$ and permit a solution for the coupled mode amplitudes
$\a$. Actually it is more convenient in practice to take
$\lambda_\perp$ as fixed and find iteratively the $\omega$ that then
makes $|\M|=0$.

As it stands, the continuum integral makes this an infinite dimension
matrix. Moreover the inner products $\bra{s}|\tilde V\ket{p}$ for the
continuum involve an infinite domain $z$-integral over which the
continuum adiabatic eigenfunctions remain finite. However, the paper shows that
the continuum contributions for any frequency are concentrated about a
specific parallel wavenumber (i.e. $p$-value) that satisfies the
sinusoidal wave dispersion relations (for kinetic electrostatic
whistler waves) in the external region. The $p$-integral over that
small range can be carried out analytically.  It is therefore possible
to approximate all the continuum contributions into a single amplitude
$a_q=\int a(p)dp$ and the entire coefficient vector as
$\a=(a_2,a_4,a_q)$ representing the two antisymmetric discrete
adiabatic eigenmodes plus the continuum, and approximate the matrix $\M$ as
3$\times$3.  The infinite $z$-integrals can be handled by subtracting
a purely multiplicative sinusoidal wave operator $V_w$.  Calculating
the matrix elements still requires numerical evaluation for the
electron hole region, but the external region integrals can be
obtained analytically. Thus, the initially overwhelming dimensionality
of the eigenmode problem is reduced to finding the zero of the
determinant of the complex 3$\times$3 matrix $\M$. When $|\M|=0$, it
has just two independent rows, and solving $\M\a=0$ determines the
direction of the vector $\a$, in the form of two values $a_2/a_4$ and
$a_q/a_4$: the coupling coefficients of the subdominant modes to
$\ket{4}$. Incidentally, the upper right matrix diagonal coefficient
is simply the shiftmode $\ket{4}$ force, assumed in the previous work
to be the total to be zeroed. The multimode analysis generalizes the
calculation. 

Completing this analysis thus provides both the corrections to the
perturbation eigenmode shape, and the coupling between the hole
shift-mode and the external ``whistler'' waves.
\begin{figure}
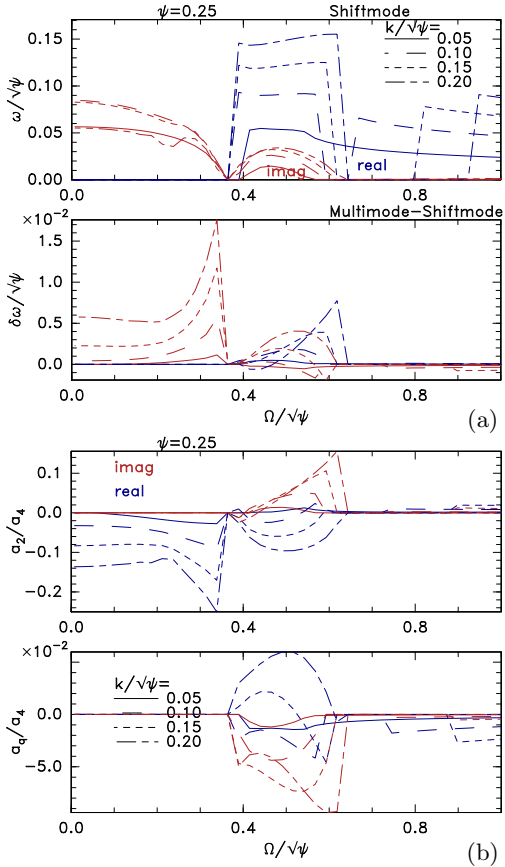

  \partwidth{0.48}{multimodeshift}\hskip-1.5em(a)\hskip.5em
  \partwidth{0.48}{multimodediff}\hskip-1.5em(b)\hskip.5em
  \caption{(a) Shiftmode frequency, and the difference multimode
    - shiftmode frequency, real (red) and imaginary (blue) parts. (b)
    Relative amplitude of the secondary discrete mode ($a_2$) and the
    continuum mode $a_q$. Reprinted with permission from
    \citep{Chen2023} copyright Cambridge University Press. \label{multimode}}
\end{figure}
Figure \ref{multimode} summarizes extensive numerical
calculations. The upper part of (a) shows shiftmode (pure $\ket{4}$
eigenmode) frequencies $\omega$ as a function of $\Omega$ for four
values of $k_\perp$, and the lower part shows the difference
$\delta \omega$ that must be added to $\omega$ to obtain the
calculated multimode value. Those differences are of order $<10$\%
growth rate increase for the low-$\Omega$ purely growing instability
(except near its upper threshold); and similar for the intermediate
oscillatory regime. In the high field regime the differences are very
small and stabilizing.  Figure \ref{multimode}(b) shows the
corresponding secondary mode relative amplitudes. The value of
$a_q/a_4$ is the coupling to the continuum mode, which is substantial
only in the oscillatory regimes, in accordance with simulation
observations.

\begin{figure}
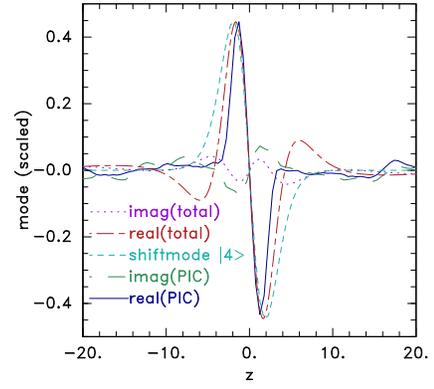
\center
  \partwidth{0.4}{modenarrowing}
  \caption{Real and imaginary parts of the unstable mode spatial
    structure, for shiftmode and total multimode, compared with
    observed mode structure in a PIC simulation.
    Reprinted with permission from \citep{Chen2023} copyright
    Cambridge University Press.\label{modenarrowing}}
\end{figure}

The contribution of the secondary discrete mode is illustrated more
intuitively in figure \ref{modenarrowing}, which shows the parallel
shape of the unstable shiftmode and multimode calculations, together
with a corresponding observed mode structure in a PIC simulation.
The parameters of this case are close to the upper end of the
intermediate (oscillatory) $\Omega$-range. 
Concentrating on the real part, it is clear that the addition of the
$\ket{2}$ discrete mode component \emph{narrows} the mode structure,
making it much closer to the observed PIC shape than the shiftmode
alone.

In summary, the multimode analysis demonstrates mathematically that
using the shiftmode slightly underestimates the growth rate of the
transverse instability near its upper threshold.  In part at least,
this is because the true unstable perturbation structure is narrower
than the shiftmode, which emphasizes the unstable contribution
of deeper trapped orbits over the stabilizing contribution of
marginally trapped orbits, in accordance with prior qualitative
considerations. 

\section{Multi-dimensional holes}
\label{Multidimensional}

This section addresses some important phenomena that are inherent to
holes of limited transverse extent.

\subsection{Equilibria}
\label{3dequil}

It is of course only an approximation to treat electron holes as
purely one-dimensional in space. Multi-dimensional equilibria have in
addition been explored for a long time in the literature. Higher
dimensionality introduces several different new effects. The first and
perhaps most obvious is that for a hole of finite transverse dimensions
Poisson's equation must be written $\nabla^2\phi=-\rho$, and the
transverse derivatives provide a new contribution. References
\citep{Schamel1979,Chen2002} take the form of the hole to be separable
and the transverse contribution to arise from the radial
derivative. Since the particle motion is taken still to conserve
parallel energy, the main effect is to require more charge deficit to
support a particular peak potential $\psi(z=0,r=0)$. The
one-dimensional motion limit is also assumed in
\citep{Chen2004,Krasovsky2004a} but with Gaussian radial variation,
and \citep{Muschietti2002} used transverse Gaussian but $\sech(z)$
parallel shape with variably flattened top. The work of
\citet{Krasovsky2004a} proceeded instead from a prescribed trapped
particle distribution that varied linearly with $-\energy$ for
$\energy<\energy_j$ but was constant for $\energy_j<\energy<0$. The
mathematical convenience of this choice is that the Poisson equation
takes the Helmholtz form $(\nabla^2+k^2)\phi=const$.

Although assuming separable potential form is mathematically
convenient, it is never in fact accurate in the distant regions of a
multidimensional hole. The reason is that when the potential is small,
Debye shielding is the dominant effect, it gives a modified Helmholtz
equation (i.e.\ with negative constant $k^2=-1/\lambda_D^2$). Thus, in
an axisymmetric hole at $z=0$ and large transverse distance $r$,
${d\phi\over dr}\simeq\phi/\lambda_D$ and ${d\phi\over dz}=0$,
while at $r=0$ and large $|z|$, ${d\phi\over dz}\simeq\phi/\lambda_D$
and ${d\phi\over dr}=0$. These limits are incompatible with a
separable form.

Aiming to discover how serious these inconsistencies are for
multidimensional holes, \citet{Hutchinson2021a} developed a more
general approach to synthesizing such holes based on a general
prescription of the trapped parallel velocity distribution
$f_t(\energy)$ allowing various parallel ($z$) hole shapes, and a
desired radial peak potential variation $\psi(r)$.  The parallel
distribution chosen allows an analytic solution of the one-dimensional
parallel differential equation incorporating an approximation of the
transverse field divergence. This solution finds the approximate
$f_t(\energy,r)$ parameters to yield the desired $\psi(r)$. The scheme
then iteratively solves the unapproximated multidimensional Poisson
equation to find the entire two-dimensional ($r,z$) variation of the
potential self-consistently with the prescribed $f_t(\energy,r)$.

\begin{figure}[ht]
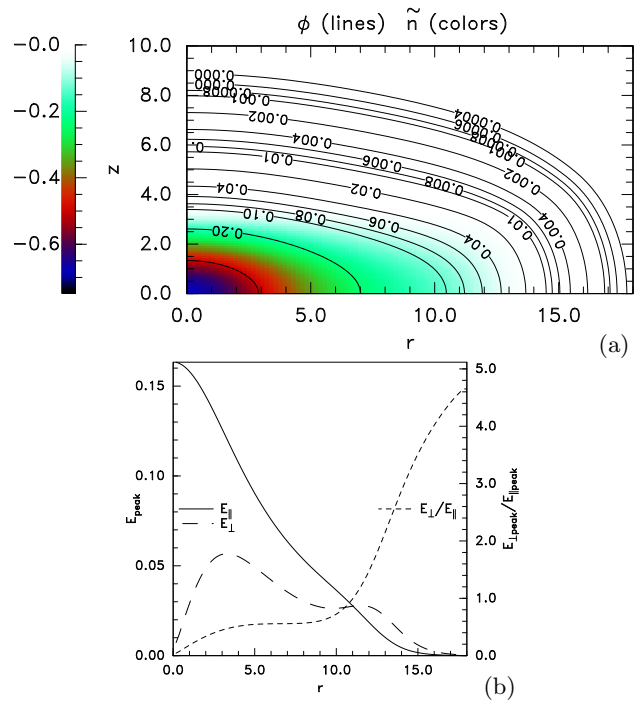
\center
  \partwidth{0.6}{phirho2}\hskip-1.5em(a)
  \partwidth{0.38}{eperpmax2}\hskip-1.5em(b)
  \caption{A fully consistent axisymmetric electron hole. (a)
    Potential and density-deficit contours. (b) Peak electric field
    variation with track radius. From \citet{Hutchinson2021a}.\label{phirho2}}
\end{figure}

Figure \ref{phirho2} shows an example.
The color contours in (a) indicate the the trapped electron density
deficit $\tilde n$ relative to a flat plateau distribution. The
passing density is set by the background velocity distribution (and
orbit energy conservation). The lines are logarithmic potential
contours. The radial and parallel positions are on identical
scales. Notice the equal spacing of the contour lines all the way round,
for distances beyond the colored region. It arises from the
predominantly passing particle Debye shielding giving effectively a
modified Helmholtz equation there with constant
$\nabla^2\phi/\phi=1/\lambda_d^2$. In the inner regions, the line
contours do not follow the color contours, because the trapped density
is a function of $r$ as well as of $\phi$. This was found to be a
general property of axisymmetric holes of limited transverse
size. Velocity distributions that depend only on $\phi$ are
self-consistent only when exactly spherical, even when limited to
purely parallel motion, demonstrating that the proposed equilibria of
\citep{Krasovsky2004a} are unphysical.

Satellite observations of electron holes generally consist of
measurements of the electric field components along a passage through
the hole in approximately the parallel direction at constant $r$,
because of the hole's high parallel velocity. Frequently the ratio
of the peak electric field in the perpendicular and parallel
directions is taken as indicative of the ratio of the hole extents in
the parallel and perpendicular directions; in other words it is
assumed $L_\parallel/L_\perp \simeq E_\perp/E_\parallel$. Although
this measure is heuristically plausible, the actual observed peak
field ratio depends on the radius of the satellite track. Figure
\ref{phirho2}(b) documents the observational variation for the oblate
hole of \ref{phirho2}(a), of the peak $E_\perp$ and $E_\parallel$, and
their ratio, as a function of track radius. The typical contour aspect
ratio, $L_\parallel/L_\perp$ is approximately $1\over 3$, but the
corresponding observed field ratio varies between zero and
approximately 0.6 in the strong field region and rises to values
substantially greater than 1 in the radial wings. This disappointingly
wide range of uncertainty in the relationship between peak $E$
ratio and hole potential shape shows that caution is required in
interpreting the field ratio; and it re-emphasizes the importance of
multi-satellite simultaneous observations of holes in order to
determine their shape.

\citet{Tong2018} report such simultaneous measurements in MMS data
from the PSBL, where all four satellites encounter the hole. Assuming
the gradients of electric field are constant over the satellite
tetrahedron, they can be obtained from the electric field
measurements, and provide an estimate of the divergence
$\nabla\cdot\E$, which is the charge density. The observed charge
densities are either tri-polar (-+-), or purely negative, consistent
with passing respectively through either the central or wing regions
of an electron hole. The hole shape is estimated by fitting a theoretical
potential profile of the form
\begin{equation}
  \label{tongfit}
  \phi=\psi\exp([z-z_0]^2/2d_\parallel^2){\cal H}({\cal R})
\end{equation}
where $z_0=v_ht$ is the moving parallel potential peak, and
${\cal R}^2=(x-x_0)^2/d_{min}^2+(y-y_0)^2/d_{max}^2$, is constant on
ellipses with a single center $x_0,y_0$, in the transverse plane,
whose minor and major axes $x$ and $y$ have been rotated to minimize
the difference between the predicted and observed field variation at
the four satellites. The most successful of the radial dependencies
explored was found to be ${\cal H}=\exp(-{\cal R}^2/2)$.  Of the
dozens of holes examined, four were able to be adequately fitted in
this way, and many others were not, for reasons not entirely
explained. Two of the fits were close to axisymmetry:
$d_{max}/d_{min}\lesssim 1.3$; but two were not: $d_{max}/d_{min}\simeq
2$ and 3. The oblateness of the fitted holes, $d_{min}/d_\parallel$ was in
the range 1 - 2, which is consistent with the empirical scaling of
\citet{Franz2000}, eq.\ (\ref{Franz}).

The study of \citet{Steinvall2019} instead fitted an axisymmetric
model $\phi=\psi\exp(-z^2/2l_\parallel^2-r^2/2l_\perp^2)$ of hole
potential to 4-satellite MMS observations of 236 slow electron holes
at the magnetopause. The $E_\perp/E_\parallel$ is small, predominantly
between 0.1 and 0.5, which is consistent with the ratio
$\Omega/\omega_p$ ($\sim$0.1 to 0.2) being small. But the scaling of
eq.\ (\ref{Franz}) is not reproduced within the data. The model
implies $l_\perp=\sqrt{\psi r/E_\perp}$, which can be used to deduce
values of $l_\perp$ if the hole center (and satellite position) is
known giving $r$. But only for 10 holes was a consistent hole center
identified, and the deduced $l_\perp/l_\parallel$ for them varied from
$<2$ to $\sim6$ without identified cause. Several instrumental and
noise factors were cited as possible reasons for the inconsistency of
the model in the rest of the cases, as well as the possibility of
non-axisymmetric hole shapes.

The observational evidence concerning hole shapes seems therefore to
indicate holes have considerable variety. Weak magnetic field strength
favors oblate electron holes, and strong field seems to permit
spherical or even prolate holes. But in nature electron holes often
seem not to be axisymmetric, and it is unclear what determines how
close they are to axisymmetry.

\subsection{Gyro orbit effects and Scattering}

Finite electron gyro-radius implies that electrons in holes do not
move purely along the magnetic field. In a spatially one-dimensional
situation (uniform in the $x$ and $y$ directions) no new effects enter
into the \emph{equilibrium}, though translational symmetry can be
broken by transverse instabilities.  There are, however, two
indubitable effects of finite gyro-radius on hole equilibria of
limited transverse size. One is an effective transverse averaging over
the gyro-radius of the parallel potential gradient, which impedes
electron trapping when the transverse hole size is small. The other is that
resonant coupling between the electron bounce frequency and the
gyro-frequency in the presence of perpendicular electric field causes
energy transfer between parallel and perpendicular electron energy
(even though the total kinetic plus potential energy is exactly
conserved). This transfer makes the parallel energy increase or
decrease, and can thereby detrap particles with negative parallel
energy near zero, and trap passing particles with small positive
parallel energy. A third proposed mechanism is sometimes attributed to
a supposed gyrokinetic modification of the effective perpendicular
plasma dielectric response; this last mechanism does not actually
apply to phase-space holes, and references to it are misleading.

The errors involved in invoking a supposed ``gyrokinetic'' dielectric
response are explained in \citet{Hutchinson2021}. The original
speculation by \citet{Franz2000} (often cited thereafter\citep{Jovanovic2002a,Berthomier2003,Wu2010,Vasko2017,Vasko2018,Holmes2018,Tong2018,Fu2020}) was that
their observations fitted by equation \ref{Franz} could be explained
by a modified Poisson equation of the form
\begin{equation}
  \left[\nabla_\parallel^2+(1+{\omega_{pe}^2\over\Omega^2})\nabla_\perp^2\right]
  \phi = (n_e-n_0),
  \label{gkpermittivity}
\end{equation}
which they attributed to gyrokinetic theory. However, the extra term
$\omega_{pe}^2/\Omega^2$ arises in the equations which they cite in a
more complicated way than a simple polarization drift. The actual
gyrokinetic expression for transverse permittivity is
$\epsilon_\perp = 1+ [1-\Gamma_0(\zeta_t^2)]/(k_\perp^2\lambda_{De}^2)$,
where the finite gyro-radius parameter is
$\zeta_t=k_\perp\sqrt{T_\perp}/\Omega=k_\perp r_L$, and
$\Gamma_0(\zeta_t^2)={\rm e}^{-\zeta_t^2} I_0(\zeta_t^2)$. It reduces
to approximately $\epsilon_\perp = 1+\omega_{pe}^2/\Omega^2$ only in
the limit that both $\zeta_t^2\ll 1$ and $k_\parallel/k_\perp \ll
1$. But neither of these parameters is small in an electron hole for
which $\omega_{pe}^2/\Omega^2$ is not small. So adopting equation
\ref{gkpermittivity} for electron holes is erroneous when
$\epsilon_\perp$ is significantly different from unity; it is an abuse
of gyrokinetic equations. Quite apart from that, the effect of a
steady electrostatic structure on the density of the attracted charge
species is a long-studied plasma problem related to probe theory
\citep{Alpert1965,Laframboise1966a,Laframboise1993,Lampe2001,Hutchinson2007a}
in which no such anisotropic shielding is introduced. The
only difference with an electron hole is the absence of particle
absorption, which is a simplification. The paper
\citep{Hutchinson2021} reports a collisionless PIC simulation of a
spherical potential structure with a wide range of applied magnetic
fields, demonstrating that there is no anisotropy in the shielding
that would be implied by equation \ref{gkpermittivity}. Equation
\ref{Franz} may be empirically useful, but the gyrokinetic shielding
explanation of it is incorrect.

\begin{figure}[ht]
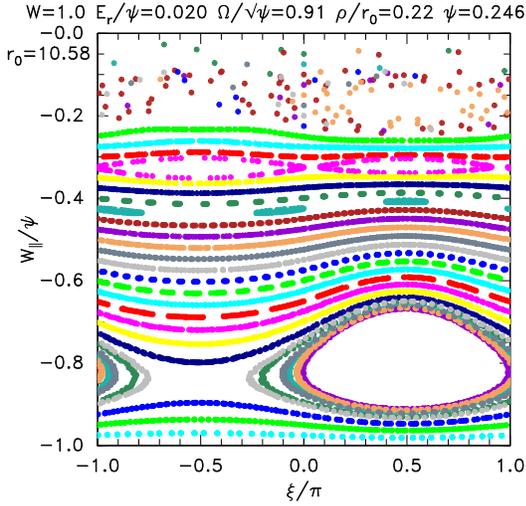
\center
  \partwidth{0.5}{stochasticorbits}
  \caption{Poincar\'e plot in the space of parallel energy versus
    gyro-phase of electron orbit evolution in a transverse electric
    field. From \citet{Hutchinson2020}.
    \label{stochasticorb}}
\end{figure}

The resonant trapping/detrapping phenomenon is explored in
\citet{Hutchinson2020}, by numerical orbit integration and by analytic
calculation of the resonant island widths in an axisymmetric
($\phi=\phi(r,z)$) electron hole. These two approaches are found to
agree remarkably well. Figure \ref{stochasticorb} shows a so-called
Poincar\'e plot of numerical orbits.
Many orbits having different starting parallel energy are
followed. Each is a
specific color, and a point is plotted on this graph each time the
orbit passes $z=0$, corresponding to its parallel energy
$W_\parallel=\energy_\parallel=v_\parallel^2/2+q\phi$ and the phase of
the gyro-motion $\xi=\tan^{-1}(v_r/v_\perp)$ as it does so. This
example has starting radius $r_0=10.58$, at which the hole equilibrium
has a chosen radial electric field $E_r=-d\phi/dr=0.02\psi$. The
potential peak at this radius is $\psi=\phi(r_0,0)=0.246$, and the
magnetic field strength is $\Omega=0.91\psi$. The total particle
energy, which is perfectly conserved, is $W=1$ and mostly consists of
perpendicular kinetic energy $v_\perp^2/2$. The resulting gyro radius
is $\rho=0.22r_0$. Orbits at the more negative $W_\parallel/\psi$,
from about -0.23 downward to the most negative possible value -1, are
observed to follow stable tracks. The open tracks correspond to
non-resonant orbits, moderately perturbed in $W_\parallel$ by transfer
to and from $W_\perp$, but never reversing their direction of phase
advance. The closed tracks surround resonances where the gyro frequency
equals an even harmonic of the bounce frequency. The resonance is
sufficiently strong to produce an island in the Poincar\'e plot on the
tracks of which the phases of bounce and gyration remain within a
limited region. At some discrete parallel starting energy there is a
separatrix between the open and closed tracks (for example the upper
pink track at $W_\parallel/\psi\simeq -0.35$) with one or more
x-points. The large island centered about $W_\parallel/\psi=-0.81$ is
the second harmonic ($m=2$) resonance, and the one at -0.35 is the
fourth ($m=4$). There are many other higher harmonic resonances, but
all those above $m=4$ are not traced out by the plot, because overlap
of successively more closely spaced islands for $m\ge 6$ causes the
orbits to become stochastic above about -0.23.  They wander
quasi-randomly in $W_\parallel$ until they become untrapped when it
becomes greater than zero. There are other orbits that start untrapped
but wander in to become trapped for a time. They can be considered to
be the illustrated by the same stochastic orbits only followed backwards
in time (by time reversal symmetry).

Thus, for a hole that is limited in transverse extent, and therefore
experiences a perpendicular field $E_r$ like this, only fairly deeply
trapped orbits are permanently trapped. The stochastic orbits stay
trapped only for a limited number of bounces before escaping the hole
as passing orbits. The anticipated consequence is that the stochastic
region has approximately constant phase space density, equal to the
phase space density of the low-energy background orbits. This
observation motivates the flat edge distribution functions of figures
\ref{c2scan} and \ref{pjscan}.  In summary, there is theoretically
always a stochastic region of only temporarily trapped orbits near the
trapping energy boundary $\energy=0$, but when $\Omega/\sqrt{\psi}>2$
a large fraction of orbits are permanently trapped. In contrast if
$\Omega/\sqrt{\psi}<1$ a large fraction are not permanent unless the
transverse scale length $\psi/E_\perp$ is rather large. The paper
\citep{Hutchinson2020} gives more quantitative description of the
trapping and detrapping criteria.  Unfortunately there seems to be, as
yet, no published experimental evidence supporting these conclusions.

Setting aside the erroneous direct invocation of a supposed
gyrokinetic permittivity explained above, the two valid
multidimensional effects discussed so far are (1) transverse field
divergence, and (2) resonant parallel energy changes leading to
stochastic orbits. There is a third, namely (3) (non-resonant)
gyro-averaging over transverse orbit.  This effect is treated for
self-consistent axisymmetric holes in \citet{Hutchinson2021b}. It
involves orbits that cannot be treated as localized in the transverse
direction because their gyro-radius is comparable to the transverse
length scale of the potential variation. This is where some of the
techniques of gyrokinetics become relevant. Of course the resonant
orbit stochasticization occurs simultaneously and its effects need to
be combined with gyro-averaging. The theory shows that gyro-averaging
and perpendicular velocity integration over a Gaussian perpendicular
velocity distribution, are together equivalent to convolving the
quantity of interest with a 2-dimensional Gaussian of width equal to
the thermal gyro-radius $r_{gt}=v_t/\Omega$ in the perpendicular
plane. The transverse orbit-averaged quantity is therefore
\begin{equation}
  \bar\phi(\x_c)=\int\phi(\x_c+\r_g){\rm
    e}^{-(\r_g/r_{gt})/2}{d^2r_g\over2\pi r_{gt}^2},
\end{equation}
where $\x_c$ is the guiding center, $\r_g$ is the (2-D) gyro-radius
vector from it to the particle position $\x_c+\r_g$, and over-bar denotes the
gyro-average of whatever quantity is under discussion ($\phi$
here). Then, because a further transverse gyro-averaging occurs in the
transformation from guiding-center density to particle density (which
determines charge density), the effective potential $\bar\phi(r,z)$
accounting for finite gyro-radius is smoothed by two equal Gaussian
transverse convolutions relative to what would be obtained for the
same trapped particle deficit without accounting for gyro-radius
(i.e.\ at high magnetic field). (Two Gaussian convolutions are the
same as one with width $\sqrt{2}r_{gt}$.)

Direct numerical orbit integration validates the intuitive idea that
parallel trapping is determined mostly by the gyro-averaged parallel
energy $\bar W_\parallel=\bar{v_\parallel^2}/2+\bar\phi$ except that
resonant orbit stochasticization can detrap the orbits near
$W_\parallel=0$.
\begin{figure}[ht]
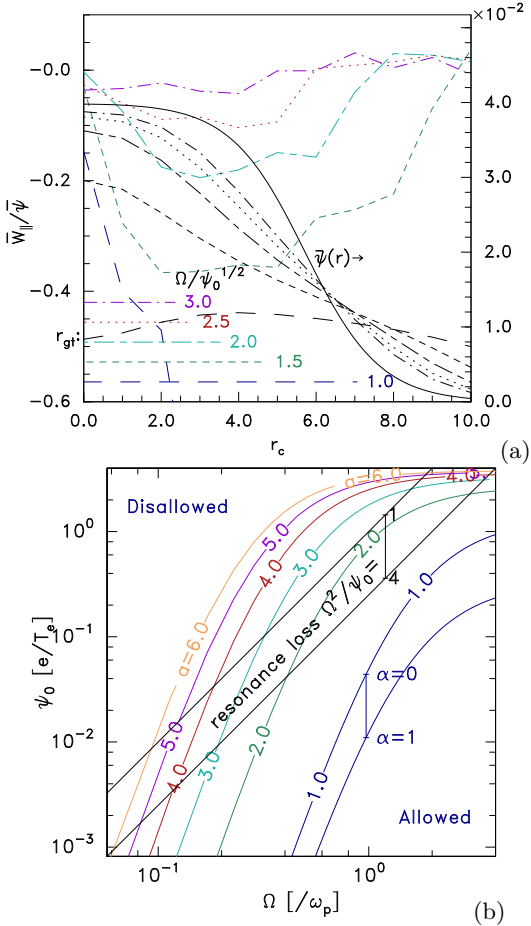

  \partwidth{0.5}{wpv.04}\hskip-1em(a)
  \partwidth{0.45}{psiomegalim}\hskip-1em(b)
  \caption{(a) Gyro-averaged potential $\bar\psi$ and lowest
    permanently trapped orbit $\bar W_\parallel/\bar\psi$ versus
    guiding center radius for a range of $\Omega$. (b) Maximum
    allowable hole amplitude $\psi_0$ permitted by distribution
    non-negativity (colors) and resonant detrapping (black).
    From \citet{Hutchinson2021b}. \label{gyroholes}}
\end{figure}

Figure \ref{gyroholes}(a) shows in black on the right-hand scale the
gyroaveraged peak potential profile $\bar\psi(r)=\bar\phi(0,r)$ for a
particular $\psi(r)$ profile having $\psi_0=0.04$. Different line
styles are for different magnetic field strengths
$\Omega/\psi_0^{1/2}$ and hence gyro-radii; the solid line is
unaveraged $\psi(r)$. The gyro-averaging effect is always to lower the
$\bar\psi(r)$ peaks and raise the wings, although the effect is
relatively small for $\Omega/\psi_0^{1/2}>2$. The colored lines
instead show the maximum value of $\bar W/\bar\psi$ for which the
orbits are permanently trapped, using the same line style notation. At the
highest $\Omega$s all orbits are permanent except for those close to
$\bar W=0$, but at the lowest field, orbits are almost entirely
detrapped.

Figure \ref{gyroholes}(b) gathers these observations to plot lines
above which axisymmetric holes are disallowed either because they
violate the requirement of non-negative $f(v)$ (colored) or because
they are seriously depleted by resonant detrapping (black). The
non-negativity constraint is made more stringent by gyro-averaging
because of the smoothing of the effective (gyro-averaged)
potential. Quantitatively that depends on the hole's effective
transverse size $a$, defined as
$2a^2=\langle r^2\rangle= \int_0^\infty
r^2\psi(r)rdr/\int_0^\infty\psi(r)rdr$, and the lines are labelled
with the value of $a$. There is some dependence also on the form
assumed for the parallel distribution deficit, which is controlled by
the parameter $\alpha$. The most forgiving value, $\alpha=0$,
corresponds to a waterbag shaped $\tilde f$ deficit and all lines use
this except the lowest: $a=1$, $\alpha=1$, which illustrates the
approximate uncertainty introduced by parallel hole shape. Resonant
orbit detrapping is represented by the straight diagonal lines for low and high
values of $\Omega^2/\psi_0$. The upper bound lies somewhere between
them. We observe that at low values of $\Omega$ (in units of
$\omega_p$) only very shallow ($\psi_0<10^{-2}$) holes are permitted,
and cannot have too small a transverse size $a$. For $\Omega\sim0.5$
though, resonant detrapping is by far the more limiting factor unless
$a\lesssim 2$. For $\Omega\gtrsim 1$ transverse size is hardly limited
and prolate holes are permitted by the criteria investigated.

A different consequence of finite transverse size of electron or ion
holes is that they can scatter passing electrons in velocity space.
Viewed in the frame of the steady hole, this process does not change
the total particle energy, but it transfers the kinetic energy between
parallel and perpendicular velocities when there is a perpendicular
component of the hole electric field (transverse potential
inhomogeneity). This transfer can be calculated by finding the total
perpendicular work done on the particle, like the corresponding
calculation for trapped particles, except considering a single hole
transit, rather than periodic bouncing. The transfer is largest when
the gyration of the transverse velocity is by approximately half a
period during the transit. That is, $\Omega \tau_t\sim \pi$, where
$\tau_t\sim L_\parallel /v_\parallel$ is the (parallel) transit
time. Depending on the gyrophase of the encounter, the transfer to
$\energy_\perp$ is either positive or negative. In contrast, if
$\Omega \tau_t\gg \pi$, then the $\v_\perp (.\E_\perp)$ power
oscillations during transit average to a small total; while conversely
$\Omega \tau_t\ll \pi$ implies the encounter is very brief because
$v_\parallel \gg L_\parallel\Omega$; so the total transfer is
small. \citet{Vasko2017c} calculate the mean square energy transfer in
a single encounter with an axisymmetric Gaussian hole of parallel and
perpendicular scales $L_\parallel$ and $L_\perp$, and deduce the
velocity diffusion coefficient as a function of particle energy and
pitch angle (i.e. of position in velocity space), when encountering a
randomly positioned assembly of equal holes. For their application to
understanding particle precipitation by scattering into the loss-cone
of a magnetic mirror trapping region, they also average over mirror
bounce motion and spatial variation. They conclude that pitch angle
scattering by electron holes can be comparable to that of ``chorus
waves'' (another leading mechanism) in the outer radiation belt. In a
follow-up paper, \citet{Vasko2018} show that their calculation of
scattering from holes is consistent with the classic quasi-linear
rates of \citet{Kennel1966}. And in a more recent calculation
\citet{Shen2021} show how allowing a broad range of hole amplitudes
broadens the extent in phase-space of the diffusivity. They give
diffusion rate expressions in terms of the spatially averaged spectrum
$\E(\k,\omega)$ of the electric field fluctuations, which satellite
observations may be able to determine more easily than trying to
determine the density, size, and shape of electron holes.

\subsection{Magnetic field perturbations}
\label{MagPert}

Ideal electron holes, viewed in the frame in which they are
stationary, are static electromagnetic structures. A guiding
assumption of the present review is that the electric field is
predominantly responsible for localizing them (at least in the
direction of any background magnetic field). In other words they are
predominantly electrostatic and the background magnetic field can be
considered uniform. Although such an electrostatic treatment is a good
starting approximation for virtually all holes that have been
observed, holes also in fact cause local perturbations to the magnetic
field, which are sometimes observable.

As has been mentioned in subsection \ref{labspace}, the earliest
observations of magnetic perturbations caused by electron holes were
attributed to a simple Lorentz transformation of the hole's electric
field from the hole frame to the frame of the satellite measuring the
electric and magnetic
fields\citep{Ergun1998a,Ergun1998,Ergun1999}. And only
later\citep{Andersson2009} did it become clear that there are parallel
magnetic field perturbations, which cannot be attributed to Lorentz
transform, and must therefore exist in the hole frame. The mechanism
proposed for enhancement of the magnetic field strength within the
hole was electric current attributable to $\E\wedge\B$ drift of electrons, and
this was given good theoretical grounding in \citep{Tao2011}. Related
ideas were proposed in \citep{Treumann2012}.  \citet{Vasko2015}
observed in data from the Van Allen probes the unusual effect instead of
reduction of $|B|$ within some holes in the outer radiation belt, and
interpreted the cause as being diamagnetic currents from energetic
anisotropic trapped electron distributions. A recent additional
theoretical development is the inclusion of the polarization drift of
electrons\citep{Yang2023}, which changes mostly the perpendicular
magnetic perturbation. In this subsection a summary
of magnetic perturbation theory for an axisymmetric
electron hole in a uniform magnetic background field is given, working
in the hole frame in dimensional (SI) units.

First, one should note that a purely one-dimensional potential
structure, in which $\E=-\hat\z d\phi/dz$ is parallel to $\B$, gives
zero magnetic field perturbation because it possesses no transverse
electric field or current. A finite electron hole transverse radius, however,
implies an $\E_\perp$ ($=\E_r$ positive) and an electron drift
electric current density
$\j_{E\wedge B}=q_en_e\E_r\wedge\B/B^2\ (=-q_en_e{\partial \phi\over
  \partial r}\hat\r\wedge\B/B^2$) in the $+\hat\btheta$ direction (for
right-handed $r,\theta,z$ coordinates). Diamagnetic current arises
from electron pressure gradient, contributing an additional current
density $\j_{\nabla p}=\nabla_\perp p_{e\perp}/B$.  If the hole is much longer
than its radial extent (highly prolate) so we can approximate it as an
infinite length cylinder, then, by Ampere's law, the azimuthal current
density generates a local magnetic field gradient
$\partial B_z/\partial r=-\mu_0j_{e\theta}$, implying that $B_z$
increases toward the hole center, that is, the current
\emph{increases} the central values of $|B_z|$. Applying these
expressions to a radial path we get
\begin{equation}\label{omegazdrift}
  {\partial\over \partial r} {B_z^2\over2\mu_0} =en_e {\partial \phi\over
    \partial r} -\nabla p_{e\perp}
\end{equation}
Integration ignoring any $n_e$ variation in the first term on the
right then yields
$\delta B_z B/\mu_o\simeq
[B(r)^2-B(\infty)^2]/2\mu_0=en_e\phi(r)-[p_{e\perp}(r)-p_{e\perp}(\infty)]=en_e\phi-\delta
p_\perp$, which can be recognized as radial force balance between
magnetic pressure, electric field and kinetic pressure. Inclusion of
diamagnetic current (arising from $\delta p_\perp$) in subsequent
expressions where it is omitted simply requires $\phi$ to be replaced
with $\phi-\delta p_\perp/en_e$.

The ion response is also neglected, which is
justified by ions' heavier mass when the hole is moving sufficiently
faster than the ion velocity distribution.
\citet{Tao2011} demonstrate
by numerical orbit integration that ignoring ion's $\E\wedge\B$ drift
current is well justified. And they observe that for a separable
Gaussian hole shape
[$  \phi(r,z)=\psi\exp(-r^2/2l_\perp-z^2/2l_\parallel)$],
the (linearized) parallel magnetic field perturbation at the origin
ignoring diamagnetic current
can be written
\begin{equation}
  \begin{split}
  \delta B_z(r=0,z=0) &= (\mu_0n_ee\psi/B)  g(l_\perp/l_\parallel)
  \pamper= B(\hat\psi/\hat\Omega)(T/m_ec^2) g(l_\perp/l_\parallel)
  \end{split}
\end{equation}
where $(\hat\psi/\hat\Omega)$ are in our normalized hole units. The effects of
finite aspect ratio $l_\perp/l_\parallel$ are expressed by the
function
\begin{equation}
  g(s)= \Re\left\{[1-s^2+s^2\cos^{-1}(1/s)\sqrt{s^2-1}]/(1-s^2)^2\right\},
\end{equation}
which is unity at $s\to 0$, and
which they plot.


\citet{Yang2023} calculate the effect of electron polarization drift,
arising from the time dependence of $E_r$ experienced by electrons
moving through the hole. It gives rise to an additional $B_\theta$
perturbation as follows.  A closed contour $C$ at $r=r_c, z=z_c$
(where $r_c$ and $z_c$ are constants) is spanned by the surface $S$:
$z=z_c, r\le r_c$. Applying Stokes' theorem to the steady Ampere's
law, using cylindrical symmetry we have the standard result
$B_\theta(r_c)=\mu_0I(r_c,z_c)/2\pi r_c$ where $I$ is the total
electric current crossing $S$.  The electric current density is
$\j=q_e(n_e\bar\v_{e}-n_i\bar\v_{i})$, where $\bar\v_e$ and $\bar\v_i$
are averages over the respective distribution functions.  Remove any
background field gradients by supposing that far away from the hole
there is no current density
$n_{e\infty} \bar\v_{e\infty}-n_{i\infty} \bar\v_{i\infty}=0$. And
again suppose that ions are sufficiently heavy and fast moving (in the
hole frame) that their density and velocity is unperturbed by the
hole: $n_{i} \bar\v_{i}=n_{i\infty} \bar\v_{i\infty}$. Then
$\j=\j_e-\j_{e\infty}=q_e(n_e\bar\v_{e}-n_{e\infty}\bar\v_{e\infty})$.
The electron guiding-center polarization drift is cumulative, such
that it results in a radial orbit displacement
$\delta r = {m_e\over q_e B^2} E_r=q_eE_r/m_e\Omega^2$, which is negative, that
is inward. Consequently the radial displacement of the guiding-center
is directly proportional to the radial electric field (and zero at
infinity, since $E_r=0$ there). Inclusion of just the polarization
drift therefore causes all the orbits that at infinity have
guiding-centers inside the radius $r_c-\delta r$ to pass through the
surface $S$ of radius $r_c$ at $z_c$, giving rise at lowest order in
$\delta r/r_c$ to a change in parallel current
$\delta I_{v_p}(r_c)= -2\pi r_c\delta r j_{e\infty}=-2\pi r_c (m_eE_r/q_eB^2)j_{e\infty}$ ($\delta r$ is
negative). And hence
\begin{equation}
  \label{eq:Btheta}
  \delta B_{\theta v_p}(r_c)
  = {\partial \phi\over\partial r} {\mu_0\over B^2} m_e n_{e}\bar
      v_{e\infty}. 
\end{equation}

In addition to the polarization drift, however, radial motion
occurs because of a magnetic field $B_r\not=0$, arising from
$\partial B_z/\partial z$ via the drift-current mechanism equation
\ref{omegazdrift} and
$\nabla\cdot\B
=0$. Parallel guiding-center motion therefore gives an additional
parallel current change $\delta I_{B_r}$ crossing $S$ proportional to
the change of the magnetic flux linked by the contour of radius
$r_c$. That is
\begin{equation}
  \delta I_{B_r}(r_c)=  j_{e\infty}\int_0^{r_c}(\delta B_z/B) 2\pi r dr,
\end{equation}
and to lowest order in prolate holes
\begin{equation}
  \delta I_{B_r}(r_c)= \int_0^{r_c} {2\mu_o\over
    B^2}en_e\phi  {r dr} q_en_e\bar v_{e\infty}.
\end{equation}
In the prolate approximation one quickly finds
\begin{equation}
  {\delta I_{Br}\over \delta I_{v_p}} \simeq {\omega_{pe}^2\over c^2}
  {\int_0^{r_c}\phi rdr\over r_c\partial\phi/\partial r|_{r_c}}
  \sim {\omega_{pe}^2r_c^2\over c^2}={T\over mc^2}{r_c^2\over\lambda_{De}^2}, 
\end{equation}
which is small unless temperatures are relativistic or
$r_c/\lambda_{De}$ is large. So ignoring the $B_r$ effect on
$\delta I_z$ and hence on $B_\theta$ is generally a good
approximation.

For the above drift analysis to be quantitatively correct, however,
holes of typical parallel length $\lambda_{De}$ require
$\omega_{pe}/\Omega\lesssim 1$; otherwise thermal electrons transit
the hole in less than one gyroperiod and more complicated
analysis is required. The approximate effect is that the radial
displacement is determined by inertial electric acceleration during the
transit, and becomes $\delta r \sim {q_eE_r\over m_e\omega_{pe}^2}$ rather
than ${q_eE_r\over m_e\Omega^2}$, which might be heuristically approximated
as $\delta r \sim q_eE_r/m_e(\Omega^2+\omega_{pe}^2)$. In any case,
evaluating the field perturbations everywhere for arbitrary aspect
ratio requires a full numerical calculation, even for separable
Gaussian potential shapes.

What is observed by the satellite in its rest frame is given by the
Lorentz transform of the hole-frame $\E$ and $\B$. In the linearized
approximation, it simply gives an additional azimuthal field
$\delta \B_{Lorentz}=\hat{\bm\theta}\delta B_{\theta Lorentz}=\hat{\bm\theta}
v_hE_r/c^2$, where $v_h$ is the parallel velocity of the hole in the
satellite frame. This azimuthal field proportional to $E_r$ is the
same shape as that of the polarization drift and their relationship
for small $\delta r/r$ ignoring diamagnetic current can be written
\begin{equation}
  \delta B_{\theta v_p} \simeq {\omega_{pe}^2\over \Omega^2}\delta B_{\theta Lorentz},
\end{equation}
(with $\Omega^2\to \sim(\Omega^2+\omega_{pe}^2)$ at low $\Omega/\omega_{pe}$). 
Estimates of hole speed obtained by assuming $B_\theta$ arises
purely from Lorentz transformation, may therefore be misleading when
$\omega_{pe}/\Omega$ is not small.

\citet{Steinvall2019} analyzed MMS data accounting for the electron
$\E\wedge\B$ drift and Lorentz transform effects but not polarization
drift, using all four spacecraft to give a 3-dimensional picture. They
measured a discrepancy in the magnetic field relative to their
Gaussian model, for high speed and amplitude holes. It consisted of an
oscillatory perpendicular component with right-handed polarization,
which they interpret as a damped whistler wave excited by the hole
through a Cerenkov mechanism previously analyzed and observed in
reconnection simulations by \citet{Goldman2014}.

There is now growing observational evidence of the local excitation of
electron holes and related electron vorticity in space plasma magnetic
reconnection (see e.g.\ \citet{Ahmadi2022}), and in reconnection
simulations (see e.g.\ \citet{Chang2022}). So magnetic field influence
on holes and hole influence on magnetic fields seems to be a topic of
increasing relevance.

\section{Prospects}

As this review has sought to convey, solitary potential structures are
a major nonlinear, coherent, constituent of many dynamic plasma
systems. They are not naturally analyzed by spatial Fourier
decomposition, and in that sense are more like the individual water
waves on a beach than the spatially periodic electromagnetic waves
used for communications or plasma heating. Our understanding of them
has developed somewhat episodically over the past sixty years, as theory,
observation, experiment, and computer simulation have all made
essential contributions. Recent progress has been remarkable, enabled
and motivated by increasingly sophisticated satellite measurements,
powerful high-performance simulation, new insights, and rigorous
analysis. Many opportunities for progress still remain, and some brief
speculative closing remarks about them are perhaps appropriate.

Electron and ion holes are anticipated to occur in situations with
kinetically unstable velocity distributions. That usually means, for
predominantly electrostatic phenomena, distributions with multiple
peaks. However, we so far have very little systematic understanding of
just how the distribution shapes relate to the parameters of generated
phase-space holes and to their long term fate. There might be much to
be learned from systematic simulation studies (going beyond the
individual examples that tend to dominate the literature) to explore
how initializing background distribution details influence the
coherent outcome, and how the velocity distributions subsequent to
instability relate to hole formation and merging processes. It is
already clear that holes can have a wide range of parallel
lengths. Therefore the extensive recent transverse stability
calculations assuming $\sech^4(z/4\lambda)$ potential shape are just a
particular choice. It would be of considerable interest to know how
hole length influences transverse stability. 

Multisatellite observations have only begun to measure the
three-dimensional shapes of electron and ion holes. There is
undoubtedly already archived data that might bring additional clarity
to the situations where holes are or are not approximately
axisymmetric, and what determines their aspect ratio
(oblateness). Future missions involving larger teams of satellites
would likely give much more comprehensive information on holes'
spatial form, from fast measurements of all three electric field
components. But managing and mining that data will require
sophisticated automated algorithms to extract it.

What electron and ion holes do to the plasma is still an area of great
uncertainty. For example, are electron holes a major player in
determining particle scattering, precipitation, reconnection rates,
anomalous resistivity, obstacle wake behavior, and so on? Or are they
mostly interesting markers, of value for diagnosing the progress of
other more dominant controlling factors? We mostly don't know.

Holes have been deliberately generated in laboratory experiments, and
observed to accompany reconnection. But how prevalent are they in
terrestrial plasmas? The uncertainty about this question arises in
large part from the difficulty of Debye-length scale measurements; can
new high-resolution diagnostics establish the degree of prevalence of
coherent solitary structures, or can basic plasma experiments of
longer Debye-length observe and measure them in a controlled way?

In closing, one might ask if there are engineering applications of
phase-space holes to accomplish human tasks. To my knowledge, not so
far. That fact probably limits the effort likely to be expended in
studying them. However, they may substantially impact human activities
in space, and they certainly help to explain a lot of what satellites
observe in space plasmas. In any case, the intellectual challenge and
adventure which lies behind all of natural science remains great in an
area of nonlinear physics like this. New knowledge and understanding
is to some degree its own reward. Solitary plasma phase-space
structures remain a topic open to new discoveries in the challenging
area of nonlinear physics.

\section*{Acknowledgments}

I am greatly indebted to my students Christian Haakonsen, Chuteng Zou,
Xiang Chen, and collaborator David Malaspina for the pleasure of
working closely with them over the years to develop a deep
understanding of electron holes. Ivan Vasko and his students have been
a source of inspiration and world-wide leadership in interpreting
current satellite data concerning electron and ion holes. I thank them
for sharing their work in progress and welcoming occasional
theoretical input from me. Ivan Vasko read the present MS in draft and
I thank him very much for this substantial help.

No external financial support was received for the writing of this
article, and no data were produced.

\bibliography{JabRef}

\end{document}